\documentclass[printer]{aa}  

\pdfoutput=1

\usepackage{graphicx}
\usepackage{txfonts}
\usepackage{hyperref}
\usepackage{amsmath} 
\usepackage{booktabs} 
\usepackage{lscape}
\usepackage{xcolor} 
\usepackage{natbib}
\bibpunct{(}{)}{;}{a}{}{,} 
\begin{document}

   \title{Angular momentum and chemical transport\\by azimuthal magnetorotational instability\\in radiative stellar interiors}

   \author{Domenico G.~Meduri$^{1,2}$
          \and
          Laur\`ene Jouve$^{1}$
          \and
          Fran\c{c}ois Ligni\`eres$^{1}$
          }

   \institute{$^1 $Universit\'e de Toulouse, CNRS, Institut de Recherche en Astrophysique et Plan\'etologie (IRAP), 14 Avenue Edouard Belin, 31400 Toulouse, France\\
                 $^2 $Leibniz-Institut f\"{u}r Astrophysik Potsdam (AIP), An der Sternwarte 16, D-14482, Potsdam, Germany\\
              \email{domenico.meduri@irap.omp.eu}
             }

   \date{Received 07 08 2023 / Accepted XX Y ZZZZ}
  \abstract
   {The transport of angular momentum and chemical elements
   within evolving stars remains poorly understood.
   Asteroseismic and spectroscopic observations
   of low mass main sequence stars and red giants
   reveal that their radiative cores rotate orders of
   magnitude slower than classical predictions from stellar evolution models
   and that their surface light elements abundances are too small.
   Magnetohydrodynamic turbulence is considered a primary mechanism
   to enhance the transport in radiative stellar interiors but its efficiency
   is still largely uncertain.
   }
   {
   We explore the transport of angular momentum and chemical elements
   due to azimuthal magnetorotational instability,
   one of the dominant instabilities expected in differentially rotating
   radiative stellar interiors.}
   {We employ 3D MHD direct numerical simulations in a spherical shell of
   unstratified and stably stratified flows under the Boussinesq approximation.
   The background differential rotation is maintained
   by a volumetric body force.
   We examine the transport of chemical elements
   using a passive scalar.}
   {We provide evidence of magnetorotational instability
   for purely azimuthal magnetic fields
   in the parameter regime expected from local and global
   linear stability analyses.
   Without stratification and when
   the Reynolds number Re and the background
   azimuthal field strength are large enough,
   we observe dynamo action
   driven by the instability at values of
   the magnetic Prandtl number Pm in the range $0.6-1$,
   the smallest ever reported in a global setup.
   When considering stable stratification at $\text{Pm}=1$, the turbulence
   is transitional and becomes less homogeneous and
   isotropic upon increasing buoyancy effects.
   The transport of angular momentum occurs radially outwards and
   is dominated by the Maxwell stresses when stratification is large enough.
   We find that the turbulent viscosity decreases when buoyancy effects
   strengthen and scales with the square root of the ratio
   of the reference rotation rate $\Omega_\text{a}$ to the Brunt-V\"{a}is\"{a}l\"{a}
   frequency $N$.
   The unstratified runs also suggest that the turbulent viscosity
   grows faster with Re than with Pm.
   The chemical turbulent diffusion coefficient scales
   with stratification similarly to the turbulent viscosity but is lower
   in amplitude so that the
   transport of chemicals is slower than the one of
   angular momentum, in agreement with recent stellar evolution
   models of low mass stars.
   }
   {We show that the transport induced by azimuthal magnetorotational
   instability scales relatively slow with stratification
   and may enforce rigid rotations of red giant cores
   on a timescale of a few thousand years.
   In agreement with recent stellar evolution models of low mass stars,
   the instability transports chemical elements less efficiently than
   angular momentum.
   }
                    
   \keywords{magnetohydrodynamics (MHD) -- instabilities -- turbulence -- methods: numerical -- stars: interiors -- stars: rotation -- stars: magnetic field}
   
   \titlerunning{Angular momentum and chemical transport by AMRI in stellar interiors}
   \authorrunning{D.~G.~Meduri et al.}
   \maketitle
%
\newcommand{\Pm}{\text{Pm}}
\newcommand{\Prt}{\text{Pr}}
\newcommand{\Ha}{\text{Ha}}
\newcommand{\Hap}{\text{Ha}_\phi}
\newcommand{\Hapmax}{\text{Ha}_{\phi}^{\text{max}}}
\newcommand{\Le}{\text{Le}}
\newcommand{\Lep}{\text{Le}_\phi}
\newcommand{\Rey}{\text{Re}}
\newcommand{\Sc}{\text{Sc}}
\newcommand{\rin}{r_{\text{i}}}
\newcommand{\rout}{r_{\text{o}}}
\newcommand{\Tin}{T_{\text{i}}}
\newcommand{\Tout}{T_{\text{o}}}
\newcommand{\Omeout}{\Omega_{\text{o}}}
\newcommand{\gout}{g_{\text{o}}}
\newcommand{\Omeref}{\Omega_{\text{a}}}
\newcommand{\Nref}{N}
\newcommand{\NOme}{\Nref/\Omeref}
\newcommand{\sg}{d}
\newcommand{\Figref}[1]{Figure~\ref{#1}}
\newcommand{\figref}[1]{Fig.~\ref{#1}}
\newcommand{\Tabref}[1]{Table~\ref{#1}}
\newcommand{\tabref}[1]{Table~\ref{#1}}
\newcommand{\Secref}[1]{Section~\ref{#1}}
\newcommand{\secref}[1]{Sect.~\ref{#1}}
\newcommand{\beq}{\begin{equation}}
\newcommand{\eeq}{\end{equation}}
\newcommand{\eqnref}[1]{Eq.~(\ref{#1})}
\newcommand{\Eqnref}[1]{Equation~(\ref{#1})}
\newcommand{\bfu}{\textbf{u}}
\newcommand{\bfB}{\textbf{B}}
\newcommand{\urf}{u_r^\prime}
\newcommand{\utf}{u_\theta^\prime}
\newcommand{\upf}{u_\phi^\prime}
\newcommand{\Brf}{B_r^\prime}
\newcommand{\Btf}{B_\theta^\prime}
\newcommand{\Bpf}{B_\phi^\prime}
\newcommand{\Bpaxi}{\overline{B}_\phi}
\newcommand{\er}{\hat{\textbf{e}}_r}
\newcommand{\etheta}{\hat{\textbf{e}}_\theta}
\newcommand{\ephi}{\hat{\textbf{e}}_\phi}
\newcommand{\am}{\mathcal{L}}
\newcommand{\mmax}{m_\text{max}}
\newcommand{\gammamax}{\gamma_\text{max}}
\newcommand{\Div}{\boldsymbol{\nabla}\cdot}
\newcommand{\Grad}{\boldsymbol{\nabla}}
\newcommand{\Curl}{\boldsymbol{\nabla}\times}
\newcommand{\azavg}[1]{\overline{#1}}
\newcommand{\Laplac}{\boldsymbol{\nabla}^{2}}
\newcommand{\Fmc}{\textbf{F}^{\text{MC}}}
\newcommand{\Frs}{\textbf{F}^{\text{RS}}}
\newcommand{\Fvd}{\textbf{F}^{\text{VD}}}
\newcommand{\Fms}{\textbf{F}^{\text{MS}}}
\newcommand{\Fmt}{\textbf{F}^{\text{MT}}}
\section{Introduction}
\label{s:intro}
A consistent description of the transport of angular momentum (AM)
and chemical elements within evolving stars is still lacking
and remains a major problem for stellar physics.
Recent asteroseismic observations
based on space photometry
have transformed our knowledge of the dynamics of stellar interiors,
offering opportunity for unprecedented advances in this field.
By uncovering the internal rotations of low mass stars
at various stages of evolution, from main sequence (MS) stars
to white dwarf remnants,
these observations unambiguously showed that
the cores of these stars rotate orders of magnitude slower than
classical predictions from stellar evolution models
and that AM is efficiently extracted from stellar cores as they evolve
\citep[for a recent review, see][]{Aerts19}.
For instance, the radiative cores of low mass subgiants
rotate slowly and do not spin up while evolving on the
red giant branch \citep{Beck11,Deheuvels14,Gehan18}.
The cores of these stars are in gravitational contraction
and, if AM was conserved, they would rotate almost
3 orders of magnitude faster and
spin up while evolving \citep[e.g.,][]{Cantiello14}.
The convective envelopes, on the other hand, expand
and should spin down leading to a strong rotation contrast with the core.
The measured envelope rotation rates of subgiants
are instead only less than 10 times
slower than those of the cores at most \citep{Deheuvels14}.
Red giants are also the sole class of stellar objects
for which, since recently, we have direct seismic
measurements of their internal magnetic fields \citep{Li22,Deheuvels23,Li23}.
The seismic detection probes a narrow region of the core
around to the hydrogen burning shell
where strong radial field strengths ranging from 30 to 600 kG
have been reported.

In order to explain all these observations, various mechanisms
to enhance the transport of AM in radiative stellar interiors
have been proposed.
The transport by atomic diffusion and standard hydrodynamical
processes, such as meridional circulation and shear
instabilities, falls short on predicting the almost
rigid rotation of the Sun's core, as well as the slow
internal rotations of red giants and white dwarfs
\citep[e.g.,][]{Eggenberger12,Marques13,Dumont21}.
Internal gravity waves, excited by convective
motions in the overlying envelope, can contribute to
transport AM in the cores of solar-type stars and subgiants,
but the process is likely negligible on the
red giant branch \citep{Fuller14,Pincon17}.

The transport by instabilities due to magnetic fields
is expected to be higher than any of
the hydrodynamical processes above and is considered
the primary mechanism to explain
the slow internal rotations observed \citep{Spruit02,Cantiello14,Spada16,Fuller19}.
In differentially rotating radiative stellar interiors,
magnetorotational instability and Tayler instability
are expected to be the two dominant magnetohydrodynamic (MHD) instabilities \citep{Spruit99}.

Magnetorotational instability (MRI) is an instability of
hydrodynamically stable shear flows
in which the magnetic field allows to release the
free energy of the shear.
For axisymmetric magnetic fields either purely azimuthal or
with both toroidal and poloidal components,
linearly unstable MRI modes
are nonaxisymmetric \citep[e.g.,][]{Balbus92,Ogilvie96,Ruediger07,Hollerbach10}.
Dominant toroidal fields are expected in
differentially rotating radiative stellar interiors,
provided that the poloidal field is weak enough \citep{Spruit99}.
Azimuthal MRI (AMRI) generally refers to
the instability of hydrodynamically stable Taylor-Couette flow, the flow of a viscous incompressible
fluid confined between two coaxial and rigidly rotating cylinders,
with imposed current-free azimuthal fields \citep{Ruediger07,Kirillov12}.
In this work, however, we will refer to this version
of the instability for generic purely or dominantly azimuthal field configurations
with a nonzero Lorentz force
that are free to evolve over time, as expected
in astrophysical situations.
Due to its nonaxisymmetric nature, AMRI can
self-sustain a magnetic field \citep[e.g.,][]{Guseva17b}.
MRI dynamo action is a highly nonlinear phenomenon
in which the turbulence due to the instability
generates large scale magnetic
fields that continuously destabilize the flow
to self-sustain the turbulence \citep{Rincon19}.

Tayler instability (TI) is instead a kink-type instability
of purely axisymmetric azimuthal fields driven
by magnetic pressure gradients \citep{Tayler73}.
This instability is expected to dominate
in radiative stellar interiors since, relying on almost
horizontal motions, is less sensitive than MRI
to stable stratification \citep{Spruit99,Bonanno12b}.
However, numerical simulations show that
the presence of a latitudinal shear may favor AMRI over TI,
even when stable stratification is relatively high \citep{Jouve20}.
While there is no asteroseismic evidence of
latitudinal differential rotation in the interior of
evolved stars so far, theoretical studies
suggest that this can be produced by gravitational contraction
when buoyancy effects are not too high, as for example
in the outer radiative regions of red giants \citep{Gouhier21,Gouhier22}.
Numerical simulations in a spherical shell
have also demonstrated that MRI can occur for
dominantly azimuthal fields generated
by shearing an initial weak poloidal field
through differential rotation,
a process known as $\Omega$-effect
and that is thought to take place in stellar interiors \citep{Jouve15,Meduri19}.

In spite of its importance, the efficiency of the AM transport
due to AMRI in radiative stellar interiors remains highly uncertain.
AMRI-induced transport is mostly investigated using shearing box
simulations, which are local numerical models of accretion
disks hardly relevant to stellar interiors \citep[e.g.,][]{Lesur07}.
Global numerical studies generally model
liquid metal laboratory experiments, hence consider
unstratified Taylor-Couette flow with imposed magnetic fields
\citep[e.g.,][]{Ruediger14,Guseva17a}.
Numerical simulations of stratified AMRI turbulence
in a spherical geometry can certainly provide
more robust constraints on the transport in stellar interiors.
However, there are only a few of these studies, which
either explore a very limited range of parameters \citep{Arlt02},
often relevant to neutron stars \citep{ReboulSalze22},
or focus only on the role played by differential rotation \citep{Jouve20}.

Stellar evolution models can provide indication
of the efficiency of the missing transport processes.
However, in the AM evolution equation of these models,
the turbulence is often parameterized
with a diffusion coefficient, which is used as a free parameter
to fit the observations.
This procedure ignores the physical origin of the transport
and how this scales with the fundamental fluid properties,
such as stratification or the molecular diffusivities,
which strongly vary in the interior of stars and during their evolution.
For example, a turbulent diffusion coefficient depending
on the ratio of the core to surface rotation rates,
attributed by analogy to the expected scaling
of AMRI turbulence with the shear,
and that increases monotonically from about $10^2\,\text{cm}^2/\text{s}$
to almost $10^6\,\text{cm}^2/\text{s}$
has been shown to reproduce the rotational evolution
of subgiants and red giants \citep{Spada16,Moyano22,Moyano23}.
As for TI turbulence, theoretical scaling laws
for the enhanced turbulent viscosity have instead been employed in
stellar evolution models \citep{Fuller19} but they
fail at capturing the rotational evolution
of subgiants and red giants simultaneously \citep{Eggenberger19}.

Advancing our understanding of the AM transport
in stellar interiors is also key to comprehend the
mixing of chemical elements.
The transport of light elements such
as lithium, beryllium and boron, which are destroyed
at temperatures as low as a few million K,
contributes to determine the chemical composition
and, consequently, the stellar evolution \citep{Deliyannis00}.
State-of-the-art stellar evolution models 
including atomic diffusion and hydrodynamical transport processes
such as those mentioned above predict
surface abundances of $^7\text{Li}$ (Li hereafter)
orders of magnitude higher than those observed
for the Sun and solar-type stars \citep{Lodders09,Dumont21},
and also for red giants \citep[see][and references therein]{Charbonnel20}.
Recent progresses on understanding the combined rotational
and chemical evolution of stars again come from considering
the transport due to MHD turbulence.
For example, stellar evolution models suggest that
TI-induced transport may reconcile the almost rigid
rotation of the Sun's core and its photospheric Li abundance \citep{Eggenberger22}
and MRI turbulence strongly influences the
chemical evolution of massive stars \citep{Wheeler15,Griffiths22}.
These studies, however, employ uncertain theoretical prescriptions of TI-driven
dynamo action or somewhat approximate estimates of
MRI-induced transport derived from accretion disk simulations.
MHD numerical simulations appropriate to model
stellar interiors and that explicitly explore the transport of chemicals
could provide better constraints but such studies are lacking so far.

In this work we investigate the transport of AM
and chemical elements due to AMRI turbulence
using 3D direct numerical simulations in a spherical shell.
We consider unstratified and stably stratified flows
under the Boussinesq approximation
where the background differential rotation is forced.
We perform a comprehensive parametric study
varying rotation rate, molecular
diffusivities and stratification.
The molecular diffusivity ratios are among
the lowest ones ever explored in a global geometry
for MRI turbulence and are relevant to
the electron-degenerate cores of red giants.
A passive scalar allows us to examine the transport
of chemical elements.

The remainder of this work is organized as follows.
In Sect.~\ref{s:model} we describe the governing equations
and the numerical model.
The unstratified axisymmetric solutions, obtained from the evolution
of an initial purely azimuthal field, are discussed in Sect.~\ref{s:axi}.
Section~\ref{s:stability} investigates their stability to
weak nonaxisymmetric perturbations.
In Sect.~\ref{s:dynamo} we describe the nonlinear evolution
of the unstable solutions where AMRI is identified
and discuss the effect of stable stratification.
Section~\ref{s:AM} explores and quantifies the transport of AM
and Sect.~\ref{s:chemicals} the one of a passive scalar.
Section~\ref{s:conclusions} closes the paper with a
discussion of the numerical results and their application
to stellar interiors.
\section{Governing equations}
\label{s:model}
We consider a stably stratified MHD flow
under the Boussinesq approximation.
The fluid is confined to a spherical shell of thickness $d=\rout-\rin$, where
$\rin$ and $\rout$ are the inner and outer boundary radii respectively, and
has uniform kinematic viscosity $\nu$, magnetic diffusivity $\eta$
and thermal diffusivity $\kappa$.
The thermal expansion coefficient is $\alpha$.
The density of the fluid $\rho$ is uniform, hence gravity varies
linearly with radius, $\textbf{g}=-\gout r/\rout \er$,
where $g_{\text{o}}$ is the gravitational acceleration at the outer boundary.
If not explicitly stated otherwise, 
$(r,\theta,\phi)$ denote dimensionless spherical coordinates hereafter
and the respective unit vectors are $\er$, $\etheta$ and $\ephi$.

The governing equations are non-dimensionalized using the shell gap $d$ as length
scale and $\tau_\Omega=1/\Omega_\text{a}$ as timescale, where $\Omega_\text{a}$
is the angular velocity at the cylindrical radius $s=r\sin\theta=0$.
The scale for temperature is $\Delta T=\Tout - \Tin > 0$,
the imposed positive temperature contrast between the isothermal outer and inner boundaries
that establishes stable stratification.
The non-hydrostatic pressure $\Pi$ is scaled by $\rho \sg \Omeref^2$ and the
magnetic field $\bfB^*$ by $(\mu_0\rho)^{1/2}d\,\Omega_\text{a}$, where $\mu_0$ is the
magnetic permeability of vacuum.
This choice makes the dimensionless magnetic field strength $B$
equal to the Lenhert number
\begin{equation}
\text{Le}=\frac{B^*}{(\mu_0\rho)^{1/2}d\,\Omega_\text{a}},
\end{equation}
which can be interpreted as the ratio of the rotation timescale $\tau_\Omega$
to the Alfv\'en travel time $d/u_\text{A}$,
where $u_\text{A}=B^*/(\mu_0\rho)^{1/2}$ is the Alfv\'en velocity.
In the following we will use $B$ and $\text{Le}$ interchangeably.
For example, the dimensionless azimuthal field strength
will be indicated with $B_\phi$ or $\text{Le}_\phi$.

In this scaling scheme, the equations governing the evolution
of the fluid velocity $\bfu$, the magnetic field
$\bfB$ and the temperature perturbations $T^\prime$ (around the
background adiabatic state) are: the momentum
equation
\beq
\label{e:NS}
\frac{\partial\bfu}{\partial t}+(\bfu\cdot\Grad)\bfu=-\Grad\Pi - \frac{\Nref^2}{\Omeref^2} T^\prime \frac{r}{\rout}\er + (\Curl \bfB)\times \bfB +\frac{1}{\Rey}\Laplac\bfu + \textbf{f},
\eeq
the induction equation
\beq
\label{e:induction}
\frac{\partial\bfB}{\partial t} = \Curl (\bfu\times\bfB) +\frac{1}{\Rey\,\Pm}\Laplac \bfB
\eeq
and the evolution equation for the temperature perturbations
\beq
\label{e:temp}
\frac{\partial T^\prime}{\partial t}+(\bfu\cdot\Grad) T^\prime = \frac{1}{\Rey\,\Pr}\Laplac T^\prime.
\eeq
The flow and the magnetic field obey to the continuity conditions
\beq
\label{e:continuity}
\Div \bfu = 0\quad\text{and}\quad \Div\bfB=0.
\eeq
The four dimensionless control parameters of the problem are:
the Reynolds number
\beq
\Rey = \frac{\Omeref\sg^2}{\nu},
\eeq
the ratio of the reference Brunt-V\"{a}is\"{a}l\"{a} frequency $\Nref$
to the reference rotation rate $\Omega_\text{a}$,
\beq
\mathcal{N}=\frac{\Nref}{\Omeref}=\left(\frac{\alpha\Delta T\gout}{\sg}\right)^{1/2}\frac{1}{\Omeref},
\eeq
the Prandtl number
\beq
\Pr=\frac{\nu}{\kappa}
\eeq
and the magnetic Prandtl number
\beq
\Pm=\frac{\nu}{\eta}.
\eeq
In this work we fix the aspect ratio of the spherical shell $\chi=\rin/\rout$ to 0.3.
We employ stress-free boundary conditions for the flow. Electrically insulating
boundary conditions are assumed for the magnetic field, which are
appropriate to match a potential field outside the fluid volume.

The azimuthal body force
\beq
\label{e:forcing}
\textbf{f}=\frac{u_\text{f} - \azavg{u}_\phi}{\tau}\,\ephi
\eeq
in the momentum equation \eqref{e:NS} imposes the background
axisymmetric differential rotation.
Here $\overline{u}_\phi$ is the axisymmetric azimuthal velocity
and $u_\text{f}=s\,\Omega_\text{f}$ is the forced contribution,
which is axisymmetric ($\partial u_\text{f}/\partial\phi =0$).
The timescale $\tau$ in \eqref{e:forcing} provides the time
on which $\overline{u}_\phi$ relaxes to $u_\text{f}$
and is fixed to $5\times 10^{-3}$.
This is the shortest timescale in the system
and practically sets the numerical integration time step
in our simulations.
\begin{figure}[t]
\centering
\resizebox{\hsize}{!}{\includegraphics{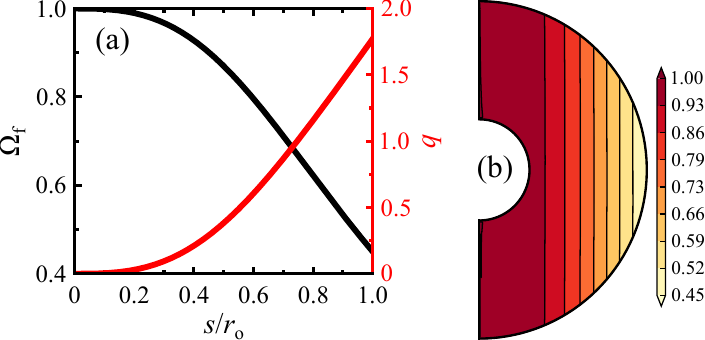}}
\caption{(a)~Forced angular velocity $\Omega_\text{f}$ (\eqnref{e:Ome_forced}
with $\mu=0.45$ and $b=2.9$) and shear parameter $q$
as a function of the cylindrical radius $s/r_\text{o}$.
(b)~Snapshot of the axisymmetric
angular velocity $\overline{\Omega}$ in the axisymmetric simulation run at
$\mathcal{N}=0$, $\text{Re}=5\times 10^4$, $\text{Pm}=1$ and $\text{Le}_0=0.1$.
}
\label{f:Ome_forced}
\end{figure}

The forced angular velocity $\Omega_\text{f}$ is
vertically invariant and is defined by
\begin{equation}
\label{e:Ome_forced}
\Omega_\text{f} = (2\mu-1)+\frac{2(1-\mu)}{1+(1-\chi)^b\,s^b},
\end{equation}
where $\mu=\Omega_\text{e}/\Omega_\text{a}<1$.
Here $\Omega_\text{e}$ is the angular velocity at the equator
on the outer boundary.
The real constant $b>0$ defines the 
steepness of the decay of $\Omega_\text{f}$ with the cylindrical radius $s$.
In this work we consider $\mu=0.45$ and $b=2.9$.
The black line in \figref{f:Ome_forced}a shows $\Omega_\text{f}$
as a function of the cylindrical radius $s$ for this parameter combination.
\Figref{f:Ome_forced}b presents a snapshot of the axisymmetric
angular velocity $\overline{\Omega}$
in one of our numerical simulations and demonstrates that
this closely follows the forced angular velocity $\Omega_\text{f}$.

The forced angular velocity that we choose here
simultaneously maximizes
the mean and the maximum of the shear parameter
$q=\left|\text{d}\ln\Omega_\text{f}/\text{d}\ln s\right|$ in the fluid domain
while still maintaining the flow stable according to
the Rayleigh criterion for inviscid flows, which prescribes
$\partial L^2/\partial s>0$, where 
$L=s \overline{u}_\phi$ is the specific AM. 
In the unstratified case, that is when $\mathcal{N}=0$
and there is no temperature equation to solve, we verified
numerically that this flow is indeed hydrodynamically
stable to weak nonaxisymmetric perturbations
by performing nonmagnetic runs
at the typical Reynolds numbers $\Rey$ explored here.
The shear parameter $q$ increases monotonically with $s$
and attains its maximum value on the outer boundary at the equator
where $q \approx 1.77$ (\figref{f:Ome_forced}a, red line). The mean value
of $q$ in the domain is $0.57$.
It is useful to define a characteristic timescale and length scale
of the shear. The shear timescale is
$\tau_{\Delta\Omega}=\Delta\Omega^{-1}$, where
$\Delta\Omega=\Omega_\text{a}-\Omega_\text{e}=(1-\mu)\,\Omega_\text{a}\approx \Omega_\text{a}/2$
is the difference between the axial and the equatorial rotation rates.
The shear length scale is $l_{\Delta\Omega}=\left(\Omega_\text{f}^{-1}\text{d}\Omega_\text{f}/ \text{d}s\right)^{-1}$, which
can be estimated as
$l_{\Delta\Omega}\sim\left(\Omega_\text{e}^{-1}\Delta\Omega/r_\text{o}\right)^{-1}=r_\text{o}/(\mu^{-1}-1)\approx r_\text{o}$.

The problem above is solved numerically using
the open-source pseudospectral MHD code MagIC\footnote{https://magic-sph.github.io} \citep{Wicht02,Schaeffer13},
which we modified to include the volumetric body force $\textbf{f}$ in the momentum equation.
The numerical technique is described in detail in \cite{Christensen07}
and we therefore mention only the essentials here.
MagIC uses a poloidal-toroidal decomposition
for the vector fields $\bfu$ and $\bfB$ and for the scalar temperature field.
Spherical harmonics are employed in the latitudinal and
azimuthal directions and Chebyshev polynomials in radius.
An implicit-explicit time stepping scheme, where
the nonlinear terms and the volumetric body force are treated explicitly, is employed.
The nonaxisymmetric simulations performed here
typically use a spatial resolution of 257 radial grid points
and a maximum spherical harmonic (SH) degree $\ell_{\text{max}}=341$.
Depending on the characteristic scales
of the turbulence to resolve, the spatial grid is sometimes coarser.
The spherical harmonic kinetic and magnetic energy spectra
of the solutions span at least three orders
of magnitude in amplitude, which ensures numerical convergence of the results
as we generally verified.

Throughout this work, we use the following notation for
azimuthal, horizontal (over a spherical
surface) and temporal averages of an arbitrary function $f=f(r,\theta,\phi,t)$,
\begin{equation*}
\overline{f}=\frac{1}{2\pi}\int_0^{2\pi}f\,\text{d}\phi,
\end{equation*}
\begin{equation*}
\langle f\rangle = \frac{1}{4\pi r^2}\int_0^{\pi}\int_0^{2\pi}f\, r^2\sin\theta\, \text{d}\phi\,\text{d}\theta,
\end{equation*}
and
\begin{equation*}
\hat{f} = \frac{1}{\Delta t}\int_{t_1}^{t_1+\Delta t} f\,\text{d}t,
\end{equation*}
respectively. Here $\Delta t$ is the time averaging period.
In the following, only when explicitly specified, the angular brackets $\langle\cdot\rangle$
denote spatial averages different from the one above.
The nonaxisymmetric flow velocity and magnetic field
are denoted by the superscript $^\prime$, e.g.~$\bfB^\prime=\bfB-\overline{\bfB}$.
\section{Unstratified axisymmetric solutions}
\label{s:axi}
We first describe the axisymmetric solutions obtained
for an unstratified flow, that is when
$\mathcal{N}=0$ and there is no temperature
equation to solve.
\subsection{Initial magnetic field condition and dimensionless parameters}
As initial condition for the magnetic field, we consider
a purely azimuthal field linearly increasing with
the cylindrical radius $s$,
\begin{equation}
\bfB (t=0)=B_0\,s\,\ephi ,
\end{equation}
where $B_0=\text{Le}_0=B_0^*/(\mu_0\rho)^{1/2}d\,\Omega_\text{a}$
is the field strength at $s=1$.
In differentially rotating radiative stellar interiors
where a weak axisymmetric poloidal field $\overline{\textbf{B}}_\text{p}$
is present, we expect dominantly azimuthal field configurations
to be generated by the $\Omega$-effect,
that is by shearing the poloidal field through the
differential rotation via the term $s(\overline{\textbf{B}}_\text{p}\cdot\boldsymbol{\nabla})\Omega$
in the azimuthal component
of the axisymmetric induction equation \citep[e.g.,][]{Spruit99}.
The azimuthal field would be stronger where the differential rotation
is higher and, given our background angular velocity profile $\Omega_\text{f}$,
this motivates the choice of the initial condition above.

To discuss the axisymmetric solutions, it is useful to rescale
the magnetic field independently of the rotation rate
by introducing the Hartmann number
\begin{equation}
\text{Ha}=\frac{B^*\,d}{(\mu_0\rho\nu\eta)^{1/2}}.
\end{equation}
The Hartmann number is related to the dimensionless parameters
introduced in the previous section
by $\text{Ha}=\text{Le}\,\Rey\,\Pm^{1/2}$
and can be interpreted as the ratio of the
geometric mean
of the viscous and magnetic diffusion times, $\tau_\nu=d^2/\nu$ and $\tau_\eta=d^2/\eta$ respectively,
to the Alfv\'en travel time $d/u_\text{A}$.
In the following $\text{Ha}_0$ and $\text{Ha}_\phi$
denote the Hartmann number based on the initial reference field strength $B_0^*$
and on the azimuthal field $B_\phi^*$ respectively.

In \secref{s:axi_evol} we discuss the flow and magnetic field
solutions obtained when varying $\text{Ha}_0$ and the Reynolds number $\text{Re}$.
The magnetic Prandtl number $\Pm$ will be fixed to 1.
For consistency with the new dimensionless magnetic field $\widetilde{B}=\text{Ha}$,
the flow velocity will be scaled with $d/(\nu\eta)^{1/2}$
and indicated with $\widetilde{u}$ in this section.
\subsection{Temporal evolution}
\label{s:axi_evol}
\begin{figure}
\centering
\resizebox{0.95\hsize}{!}{\includegraphics{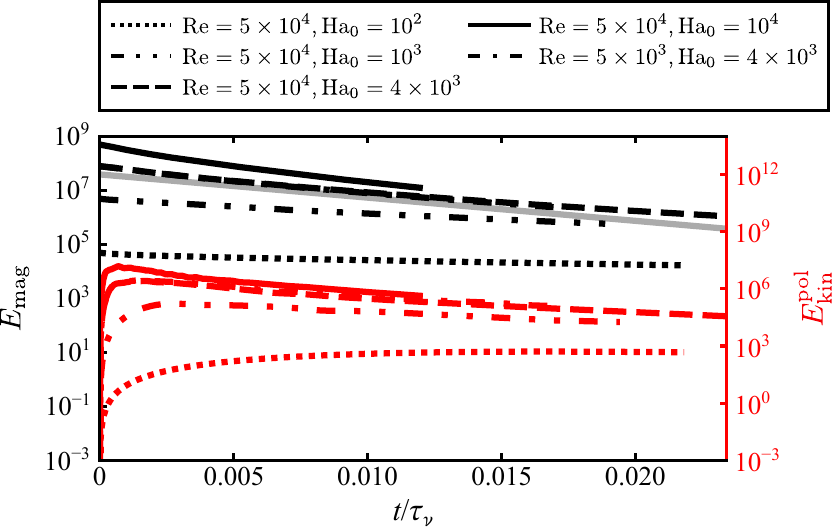}}
\caption{Temporal evolution of the magnetic energy $E_{\text{mag}}$
(black lines) and of the poloidal kinetic energy $E_{\text{kin}}^{\text{pol}}$
(red lines) in five unstratified  axisymmetric runs
at different $\text{Re}$ and $\text{Ha}_0$ (see the legend at the top).
The magnetic Prandtl number $\text{Pm}$ is 1.
Time is scaled with the viscous diffusion time
$\tau_\nu$ here. Note that the left and right vertical axes
span different ranges.
The gray line displays the exponential Ohmic decay rate
based on $\eta/l_{\overline{B}}^2$ for
the run at $\Rey=5\times 10^4$ and $\text{Ha}_0=4\times 10^3$
(see the main text for details).}
\label{f:ener_axi}
\end{figure}
\begin{figure*}
\centering
\includegraphics[width=12cm]{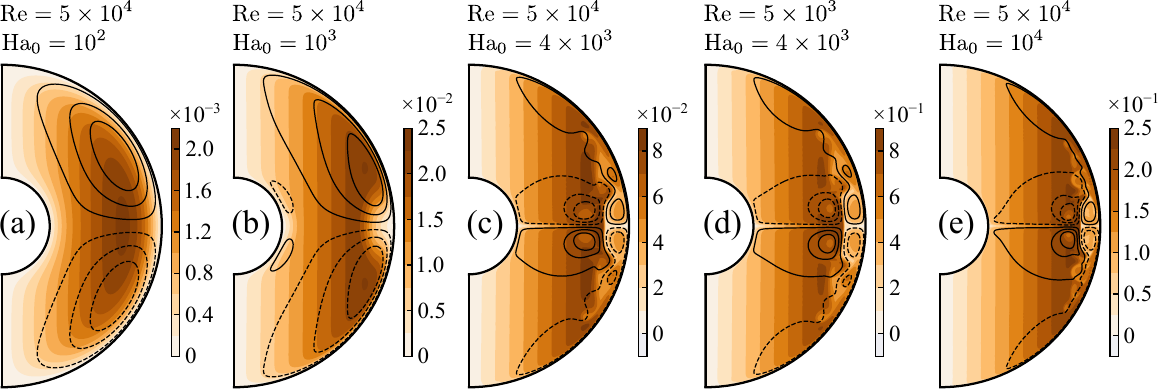}
\caption{Unstratified axisymmetric solutions
of the runs in \figref{f:ener_axi}
at the times when the poloidal kinetic energy $E_{\text{kin}}^{\text{pol}}$ reaches its maximum.
The Reynolds number $\text{Re}$ and the Hartmann number
$\text{Ha}_0$ are shown at the top of each panel. The
magnetic Prandtl number $\text{Pm}$ is 1.
The colour contours show the azimuthal magnetic field $\text{Le}_\phi$
and the black isocontour lines the meridional circulation
(solid and dashed for clockwise and counterclockwise, respectively).}
\label{f:snapsh_axi}
\end{figure*}
The black lines in \figref{f:ener_axi} illustrate
the temporal evolution of the volume integrated
magnetic energy $E_\text{mag}=\frac{1}{2}\int \widetilde{B}^2\text{d}V$
in four runs at $\Rey=5\times 10^4$ with $\text{Ha}_0$
in the range $10^2-10^4$ and in one run at the lower $\Rey$ of $5\times 10^{3}$.
Since no poloidal field is initialized, the 
solution remains purely toroidal.
Axisymmetric magnetic fields
cannot be maintained by dynamo action (Cowling's antidynamo theorem)
and the toroidal field decays exponentially due
to Ohmic diffusion. The magnetic energy evolution is compatible with
a field decay rate of $\eta/l_{\overline{B}}^2$, where
$l_{\overline{B}}$ is the characteristic azimuthal field length scale,
which we calculated as a meridional average of $\left|\widetilde{B}_\phi \right|\big/\left|\boldsymbol{\nabla}\widetilde{B}_\phi \right|$
at the last numerical integration time step
of these runs (the gray line in \figref{f:ener_axi} shows the
exponential Ohmic decay rate of the
run at $\Rey=5\times 10^4$
and $\text{Ha}_0=4\times 10^3$ as an example). 

Due to the azimuthal flow forcing, the toroidal kinetic energy
$E_\text{kin}^\text{tor}=\frac{1}{2}\int \widetilde{u}_\phi^2\text{d}V\approx \frac{1}{2}\int \widetilde{u}_\text{f}^2\text{d}V$,
which is approximately 
$5.75\times 10^9$ and $5.75\times 10^7$ in the
runs at $\Rey=5\times 10^4$ and $5\times 10^3$ respectively,
is the dominant energy contribution and
remains constant over time (not shown).
Boundary driven flows produced by the Lorentz force
induce a global meridional circulation
in the first few rotation times $\tau_\Omega$, as evidenced by
the initial rapid growth of
the poloidal kinetic energy $E_\text{kin}^\text{pol}$
(red lines in \figref{f:ener_axi}).
The peak amplitude of $E_\text{kin}^\text{pol}$
scales with $\text{Ha}_0^2$
and, on longer times, $E_\text{kin}^\text{pol}$ decays at a rate
comparable to the one of the field, confirming
the magnetic origin of the meridional flow.
We now describe qualitatively how
this global circulation is generated
and how its morphology changes when varying $\text{Ha}_0$.

The jump between the interior azimuthal field
solution $\widetilde{B}_\phi\propto s$ and the electrically insulating
boundary conditions, which impose
$\widetilde{B}_\phi=0$ at $r=\rin/d$ and $\rout/d$,
is accommodated in thin layers close to the boundaries,
where the radial Lorentz force is strong
due to the high radial gradients of the field.
Due to the form of the interior field solution,
the radial Lorentz force is stronger
in the outer boundary layer than in the inner one.
The Lorentz force in these boundary layers induces radial flows  
which, by mass conservation, generate
return flows in the latitudinal direction.
These return flows remain high in amplitude due to
the stress-free boundary conditions and
induce the large scale circulation observed.
We note that the Lorentz force plays a similar dynamical role
in modifying viscous boundary layers
when a transverse magnetic field is applied
\citep[Hartmann boundary layers; see, e.g.,][]{Dormy07}.

The boundary driven circulation is equatorially antisymmetric and,
at the time when the poloidal kinetic energy peaks, is
characterized by one cell in each hemisphere
for the run at the lowest $\text{Ha}_0$
of $10^2$ (black isocontours in \figref{f:snapsh_axi}a).
Increasing $\text{Ha}_0$ to $10^3$, the outer boundary layer
thickness decreases as expected (black isocontours in \figref{f:snapsh_axi}b)
and the meridional circulation amplitude strengthen (\figref{f:ener_axi}, dash double-dot red line).
The radial Lorentz force is now
important in the inner boundary layer as well,
producing secondary circulation cells in the inner
fluid regions (\figref{f:snapsh_axi}b).
These secondary circulation cells are characterized
by flows in the opposite direction to those of the primary ones,
due to the opposite sign
of the driving radial Lorentz force in the
inner and outer boundary layers.

At $\text{Ha}_0\gtrsim 4\times 10^3$,
the circulation driven by the inner boundary layer extends
towards the outer fluid regions and becomes
comparable in amplitude to the outer one (black isocontours in \figref{f:snapsh_axi}c,e).
The azimuthal field is efficiently advected by
such strong flows and complex configurations
with locally strong field gradients are produced
(\figref{f:snapsh_axi}c,e).

Further evidence that the meridional flow in our simulations
is of magnetic origin is given by the fact that
the solution does not depend on $\Rey$.
The evolution of the magnetic and kinetic energies
of the two runs at $\text{Ha}_0=4\times 10^3$ in \figref{f:ener_axi}
with $\Rey=5\times 10^3$ (dotted-dashed line) and $5\times 10^4$ (dashed line)
are indeed almost identical and 
snapshots of the two solutions
are nearly indistinguishable (\figref{f:snapsh_axi}c,d).

While the axisymmetric solutions above
may be regarded as peculiar,
they have to be considered only as initial states
prone to AMRI supporting turbulence
for a period long enough to study the transport of
AM and chemical elements, as we shall see in the following.
\section{Linear stability}
\label{s:stability}
We now investigate the stability of the unstratified
axisymmetric solutions
discussed in the previous section.
We introduce nonaxisymmetric perturbations
after the magnetically driven meridional flows develop
and the azimuthal field slowly decays due to Ohmic diffusion.
The perturbations consist of a spatially uncorrelated
nonaxisymmetric toroidal
field of weak amplitude.
This is obtained by adding
to the spherical harmonic decomposition
of the toroidal field and for each radii small amplitude coefficients
$\delta h_{\ell,m}(r)$ drawn from a uniform distribution
for all degrees $\ell$ and orders $m>0$.
As in the previous section, the magnetic Prandtl
number $\text{Pm}$ is fixed to 1 here.
The Reynolds number $\text{Re}$ is varied
in the range $10^3-10^5$.
At a fixed $\text{Re}$, to explore the effect of the
background azimuthal field strength
on the stability, we either perform runs with different $\text{Ha}_0$
or we introduce the perturbations
at successive times in a single run.
The azimuthal field strength of the perturbed
axisymmetric solution is characterized by the
maximum value of $\text{Ha}_\phi$ in the fluid domain
at the perturbation time $t=t_\text{pert}$.
This is indicated as $\Hapmax$ hereafter and
ranges from $60$ to about $10^4$.
\subsection{Parameter space for stable and unstable regimes}
\label{s:unstrat_stability}
Hydrodynamically stable shear flows
with purely axisymmetric azimuthal fields as those considered here
are stable to axisymmetric perturbations \citep{Velikhov59}
but they can be unstable to AMRI and to TI.

At fixed $\Rey$, AMRI can develop only if the azimuthal field is nor too weak
nor too strong, otherwise AMRI modes are stabilized by
diffusive effects or by magnetic tension respectively.
We provide order of magnitude estimates of these stability limits
using results from the local linear stability analysis
of \cite{Masada07}. The authors employ cylindrical coordinates
and consider classical local harmonic
perturbations with an azimuthal wavelength much larger
than the meridional ones. The azimuthal velocity $u_\phi=s\Omega$
and the toroidal field $B_\phi$ of the basic, purely toroidal axisymmetric state
are assumed to depend only on the cylindrical radius $s$.
For an unstratified flow with strong differential rotation,
as in most of our numerical simulations where $q\gg \text{Le}_\phi^2$,
the most unstable azimuthal mode predicted by the
local linear analysis is
\beq
\label{e:mmax}
m_\text{max}^\text{AMRI}=\frac{(4q-q^2)^{1/2}\Omega}{2\,\omega_\text{A}}
\eeq
in the absence of diffusive effects. Here
$\omega_\text{A}=B_\phi/(\mu_0\rho)^{1/2}s$
is the local azimuthal Alfv\'en frequency.
The adiabatic growth rate of the most unstable mode is
\beq
\label{e:grmax}
\gamma_\text{max}^\text{AMRI}=\frac{q\,\Omega}{2}.
\eeq
AMRI can develop only when its most unstable mode
grows faster than it decays by resistive effects (or equivalently
by viscous effects since $\Pm=1$ here)
\beq
\gamma_\text{max}^\text{AMRI}>\eta\,\left(k_\text{max}^\text{AMRI}\right)^2,
\eeq 
where $k_\text{max}^\text{AMRI}=m_\text{max}^\text{AMRI}/s$ is
the most unstable azimuthal wavenumber.
Substituting Eqs.~(\ref{e:mmax}) and (\ref{e:grmax})
in the relation above yields
\begin{equation}
\label{e:HaLow}
\text{Ha}_\phi^2 > 2\left(1-\frac{q}{4}\right)\,\text{Re}
\end{equation}
when considering $\Omega\approx\Omega_\text{a}$,
$s\approx d$ and $\Pm=1$.

For increasing field strengths, large azimuthal modes
are stabilized by magnetic tension and the spectrum
of unstable AMRI modes shifts towards lower $m$.
The condition $m_\text{max}^\text{AMRI}\geq 1$ provides
an upper limit on the field strength for instability which, for $\Pm=1$
and using the same values for $\Omega$ and $s$ as above, reads
\begin{equation}
\label{e:HaUp}
\text{Ha}_\phi\leq \left( q-\frac{q^2}{4}\right)^{1/2}\text{Re}.
\end{equation}
Combining Eqs.~\eqref{e:HaLow} and \eqref{e:HaUp}, we obtain
the instability condition
\begin{equation}
\label{e:AMRI_cond}
2\left(1-\frac{q}{4}\right)\,\text{Re}<\text{Ha}_\phi^2 \leq q\left(1-\frac{q}{4}\right)\,\text{Re}^2
\end{equation}
when $q<4$.
For our simulations where $q\approx 1$, (\ref{e:AMRI_cond}) reduces to
\begin{equation}
\label{e:HaAMRI}
\frac{3}{2}\text{Re}<\text{Ha}_\phi^2\leq \frac{3}{4}\text{Re}^2
\end{equation}
or equivalently
\begin{equation}
\label{e:LeAMRI}
\sqrt{\frac{3}{2}}\,\frac{1}{\text{Re}^{1/2}} < \text{Le}_\phi \leq \frac{\sqrt{3}}{2}
\end{equation}
since $\text{Ha}_\phi=\text{Le}_\phi\,\text{Re}$ when $\text{Pm}=1$.

Another prediction from the local linear theory of \cite{Masada07}
that will be used in the following is the critical AMRI azimuthal mode
\begin{equation}
\label{e:mcrit}
m_\text{crit}^\text{AMRI}=(2\,q)^{1/2}\Omega/\omega_\text{A}.
\end{equation}

As mentioned above, the axisymmetric solutions explored here
are in principle unstable to both AMRI and TI.
However, in the AMRI regime defined by (\ref{e:LeAMRI}),
the expected most unstable TI mode is virtually always growing slower
than the corresponding AMRI mode.
The most unstable TI mode is $m_\text{max}^{\text{TI}}=1$
and its adiabatic growth rate, in the presence of rotation,
is $\gamma_\text{max}^\text{TI}\sim\omega_\text{A}^2/\Omega$ \citep{Pitts85,Spruit99,Bonanno13b}.
AMRI dominates over TI when
$\gamma_\text{max}^\text{AMRI}> \gamma_\text{max}^\text{TI}$, that is
\begin{equation}
\frac{\omega_\text{A}}{\Omega} < \left( \frac{q}{2}\right)^{1/2}.
\end{equation}
When assuming $\Omega\approx\Omega_\text{a}$,
$s\approx d$ and $q\approx 1$ as done above, this condition reads
\begin{equation}
\label{e:LeTI}
\text{Le}_\phi < 1\big/\sqrt{2}
\end{equation}
which is slightly smaller than the upper bound of (\ref{e:LeAMRI}).
Similarly to what we obtain here using simple
order of magnitude estimates,
\cite{Jouve15} showed that TI dominates over AMRI
when $\text{Le}_\phi \gtrsim 1$ by applying a local
linear dispersion relation to dominant azimuthal
field configurations obtained from direct numerical simulations.

\begin{figure}
\centering
\resizebox{\hsize}{!}{\includegraphics{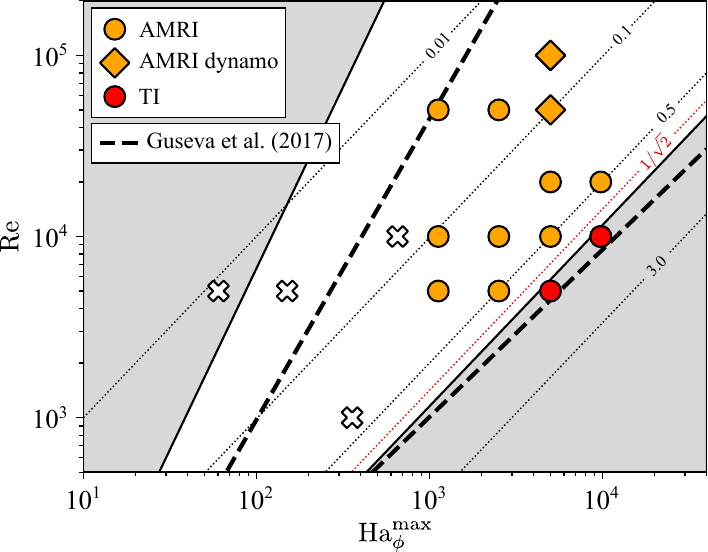}}
\caption{Instability domain of the unstratified ($\mathcal{N}=0$)
axisymmetric solutions at $\Pm=1$ and comparison with local
and global linear stability analysis results.
The Reynolds number $\Rey$ is shown as a function of
$\Hapmax$, the maximum azimuthal Hartmann number
in the fluid domain at the perturbation time $t_\text{pert}$.
Crosses denote stable runs, i.e.~simulations where the applied
nonaxisymmetric perturbations decay.
Orange (red) symbols display runs
where AMRI (TI) is identified.
Circles (squares) denote transient (self-sustained)
turbulence for the nonlinear instability evolution.
The white area shows the region unstable
to AMRI according to order of magnitude estimates
based on a local linear analysis (Eq.~(\ref{e:HaAMRI})).
AMRI is stabilized by diffusive effects at lower
magnetic field strengths (gray area on the left)
and by magnetic tension
at larger field strengths (gray area on the right).
The thin dotted lines are isocontours of the
azimuthal Lehnert number $\text{Le}_\phi^\text{max}=\Hapmax/\text{Re}$.
For $\text{Le}_\phi^\text{max}$ up to $1/\sqrt{2}$ (red dotted line)
the expected maximum growth rate of AMRI
is larger than the one of TI (Eq.~(\ref{e:LeTI})).
The thick dashed lines are reproduced from Fig.~1a of \cite{Guseva17a}
and show the lower and upper neutral stability
lines of AMRI from a global linear analysis
of Taylor-Couette flow for a quasi-Keplerian shear.
}
\label{f:ReHa}
\end{figure}

\Figref{f:ReHa} compares our numerical simulation results
with the local linear analyses predictions just discussed.
Crosses show runs where
the applied nonaxisymmetric perturbations
decay and no instability is found; for circles and squares,
the perturbations grow exponentially over time.
Runs where, as demonstrated in the next section,
we observe AMRI (orange symbols)
fall within the region of the parameter
space predicted by the
instability condition \eqref{e:HaAMRI} (white background)
as expected.
For low azimuthal field strengths $(\text{Ha}_\phi^{\text{max}})^2<3\text{Re}/2$
(gray background on the left) where
AMRI is stabilized by diffusive effects
according to the estimates above, 
we found no unstable run.

For larger azimuthal field strengths
$(\text{Ha}_\phi^{\text{max}})^2 > 3\text{Re}^2/4$ (gray background
on the right),
AMRI is expected to be suppressed by magnetic tension
and we indeed observe TI (red circles).
Evidence of TI in these simulation runs
is provided in Appendix~\ref{s:appendix_Tayler}.
Note that the critical $\text{Le}_\phi^{\text{max}}$
of $1/\sqrt{2}$ for which the maximum TI growth rate
becomes larger than the one of AMRI (red dotted line; cf.~Eq.~(\ref{e:LeTI}))
lies close to the boundary where AMRI is suppressed.
In other words, in the AMRI regime, the maximum TI growth rate
is virtually always lower than the one of AMRI.

\cite{Guseva17a} performed a global linear stability analysis
of AMRI in Taylor-Couette flow with
a quasi-Keplerian shear that resembles
the one employed here.
The lower and upper neutral instability lines
calculated by the authors (thick dashed lines in \figref{f:ReHa})
agree remarkably well
with our numerical simulation results.

In the reminder of this work, we focus on the AMRI regime only.
In \secref{s:dynamo} we discuss the nonlinear evolution of
the instability but we anticipate here that
transient AMRI turbulence (orange circles in Fig.~\ref{f:ReHa}) is generally observed
and only two runs at $\Rey \geq 5\times 10^4$
show self-sustained turbulence (orange squares).
\subsection{Evidence of AMRI}
\label{s:lin_AMRI}
We now analyze the linear evolution of the instability observed
in our numerical simulations and provide evidence of AMRI.
Here we focus on the run at
$\text{Re}=5\times 10^4$ and $\text{Ha}_{\phi}^{\text{max}}=5012$
as an example. The perturbations are introduced at time
$t_{\text{pert}}=326.0$.
Similar results are obtained for the other unstable runs of
\figref{f:ReHa} and we therefore do not discuss them in detail here.

\begin{figure}
\centering
\resizebox{0.7\hsize}{!}{\includegraphics{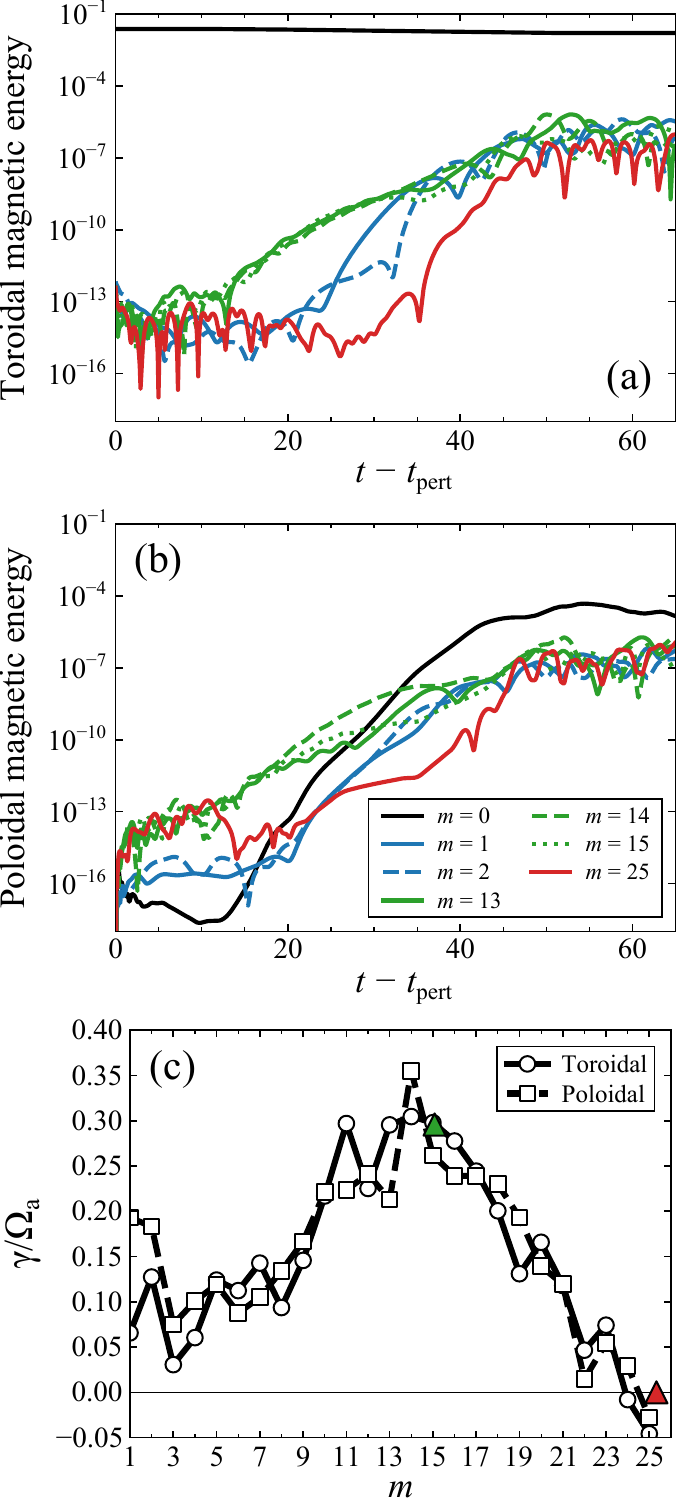}}
\caption{
(a)~Toroidal and (b)~poloidal magnetic energies of
various linearly unstable azimuthal modes $m$
as a function of time for the unstratified run at
$\Rey=5\times 10^4$, $\text{Ha}_{\phi}^{\text{max}}=5012$ and $\Pm=1$.
The energies are calculated over the spherical surface
at radius $r/r_\text{o}=0.9$.
The most unstable linear mode is $m=14$.
The first subcritical mode observed for both
energy components is $m=25$.
(c)~Linear growth rates $\gamma/\Omega_\text{a}$ of
the azimuthal modes $1\leq m\leq 25$.
The growth rates are calculated by fitting the
toroidal (solid line) and poloidal (dashed line)
magnetic energy evolution shown in (a,b)
during the period $t-t_{\text{pert}}=9-24$.
The green and red triangles show, respectively, the most unstable and the critical
AMRI modes predicted by the local linear theory
and evaluated as explained in the main text.
}
\label{f:lin_inst_emag}
\end{figure}
\begin{figure}
\centering
\resizebox{0.8\hsize}{!}{\includegraphics{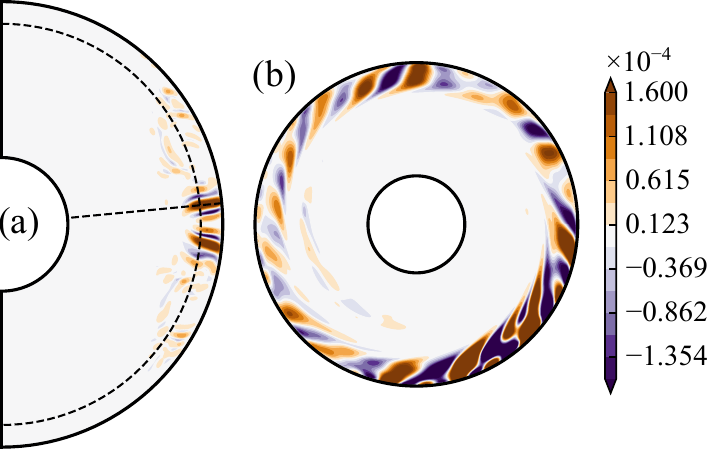}}
\caption{
Snapshot of the nonaxisymmetric azimuthal field $B_\phi^\prime$
during the linear growth of the instability ($t-t_{\text{pert}}=22$)
for the run in \figref{f:lin_inst_emag}.
(a)~Meridional cut at longitude $\phi=90^\circ$. The curved dashed
line shows radius $r/r_\text{o}=0.9$ where the magnetic energies
of \figref{f:lin_inst_emag}a,b are calculated.
(b)~Azimuthal cut at colatitude $\theta=85^\circ$
passing through a local maximum of $B_\phi^\prime$
(shown as an oblique dashed line in~(a)).
}
\label{f:lin_inst_snap}
\end{figure}
\Figref{f:lin_inst_emag}a,b present the initial temporal evolution
of the toroidal and poloidal magnetic energies
of various azimuthal modes $m$ in this run.
The energies are calculated over the
spherical surface at radius $r/r_\text{o}=0.9$ and map
the outer fluid regions
where the instability first develops (\figref{f:lin_inst_snap}a).
As expected for AMRI, the instability fluctuations
are larger in the outer equatorial region
where the unstable modes have higher growth rates
due to the stronger background shear (\figref{f:Ome_forced}a).

Nonaxisymmetric azimuthal modes $m<25$
grow exponentially typically after about 10
system rotations from the perturbation time
until $t-t_{\text{pert}}\approx 25$ (\figref{f:lin_inst_emag}a,b).
The background axisymmetric ($m=0$) azimuthal field (black line in \figref{f:lin_inst_emag}a)
slowly decays due to Ohmic diffusion
and its evolution is practically stationary
during the linear instability growth.
We also observe the generation of an initially weak and decaying
axisymmetric poloidal field (black line in \figref{f:lin_inst_emag}b).
This field is produced from weak nonlinear correlations of the 
initial flow and field instability fluctuations.

\Figref{f:lin_inst_emag}c displays the growth rates
of the nonaxisymmetric azimuthal modes $m$ from 1 to 25,
calculated from the evolution of the poloidal (dashed line) and toroidal (solid line)
magnetic energies above
during the period $t-t_{\text{pert}}=9-24$
of the linear instability growth.
The two growth rates spectra are in good agreement
between each other.
According to the toroidal spectrum,
modes $m< 24$ are linearly unstable.
The most unstable mode is $m_\text{max}=14$, although
very similar growth rates are observed for
the two adjacent modes $m=13$ and $15$.
The longitudinal structure of the nonaxisymmetric azimuthal field $B_\phi^\prime$
is compatible with such dominant modes (\figref{f:lin_inst_snap}b).

The observed most unstable mode $m_\text{max}$
and its growth rate $\gamma_\text{max}$
closely agree with the ones expected from the
local linear theory
discussed in the previous section.
Equation~\eqref{e:mmax} indeed provides
$\langle m_\text{max}^\text{AMRI}\rangle \approx 15$, which 
is very close to the observed most unstable mode.
Here and in the remainder of this section
the angular brackets $\langle\cdot\rangle$
denote average over a spherical surface
at radius $r/r_\text{o}=0.9$
where the azimuthal modes spectra above are calculated.
Based on Eq.~(\ref{e:grmax}),
the expected maximum growth rate of AMRI
is $\langle \gamma_\text{max}^\text{AMRI}\rangle/\Omega_\text{a}\approx 0.29$,
which also closely agrees with the one observed from
the toroidal spectrum at $0.30$
(\figref{f:lin_inst_emag}c).

The toroidal spectrum in \figref{f:lin_inst_emag}c
shows that azimuthal modes $m\geq 24$ are initially subcritical.
This is again compatible with the local linear theory
which predicts a critical azimuthal mode
$\langle m_\text{crit}^\text{AMRI}\rangle \approx 25$
(Eq.~(\ref{e:mcrit}); red triangle in \figref{f:lin_inst_emag}c).
The linearly stable modes
start to grow only at $t-t_{\text{pert}}\gtrsim 25$ due to nonlinear mode
energy transfers (the red line
in \figref{f:lin_inst_emag}a,b displays the evolution of $m=25$ as an example),
when we verified
that the axisymmetric magnetic energy becomes comparable
in amplitude to the nonaxisymmetric one.
The nonlinear growth rates of the modes $m=1$ and $m=2$
(blue lines in \figref{f:lin_inst_emag}a,b) are roughly
twice the growth rate of $m_\text{max}$, possibly because of mode interactions
between the faster growing linear modes.
Finally, the instability saturates at $t-t_{\text{pert}}\approx 55$
with the nonaxisymmetric poloidal and toroidal energies
roughly in equipartition.

\begin{figure}
\centering
\resizebox{\hsize}{!}{\includegraphics{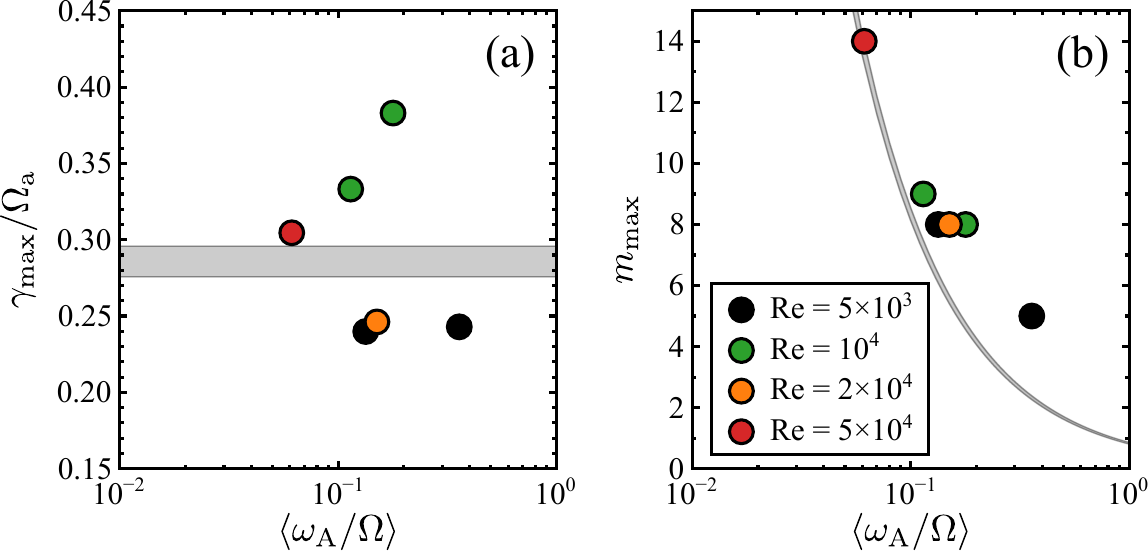}}
\caption{
(a)~Observed maximum growth rates $\gamma_\text{max}$
and (b)~most unstable azimuthal modes $m_\text{max}$
as a function of $\langle \omega_\text{A}/\Omega\rangle$
for the unstable AMRI runs of \figref{f:ReHa}
at $(\text{Re},\text{Ha}_\phi^\text{max})=(5\times 10^3, 1125)$,
$(5\times 10^3, 2520)$, $(10^4, 1125)$, $(10^4, 2520)$, $(2\times 10^4, 5012)$
and $(5\times 10^4, 5012)$.
For each run, $\gamma_\text{max}$ and $m_\text{max}$
are calculated from the evolution of
the toroidal magnetic energy of the linearly unstable modes
as explained in the main text.
The gray shaded regions display the range of maximum
growth rates and most unstable modes predicted
by the local linear theory for these runs.
}
\label{f:inst_theo}
\end{figure}
\Figref{f:inst_theo} displays the observed maximum growth rates $\gamma_\text{max}$
and the most unstable azimuthal modes $m_\text{max}$
in some of the unstable runs of \figref{f:ReHa} in the AMRI regime.
The simulation data points are shown as a function
of $\langle \omega_\text{A}/\Omega\rangle$ at the perturbation time.
The numerical results are in good agreement with the
local linear analysis predictions for AMRI (Eqs.~(\ref{e:grmax}) and (\ref{e:mmax})), which
are indicated by the gray shaded regions.
The largest differences between the numerical and
theoretical values are of about $20\%$ in the maximum growth rate
and of a factor 2 in the most unstable mode.
Such differences are surprisingly very limited
considering all the simplifying assumptions
of the local linear analysis, which can only
roughly describe the global modes excited in our numerical
simulations.
\section{Nonlinear solutions}
\label{s:dynamo}
In this section we discuss the nonlinear evolution
of AMRI and describe the transient or self-sustained
turbulent solutions
obtained for unstratified and stratified flows.
First, we consider the unstratified runs at $\Pm=1$
of the previous section.
In \secref{s:dyn_fiducial} we describe in detail
the fiducial dynamo run at $\text{Re}=5\times 10^4$
and $\text{Ha}_{\phi}^{\text{max}}=5012$ and we
examine the dynamo onset in \secref{s:dyn_onset}.
The effect of $\Pm$ on the dynamo onset is presented in \secref{s:Pm}.
Finally, \secref{s:stable_strat} analyzes the nonlinear solutions
obtained when including stable stratification.
\subsection{Fiducial dynamo run}
\label{s:dyn_fiducial}
\begin{figure}[b]
\centering
\resizebox{\hsize}{!}{\includegraphics{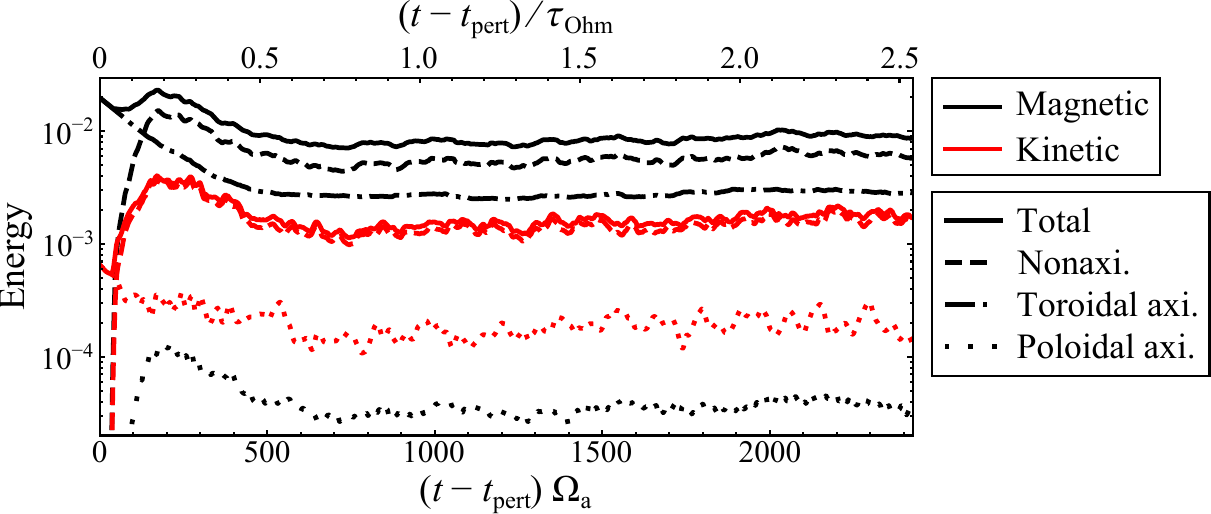}}
\caption{
Temporal evolution of the kinetic (red lines)
and magnetic (black lines) energies
of the fiducial dynamo run U0
($\mathcal{N}=0$, $\text{Re}=5\times 10^4$, $\Pm=1$ and $\Hapmax=5012$).
The axisymmetric toroidal kinetic energy
of the forced azimuthal flow is the dominant
energy contribution (not shown) and has been
subtracted from the
calculation of the total kinetic energy.
The lower (upper) horizontal axis shows time scaled in
units of the rotation timescale $\tau_\Omega=1/\Omega_{\text{a}}$
(Ohmic diffusion time $\tau_\text{Ohm}$).
}
\label{f:dyn_ener}
\end{figure}
\begin{figure*}[t]
\centering
\includegraphics[width=12cm]{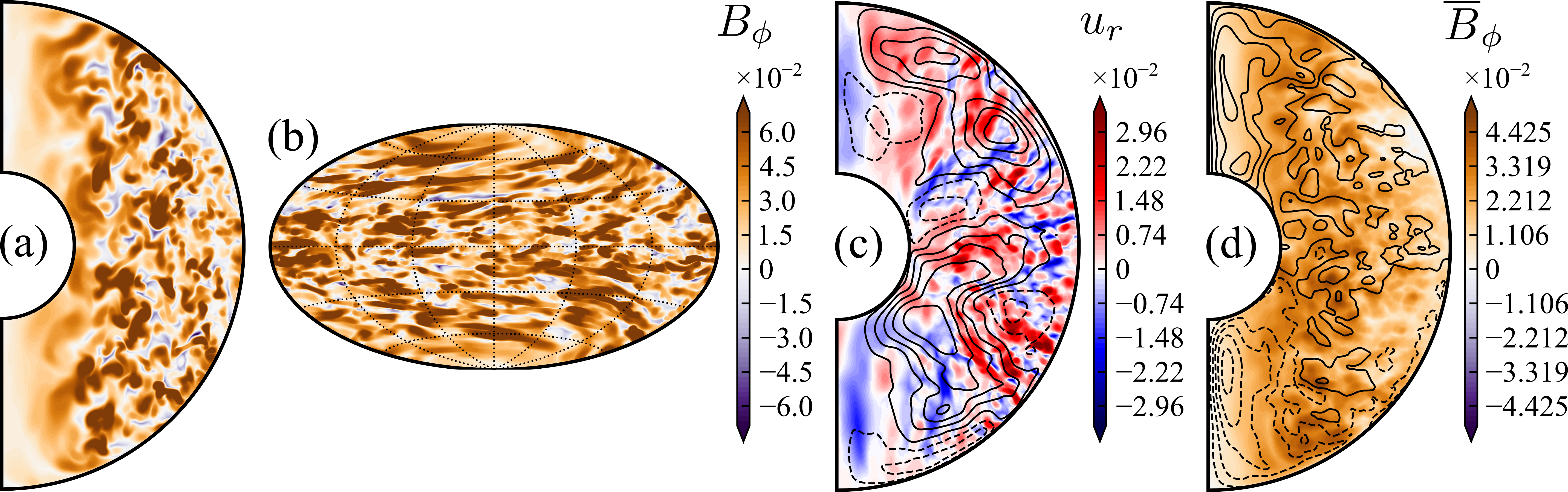}
\caption{
Steady flow and magnetic field solution
of the fiducial dynamo run U0 at time $t_\text{s}=1241$
(or $t_\text{s}-t_\text{pert}=915$).
(a,b)~Meridional cut and surface projection at
$r/r_{\text{o}}=0.8$ of the azimuthal field $B_\phi$.
(c,d)~Meridional cuts of the radial flow velocity $u_r$
and of the axisymmetric azimuthal field $\overline{B}_\phi$.
Black isocontours in (c) and (d) show the meridional circulation
and the axisymmetric poloidal field respectively.
}
\label{f:dyn_snap}
\end{figure*}
We describe the self-sustained turbulent solution
of run U0 at $\mathcal{N}=0$, $\Rey=5\times 10^4$, $\Pm=1$ and
$\Hapmax=5012$, which we refer to as the fiducial dynamo run hereafter.
The linear phase of the instability growth in this run
has been discussed in \secref{s:lin_AMRI}.

\Figref{f:dyn_ener} presents the temporal evolution of
the various components of the kinetic and magnetic energies.
The dominant (stationary) toroidal axisymmetric kinetic energy
of the background flow is
$\frac{1}{2}\int \overline{u}_\phi \,\text{d}V\approx\frac{1}{2}\int u_\text{f} \,\text{d}V\approx 2.3$
(not shown) and has been subtracted from the total kinetic energy contribution.
After around 700 system rotations from the perturbation time $t_{\text{pert}}=326.0$,
a stationary regime is reached.
The magnetic energy of this state is dominantly nonaxisymmetric (dashed black line).
The axisymmetric toroidal magnetic energy (dot dashed black line)
is the second largest contribution,
with an amplitude about 3 times smaller than the total magnetic energy.
The axisymmetric poloidal field $\overline{B}_\text{p}$, generated by the instability fluctuations
through the azimuthal component of the mean electromotive force (EMF)
$\overline{\mathbf{\mathcal{E}}}=\overline{\mathbf{u}^\prime\times\mathbf{B}^\prime}$,
is decisively weaker and saturates at an energy (dotted black line)
roughly 2 orders of magnitude lower than
the one of the axisymmetric toroidal field.
Although weak, a positive EMF feedback on $\overline{B}_\text{p}$
as the one observed here
is crucial for MRI dynamos to operate.
The $\Omega$-effect obtained by
shearing this weak $\overline{B}_\text{p}$
through the background differential rotation provides indeed
the required closure for self-sustained dynamo action \citep{Rincon07,Rincon19}.
The time averaged nonaxisymmetric magnetic energy
of the steady state is about 4 times larger than the kinetic one (dashed red line).
Similar turbulent magnetic to kinetic energy ratios are observed
in global and local simulations of MRI turbulence
with zero net flux \citep{ReboulSalze21}.

\Figref{f:dyn_snap} illustrates the complex turbulent character
of the solution in the steady state.
The magnetic field is small scaled in the meridional
directions, while it presents elongated structures in the azimuthal
direction, which are typical of MRI turbulence and are due to shear effects (\figref{f:dyn_snap}a,b).
Inside the tangent cylinder, the imaginary vertical cylindrical
surface tangent to the inner boundary at the equator,
the magnetic field is weak
since AMRI is less active due to the low background shear (\figref{f:Ome_forced}).
Nonetheless, these regions host large scale axisymmetric poloidal
fields confined at high latitudes by the meridional flow (black isocontours in \figref{f:dyn_snap}c,d).
Outside the tangent cylinder, both $\overline{B}_\text{p}$
and $\overline{B}_\phi$
are concentrated in small flux patches generated by the instability fluctuations (\figref{f:dyn_snap}d).
The structures of the flow velocity components
are similar to those of the magnetic field, except for $u_r$
where the large scale meridional circulation
produces radial plumes (\figref{f:dyn_snap}c).

Dynamo action occurs when the magnetic field is sustained for
a period longer than its characteristic Ohmic diffusion time $\tau_\text{Ohm}=l_B^2/\eta$,
where $l_B$ is a characteristic magnetic field length scale.
Since the typical radial and latitudinal length scales
of the magnetic field in the stationary state are similar, we estimate
$l_B$ based on the horizontal length scales only.
To this end, we define the instantaneous horizontal half wavelength
of the field
\begin{equation}
l_{B,\perp}=\frac{\pi\,d}{\ell_B},
\end{equation}
where
\beq
\ell_B=\frac{\sum_{\ell,m} \ell\, \langle B_{\ell,m}^2\rangle}{\langle B^2\rangle}
\eeq
is the mean SH degree of the field \citep{Christensen06}.
Here $B_{\ell,m}$ is the coefficient at degree $\ell$ and order $m$
of the SH expansion of the field and
the angular brackets denote a volume integral over the fluid domain.
The time average of $l_{B,\perp}$
during the steady state of the fiducial dynamo run
is $\hat{l}_{B,\perp}=0.14\,d$, which yields an Ohmic diffusion time
$\tau_\text{Ohm}\approx 955\,\Omega_\text{a}^{-1}$.
The top horizontal axis of \figref{f:dyn_ener} shows that the stationary evolution
of this run, which we define to start
after about $670\,\Omega_\text{a}^{-1}$ from the perturbation time,
covers $1.84\,\tau_\text{Ohm}$
and therefore self-sustained dynamo action is at work.

Hereafter we classify a simulation run as a dynamo if
its quasi steady evolution lasts longer than $\tau_\text{Ohm}$.
The simulation run presented here
is the first MRI dynamo at a value of the magnetic
Prandtl number Pm as low as 1 ever reported in a global setup.
Previous global numerical studies have shown
self-sustained MRI turbulence only for $\Pm\geq 10$ \citep{Guseva17a,ReboulSalze21}.
We remind here that these solutions have to be
interpreted as small scale dynamos, in the sense that
the generated flow and magnetic fields
are at scales smaller than the forcing scale of the background flow,
which is of the order of $r_\text{o}$ in our runs \citep[e.g.,][]{Rincon19}.

Similarly to the above, we define the horizontal half wavelength
of the flow
$l_{u,\perp}=\pi\,d/\ell_u$, where $\ell_u$ is the mean SH degree
of $\left|\mathbf{u}-\mathbf{u}_\text{f}\right|$, the flow velocity amplitude
after subtracting the contribution of the forcing.
For the simplicity of notation, we refer to $l_{u,\perp}$ and $l_{B,\perp}$
as their time averaged values over a period
of stationary evolution of the solution hereafter.
For run U0, $l_{u,\perp}=0.19\,d$, which is slightly larger than $l_{B,\perp}$.
These horizontal length scales are listed, together with the control parameters
and other output measures, in 
\tabref{t:runs_dyn} for run U0
and for the other unstratified dynamos and
stratified runs that we shall discuss in the next sections.
\begin{landscape}
\begin{table}[]
\caption{Input parameters and output diagnostics of the unstratified ($\mathcal{N}=0$) dynamo
runs and of the stratified ($\mathcal{N}>0$) runs.
$\text{Re}$, $\text{Pm}$ and $\text{Pr}$ are the Reynolds number,
magnetic Prandtl number and Prandtl number respectively.
$\text{Ha}_\phi^\text{max}$ is the maximum value of the azimuthal Hartmann number
in the fluid domain at the perturbation time $t_\text{pert}$.
The initial condition of the stratified runs is the solution of fiducial dynamo run U0
at time $t_\text{s}=1241$ (\figref{f:dyn_snap}).
$l_\text{c}/d$ is the critical radial flow length scale
below which thermal diffusion is expected to weaken the stabilizing buoyancy.
For the run regime, D, T and S stand for dynamo, transient turbulence and stable, respectively.
$\text{Rm}_\text{eff}$ is the effective magnetic Reynolds number
based on the rms flow velocity after subtracting the forced azimuthal flow.
$\tau_\text{eddy}$ and $\tau_\text{Ohm}$ are the eddy turnover time and
the Ohmic diffusion time respectively and are
reported in units of the rotation timescale $\tau_\Omega=1/\Omega_\text{a}$.
$\Delta t$ is the period of quasi steady evolution and over which all time averages are calculated.
The horizontal length scales based on the (turbulent) flow velocity and magnetic field are
($l_{u,\perp}^\prime$ and $l_{B,\perp}^\prime$) $l_{u,\perp}$ and $l_{B,\perp}$ respectively.
See the main text for definitions of all these output diagnostics.
}
\centering
\begin{tabular}{@{}lcccccccccccccccr@{}}
\toprule
Name & $\mathcal{N}$ & $\text{Re}/10^4$ & Pm & $\text{Ha}_\phi^\text{max}$  & Pr & $t_\text{pert}$ & $l_\text{c}/d$ & Regime & $\text{Rm}_\text{eff}$ & $\tau_\text{eddy}/\tau_\Omega$ & $\tau_\text{Ohm}/\tau_\Omega$ & $\Delta t/\tau_\text{Ohm}$ & $l_{u,\perp}/d$ & $l_{B,\perp}/d$ & $l^\prime_{u,\perp}/d$ & $l^\prime_{B,\perp}/d$ \\ \midrule\midrule
U0  & 0 & $5$   & 1    & 5012      & - & 326 & -  & D                  & $823$   & 5.8 & $955$ & 1.84 & $0.190$ & $0.138$ & $0.090$ & $0.075$\\
U1  & 0 & $5$   & 2    & 5012       & - & 566 & - & D                  & $2136$  & 3.3 & $828$ & 0.82 & $0.142$ & $0.091$ & $0.067$ & $0.057$\\
U2  & 0 & $10$ & 1    & 5012       & - & 654 & -      & D             & $2527$  & 2.2 & $956$ & 1.07 & $0.114$ & $0.098$ & $0.054$ & $0.059$\\
\vspace{0.5ex}
U3  & 0 & $10$          & 0.6 & 5012   & - & 412 & - & D             & $1035$  & 4.2 & $935$ & 1.60 & $0.144$ & $0.125$ & $0.068$ & $0.067$\\
S0 & 1 & $5$ & 1 & -  & $10^{-3}$  & - & 0.141 & T                  & $677$ & 6.3    & $1206$ & 0.95 & $0.169$ & $0.155$ & $0.083$ & $0.076$\\
S1 & 1 & $5$ & 1 & -  & $10^{-2}$  & - & 0.045 & T                  & $669$ & 6.4    & $1878$ & 0.40 & $0.169$ & $0.194$ & $0.084$ & $0.084$\\
\vspace{0.5ex}
S2 & 1 & $5$ & 1 & -  & $10^{-1}$ & -  & 0.014   & T                & $770$ & 6.7   & $2946$ & 0.17 & $0.204$ & $0.243$ & $0.102$ & $0.083$\\
S3 & 10 & $5$ & 1 & -   & $10^{-4}$ & -  & 0.141 & T              & $490$ & 9.3    & $2052$ & 0.37 & $0.173$ & $0.203$ & $0.086$ & $0.083$\\
S4 & 10 & $5$ & 1 & -   & $10^{-3}$ & - & 0.045  & T               & $424$  & 13.0 & $9831$ & 0.19  & $0.216$ & $0.443$ & $0.107$ & $0.091$\\
S5 & 10 & $5$ & 1 & -   & $2\times 10^{-3}$ & -  & 0.032  & T & $438$  & 13.6 & $17\,912$ & 0.09  & $0.238$ & $0.599$ & $0.117$ & $0.103$\\
\vspace{0.5ex}
S6 & 10 & $5$ & 1   & -   & $10^{-2}$ & - & 0.014 & S               & -   & - & - &  - & - & - & -  & -\\
S7 & 20 & $5$ & 1   & -  & $10^{-4}$  & - & 0.100 & T               & $415$  & 12.5 & $4076$ & 0.17 &  $0.197$ & $0.286$ & $0.098$ & $0.087$\\
S8 & 20 & $5$ & 1   & -  & $5\times 10^{-4}$ & -  & 0.045 & T  & $373$  & 17.1 & $18\,273$ & 0.05 & $0.254$ & $0.605$ & $0.125$ & $0.110$\\
S9 & 20 & $5$ & 1   & -  & $10^{-3}$ & -  & 0.032 & T              & $429$ & 15.6   & $29\,011$& 0.04  & $0.265$ & $0.762$ & $0.131$ & $0.121$\\
S10 & 20 & $5$ & 1 & -   & $5\times 10^{-3}$ & -  & 0.014 & S & -  & - & - &  - & - & - & -  & -\\
\bottomrule
\end{tabular}
\label{t:runs_dyn}
\end{table}
\end{landscape}
\begin{landscape}
\begin{table}[]
\caption{Time averaged diagnostics for the flow, magnetic field
and angular momentum transport for all unstable runs of
Table~\ref{t:runs_dyn}. The averages are evaluated over the periods $\Delta t$
reported in Table~\ref{t:runs_dyn}.
$\overline{B}_{\phi,\text{rms}}/\overline{B}_\text{p,rms}$
is the ratio of the rms axisymmetric azimuthal field to the rms axisymmetric poloidal field.
$B_{\phi,\text{rms}}/B_{r,\text{rms}}$ is the ratio of the rms azimuthal field to the rms radial field.
The rms turbulent (nonaxisymmetric) flow velocity and magnetic field are $u^{\prime}_{\text{rms}}$
and $B^{\prime}_{\text{rms}}$ respectively. The rms meridional flow velocity is $\overline{u}_{\text{M},\text{rms}}$.
$\text{R}$ denotes the ratio of the volume averaged radial Maxwell stresses
$(\mu_0\rho)^{-1}\langle B_r^\prime B_\phi^\prime\rangle$
to the volume averaged radial Reynolds stresses $\langle u_r^\prime u_\phi^\prime\rangle$.
$\widetilde{\nu}_{\text{T}}$ is the dimensionless turbulent viscosity
estimated as explained in the main text; the error indicates 1 standard deviation
in time.
}
\centering
\begin{tabular}{@{}lcccccccccccccccr@{}}
\toprule
Name & $\mathcal{N}$ & $\text{Re}/10^4$ & Pm  & Pr & $\overline{B}_{\phi,\text{rms}}/\overline{B}_\text{p,rms}$ & $B_{\phi,\text{rms}}/B_{r,\text{rms}}$ & $u^{\prime}_{\text{rms}}/10^{-3}$ &  $B^{\prime}_{\text{rms}}/10^{-3}$ & $\overline{u}_{\text{M},\text{rms}}/10^{-3}$ & $\text{R}$ & $\widetilde{\nu}_{\text{T}}\times 10^{5}$  \\ \midrule\midrule
U0  & 0 & $5$ & 1     & -                               & 8.9 & 3.9 & $15.5$ & $30.8$ & $5.6$ & $11.9$ & $37.1\pm 4.5$ \\
U1  & 0 & $5$ & 2      & -                               & 7.4 & 3.3 & $20.3$ & $42.2$ & $6.7$  & $13.0$ & $74.0\pm 8.1$ \\
U2  & 0 & $10$              & 1      & -                               & 7.4 & 3.1 & $24.3$  & $46.1$ & $7.1$ & $11.8$ & $88.4\pm 7.7$ \\
\vspace{0.5ex}
U3  & 0 & $10$              & 0.6  & -                               & 9.0 & 3.6 & $16.5$  & $30.5$ & $5.2$ & $10.4$ & $38.1\pm 8.0$ \\
S0 & 1 & $5$ & 1        & $10^{-3}$                 & 9.6 & 4.1 & $13.1$ & $28.0$ & $3.5$ & $11.1$ & $30.8\pm 3.4$\\
S1 & 1 & $5$ & 1        & $10^{-2}$                 & 10.2 & 3.9 & $13.2$ & $26.5$ & $2.0$ & $9.4$ & $30.6\pm 3.7$\\
\vspace{0.5ex}
S2 & 1 & $5$ & 1        & $10^{-1}$                 & 12.2 & 4.7 & $15.3$ & $28.5$ & $2.0$ & $6.8$ & $31.3\pm 1.9$\\
S3 & 10 & $5$ & 1         & $10^{-4}$              & 10.9 & 4.4 & $9.6$ & $22.1$ & $1.9$ & $10.1$ & $20.3\pm 5.3$\\
S4 & 10 & $5$ & 1         & $10^{-3}$              & 16.2 & 6.9 & $8.4$ & $18.3$ & $1.3$ & $6.6$ & $11.8\pm 2.2$\\
\vspace{0.5ex}
S5 & 10 & $5$ & 1         & $2\times 10^{-3}$ & 19.9 & 10.0 & $8.6$ & $18.3$ & $1.6$ & $5.7$ & $8.2\pm 1.1$ \\
S7 & 20 & $5$ & 1        & $10^{-4}$               & 13.0 & 5.2 & $8.2$ & $18.8$ & $1.3$ & $8.1$ & $14.3\pm 4.3$\\
S8 & 20 & $5$ & 1        & $5\times 10^{-4}$  & 20.3 & 10.9 & $7.3$ & $16.2$ & $1.4$ & $6.2$ & $5.8\pm 0.8$\\
S9 & 20 & $5$ & 1        & $10^{-3}$               & 24.6 & 14.8 & $8.5$ & $17.6$ & $1.5$ & $5.2$ & $4.9\pm 0.6$\\
\bottomrule
\end{tabular}
\label{t:runs_fields}
\end{table}
\end{landscape}
\subsection{Dynamo onset at $\Pm=1$}
\label{s:dyn_onset}
Dynamo action driven by MRI is
an inherently nonlinear phenomenon: it requires instability fluctuations
that transiently grow to finite amplitudes, leading to nonlinear
effects that eventually sustain the turbulence \citep{Rincon08}.
Linear MRI modes can easily reach finite amplitudes
since they exhibit nonmodal growth, that is they
can transiently grow faster than the least stable eigenmode
on short timescales \citep[e.g.,][]{Balbus92,Squire14,Mamatsashvili20}.
Nonmodal effects are a well known property
of non self-adjoint linear systems, which include not only those prone to MRI
but also purely hydrodynamical shear flows \citep{Schmid07}.

\begin{figure}
\centering
\resizebox{\hsize}{!}{\includegraphics{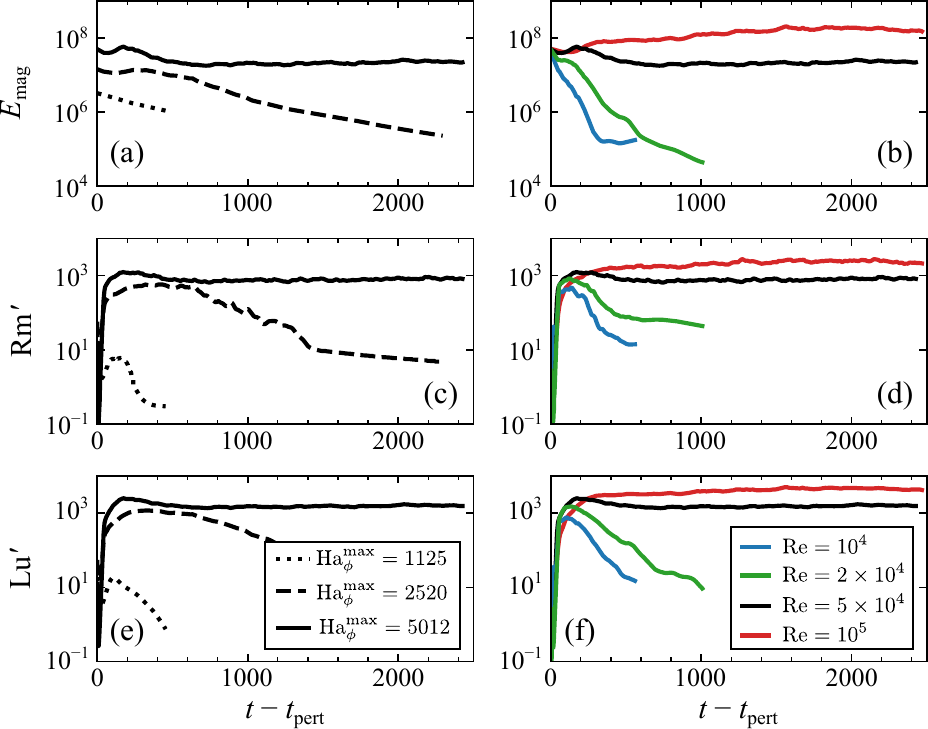}}
\caption{
Temporal evolution of the (a,b) magnetic energy $E_\text{mag}$,
the (c,d) turbulent magnetic Reynolds number $\text{Rm}^\prime$
and the (e,f) turbulent Lundquist number $\text{Lu}^\prime$
for six unstratified runs at $\Pm=1$.
The solid black line shows the fiducial dynamo run U0.
The left (right) panels present runs at $\text{Re}=5\times 10^4$ ($\text{Ha}_\phi^\text{max}=5012$)
and different $\Hapmax$ ($\text{Re}$) as indicated in the legend of the bottom panel.
}
\label{f:Emag_cf}
\end{figure}
As a consequence of their nonmodal character, the excitation of
MRI dynamos strongly depends on the amplitude
and morphology of the initial condition \citep{Riols13,Guseva17b}.
Consistently with these findings, as anticipated
in \secref{s:unstrat_stability}, we observe self-sustained dynamo action
in our unstratified simulations at $\Pm=1$ and fixed $\Rey$
only when the
field strength of the perturbed axisymmetric solution is large enough.
\Figref{f:Emag_cf}a compares the temporal evolution
of the magnetic energy $E_\text{mag}$, scaled as in \secref{s:axi},
of the fiducial dynamo run U0 (solid black line)
with two runs at lower $\text{Ha}_{\phi}^{\text{max}}$.
These latter runs show transient turbulence
where the instability decays after saturating for a period which
shortens when $\text{Ha}_{\phi}^{\text{max}}$ decreases.

The turbulent magnetic Reynolds number
\beq
\text{Rm}^\prime = \frac{u^\prime_{\text{rms}}\,d}{\eta}
\eeq
and the turbulent Lundquist number
\beq
\text{Lu}^\prime = \frac{B^\prime_{\text{rms}}\,d}{(\mu_0\rho)^{1/2}\eta},
\eeq
where $u^\prime_{\text{rms}}$ and $B^\prime_{\text{rms}}$
are the root mean square (rms)
nonaxisymmetric flow and magnetic field strengths respectively, 
can be used to compare the instability fluctuation amplitudes
of the flow and the field between these runs.
The peak values reached by $\text{Rm}^\prime$ and $\text{Lu}^\prime$
increase with $\text{Ha}_\phi^{\text{max}}$ (\figref{f:Emag_cf}c,e),
arguably producing a stronger mean EMF $\overline{\mathbf{\mathcal{E}}}$
which sustains the axisymmetric field against resistive effects
at the larger $\text{Ha}_\phi^{\text{max}}$ of $5012$
of run U0.
As already mentioned above, this instability-driven EMF
is key to generate the axisymmetric poloidal field
required for MRI dynamos to operate.
We note that studies of AMRI in Taylor-Couette flow also show
that both the turbulent flow and field amplitudes
increase with the imposed background field strength
as we observe here \citep{Ruediger14}.

As anticipated in \secref{s:unstrat_stability},
we find dynamo action for $\Rey\geq 5\times 10^4$,
that is when the azimuthal flow forcing is strong enough.
\Figref{f:Emag_cf}b indeed demonstrates that $E_\text{mag}$
in runs at a fixed $\text{Ha}_\phi^{\text{max}}$ of $5012$
decays for $\text{Re}\leq 2\times 10^4$, while a 
quasi steady state is reached in the two other runs at larger Re.
The peak amplitudes of $\text{Rm}^\prime$ and $\text{Lu}^\prime$
increase with the Reynolds number (\figref{f:Emag_cf}d,f),
which again suggests that dynamo action occurs
when the mean EMF becomes strong enough.
The background differential rotation contrast
from pole to equator $\Delta\Omega$
scales roughly as $\Rey/2$ (\secref{s:model}).
This produces a stronger $\Omega$-effect
in the two runs at larger $\Rey$
which also helps to sustain the axisymmetric azimuthal field $\overline{B}_\phi$.

We remark here that the perturbed axisymmetric azimuthal field
solutions of these runs at fixed $\text{Ha}_\phi^{\text{max}}$
are very similar but not exactly identical.
Although mild, these differences may contribute
to explain the regime change from decaying
to self-sustained turbulence when increasing $\Rey$.
The critical magnetic Reynolds number for MRI dynamo onset
indeed strongly depends on the initial condition itself \citep{Riols13}.
\subsection{Dependence of dynamo onset on $\text{Pm}$}
\label{s:Pm}
\begin{figure}[b]
\centering
\resizebox{0.7\hsize}{!}{\includegraphics{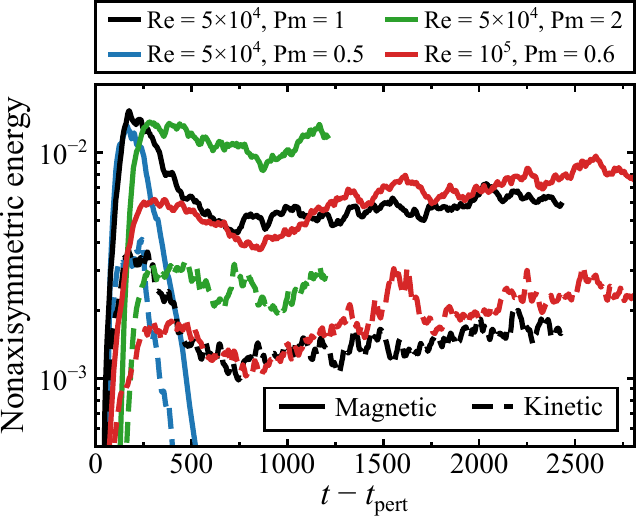}}
\caption{
Temporal evolution of the nonaxisymmetric kinetic (dashed lines)
and magnetic (solid lines) energies
for four unstratified runs at $\Rey=5\times 10^4$ and $\Rey=10^5$ and
different values of $\Pm$ as indicated
in the legend at the top.
The black lines show the fiducial dynamo run U0.
The perturbed axisymmetric azimuthal field solutions
of these runs are different.
}
\label{f:en_Pm}
\end{figure}
Local shearing box simulations with zero or nonzero net magnetic fluxes show
that viscous and resistive effects strongly affect
the saturated state of MRI turbulence \citep{Lesur07,Fromang07,Simon09}. 
In particular, magnetic diffusivity limits
the turbulence saturation level such
that, at fixed $\Rey$, dynamo action is lost
below a critical value of $\Pm$, a behavior that we
also find here.

\Figref{f:en_Pm} presents the temporal evolution of the
nonaxisymmetric kinetic and magnetic energies
in run U0 (black lines) and
in two other runs at the same $\text{Re}$ of $5\times 10^4$
but with $\text{Pm}=0.5$ and 2 (run U1).
After a few hundreds of system rotations,
run U1 (green lines) reaches a quasi steady state
covering a period $\Delta t=0.82\,\tau_\text{Ohm}$ (\tabref{t:runs_dyn}).
Given that the various energy components
are stationary and that the period of quasi steady
evolution is close to one Ohmic diffusion time, we consider this run as a dynamo.
The time averaged turbulent kinetic and magnetic
energies in the quasi steady state
increase by about a factor 2 relative to run U0,
in qualitative agreement
with the shearing box studies mentioned above.

When decreasing $\Pm$ to $0.5$, AMRI turbulence is maintained
only for some tens of system rotations and
then it decays away (blue lines in \figref{f:en_Pm}).
We note that the perturbed axisymmetric azimuthal field configurations
are different in all these runs, but their mean strength
increases when $\Pm$ lowers which, in principle, favors
dynamo action as discussed in the previous section.
While this cannot justify the observed dynamo behavior,
it may explain the large peak values of the 
turbulent energies of the nondynamo run
at $\Pm=0.5$, which are comparable to those
of runs U0 and U1.

The shearing box MRI studies mentioned above also indicate that
the critical $\Pm$ for dynamo onset
decreases when $\Rey$ increases \citep[e.g.,][]{Fromang07}.
At the largest $\Rey$ of $10^5$ that we explored here,
we indeed observe dynamo action for a value of $\Pm$
as low as 0.6 (run U3; red lines in \figref{f:en_Pm}).
After an initial transient of about 340 system rotations
the turbulence reaches a quasi steady
evolution lasting for $1.6\,\tau_\text{Ohm}$.
MRI dynamos at $\Pm<1$ were reported
in local shearing box simulations
with net magnetic fluxes \citep{Simon09,Kapyla11}
but never so far in global setups as the one explored here.
\subsection{Effect of stable stratification on AMRI turbulence}
\label{s:stable_strat}
We now study how stable stratification modifies
unstratified AMRI turbulence
at $\text{Re}=5\times 10^4$ and $\Pm=1$.
To this end, we solve Eqs.~\eqref{e:NS}-\eqref{e:temp} using as initial condition
the quasi steady solution of the fiducial dynamo run U0
at time $t_\text{s}=1241$ (\figref{f:dyn_snap}).
\begin{figure}
\centering
\resizebox{0.7\hsize}{!}{\includegraphics{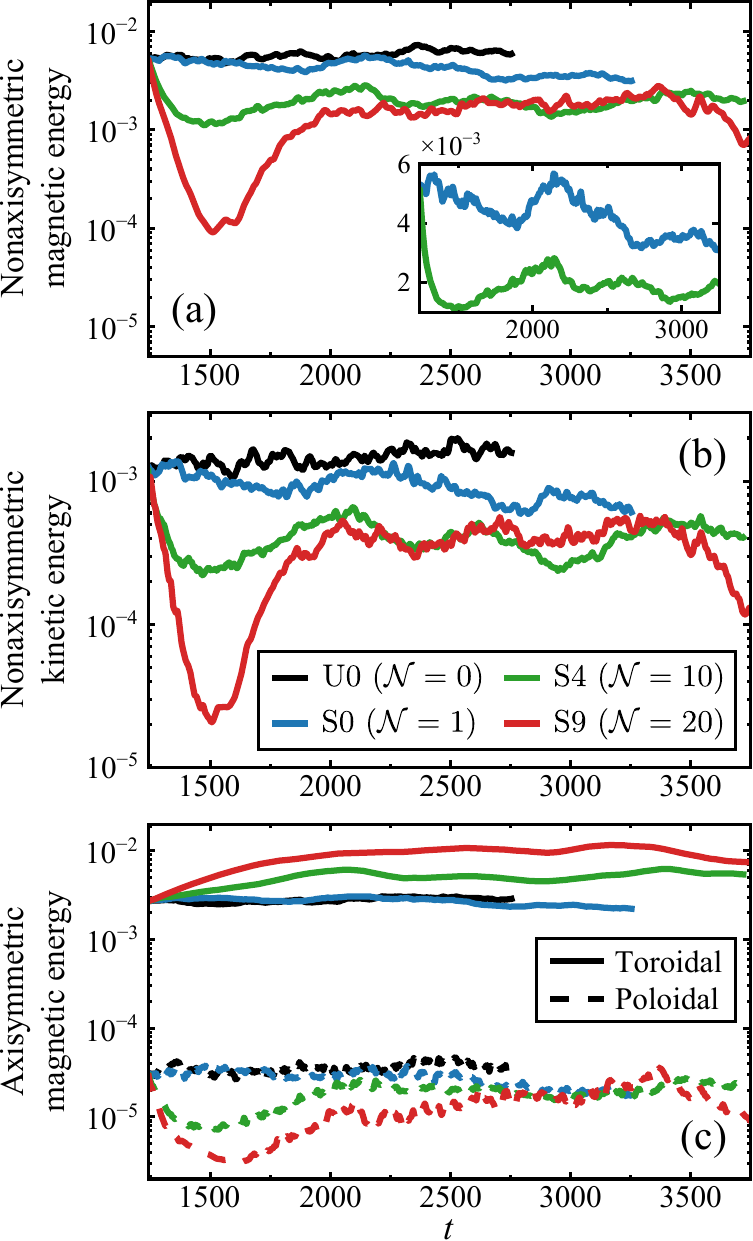}}
\caption{
(a)~Nonaxisymmetric magnetic energy, (b)~nonaxisymmetric kinetic energy,
(c)~axisymmetric toroidal (solid lines) and poloidal (dashed lines)
magnetic energies as a function of time
for the unstratified fiducial dynamo run U0 (black) and the three runs
S0, S4 and S9 at $\mathcal{N}=1$, $10$ and $20$ respectively
(see the legend in the middle panel).
All runs are at $\text{Re}=5\times 10^4$ and $\text{Pm}=1$
(and $\text{Pr}=10^{-3}$ for the stratified cases).
The inset in~(a) shows the nonaxisymmetric magnetic
energies of runs S0 and S4 on a linear scale.
The initial condition of the stratified runs is the steady state
solution of run U0 at time $t_\text{s}=1241$ shown in \figref{f:dyn_snap}.
}
\label{f:ener_strat}
\end{figure}

We first discuss the effect of stable stratification by varying
$\mathcal{N}$ at a fixed Prandtl number $\text{Pr}$ of $10^{-3}$.
Stable stratification strongly limits radial motions,
modifying the characteristics of the turbulence.
\Figref{f:ener_strat}a,b present the temporal evolution, starting at $t=t_\text{s}$,
of the turbulent (nonaxisymmetric)
magnetic and kinetic energies in run U0 (black lines)
and in the three stratified runs S0 ($\mathcal{N}=1$), S4 ($\mathcal{N}=10$)
and S9 ($\mathcal{N}=20$),
showing that they lower when increasing stratification.
After an initial transient which gets longer when increasing $\mathcal{N}$,
the quasi steady evolutions of runs S4
(green line) and S9 (red line)
cover 0.95 and $0.19\,\tau_\text{Ohm}$ respectively.
The turbulence in the weakly stratified run S0 (blue line)
slowly decays following the
resistive decay of the background axisymmetric field (blue lines in \figref{f:ener_strat}c).
By limiting radial motions, stable stratification
lowers the effective magnetic Reynolds number
$\text{Rm}_\text{eff} = u_\text{rms}\,d\big/ \eta$
below the critical value for dynamo onset in our runs,
which is of about 820 based on the unstratified simulations (\tabref{t:runs_dyn}).
Here $u_\text{rms}=\left(V^{-1}\int \left| \textbf{u}-\textbf{u}_\text{f}\right|^2\text{d}V\right)^{1/2}$
is the rms flow velocity after
subtracting the forced azimuthal flow.
The energies of all stratified runs explored here
display an oscillatory behavior (see the inset
in \figref{f:ener_strat}a) which is characteristic
of stratified MRI turbulence and is often attributed
to an $\alpha\Omega$-dynamo process \citep{Gressel10,ReboulSalze22}.

\begin{figure}
\centering
\resizebox{\hsize}{!}{\includegraphics{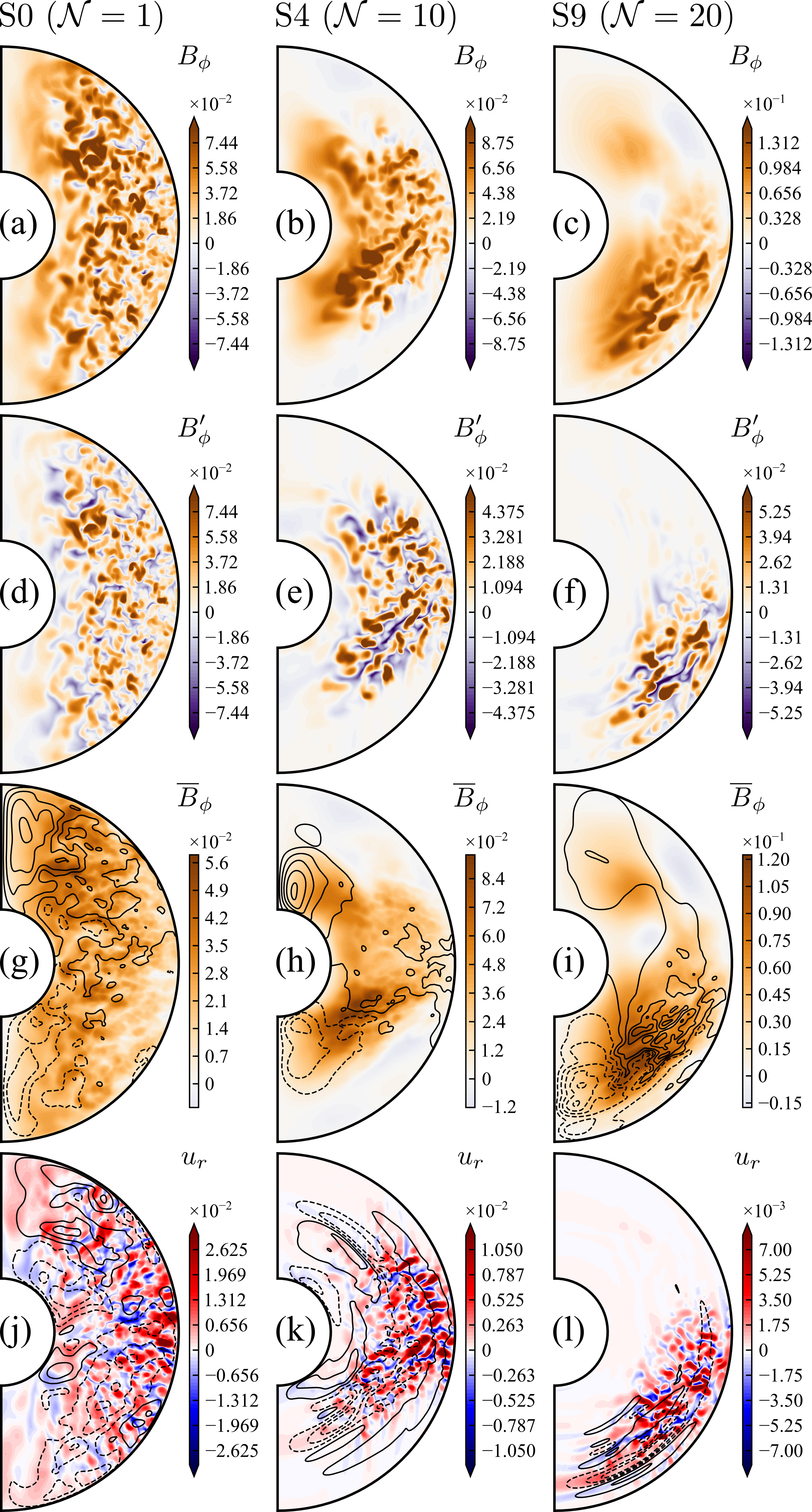}}
\caption{
Snapshots of the flow and magnetic field solutions
of runs S0, S4 and S9 shown in \figref{f:ener_strat}.
From top to bottom: meridional cuts of the azimuthal field $B_\phi$, nonaxisymmetric
azimuthal field $B_\phi^\prime$, axisymmetric azimuthal field $\overline{B}_\phi$
and radial flow velocity $u_r$.
Black isocontours superimposed on $\overline{B}_\phi$ and $u_r$
show the axisymmetric poloidal field and the meridional circulation respectively.
}
\label{f:snap_strat}
\end{figure}
The location and structure of the turbulence also
change with stratification.
The snapshots of $B_\phi^\prime$
in \figref{f:snap_strat}d-f reveal that
the turbulence in the weakly stratified run S0
remains homogeneous outside the tangent cylinder
as in the unstratified fiducial dynamo,
while the instability is confined
up to intermediate latitudes in run S4
and is almost entirely located in the southern hemisphere
in the most stratified run S9.
The unstable regions in the latter two runs
correlate with the locations where the axisymmetric
azimuthal field $\overline{B}_\phi$ is stronger (\figref{f:snap_strat}h,i)
and unstable to AMRI according to
the local linear analysis, which predicts $5.5\times 10^{-3}<\overline{B}_\phi\leq 0.87$
based on the instability condition \eqref{e:LeAMRI}.
Outside these locations, $\overline{B}_\phi$ is too weak, with typical absolute
amplitudes of $10^{-3}$ or smaller, and AMRI modes are
damped by diffusive effects.

In the most stratified run S9, the turbulence becomes weakly anisotropic,
with structures
somewhat elongated in the latitudinal direction as expected (\figref{f:snap_strat}f).
The horizontal turbulent field length scale
$l^\prime_{B,\perp}/\sg$ is
0.12 in this run, which is $60\%$ larger
than the value of the weakly stratified run S0 (\tabref{t:runs_dyn}).
In the latitudinal direction, the structures of the flow are less
stretched than those of the field, as evidenced by
the snapshots of $B_\phi$ (\figref{f:snap_strat}c) and $u_r$ (\figref{f:snap_strat}l).
The horizontal field length scale $l_{B,\perp}/\sg$
is indeed as large as $0.76$, whereas the one of the
flow is lower at $l_{u,\perp}/\sg=0.27$.

It is interesting to note that the axisymmetric
toroidal magnetic energy increases with stratification (solid lines in \figref{f:ener_strat}c).
This energy contribution dominates over
the nonaxisymmetric one for $\mathcal{N}>1$, so that
$B_\phi$ is characterized
by larger spatial scales as stratification increases (\figref{f:snap_strat}a-c).
On the contrary, the axisymmetric poloidal magnetic energy
weakly lowers with stratification (dashed lines in \figref{f:ener_strat}c).
As a consequence, the ratio $\overline{B}_{\phi,\text{rms}}/\overline{B}_{\text{p,rms}}$,
where $\overline{B}_{\phi,\text{rms}}$ ($\overline{B}_{\text{p,rms}}$)
is the time averaged rms axisymmetric azimuthal (poloidal) field,
increases when increasing $\mathcal{N}$ and ranges
between 10 and 25 in our stratified simulations (\tabref{t:runs_fields}).
In the mean induction equation, the source terms that generate
the axisymmetric toroidal field $\overline{B}_\phi$
are the radial mean EMF $\overline{\mathcal{E}}_r$
and the term associated to the $\Omega$-effect, $s(\overline{\textbf{B}}_\text{p}\cdot\boldsymbol{\nabla})\Omega$,
where $\overline{\textbf{B}}_\text{p}$ is the axisymmetric poloidal field.
In runs S4 and S9,
$\overline{B}_\phi$ is mostly concentrated in large flux patches
located where the turbulence is most active
which suggests a positive feedback of $\overline{\mathcal{E}}_r$
on its generation.
The more coherent structure of the turbulence
obtained when increasing stratification
may yield locally stronger $\overline{\mathcal{E}}_r$,
which then causes the observed increase in $\overline{B}_\phi$.
Although $\overline{B}_\text{p}$
is weak in all runs explored, the $\Omega$-effect
-- a key ingredient of MRI dynamos as mentioned before --
is still of leading order due to the strong background shear.
The positive feedback of the $\Omega$-effect 
on the generation of $\overline{B}_\phi$
is suggested, for example,
by the positive flux patch in the northern hemisphere
observed in run S9 (\figref{f:snap_strat}i).
The maximum of this flux patch is indeed located
where $\overline{B}_\text{p}$ (black
isocontours in \figref{f:snap_strat}i) is roughly antiparallel to
the shear direction $\hat{\mathbf{e}}_s$ as expected.

We note that the ratio of the time averaged rms azimuthal field to
the radial one $B_{\phi,\text{rms}}/B_{r,\text{rms}}$
ranges between 3 and 5 in the unstratified and weakly
stratified runs at $\mathcal{N}=1$ and increases
when stable stratification strengthens.
At $\text{Pr}=10^{-3}$, for example, $B_{\phi,\text{rms}}/B_{r,\text{rms}}$
grows from 4.1 to 14.8 when increasing $\mathcal{N}$
from 1 to 20.
This field ratio is in the range $2-6$
in global compressible simulations of stratified MRI turbulence
in accretion disks \citep{Hawley11,Hawley13}, while
local shearing box simulations obtain values around 2 \citep{Shi10}.

By reducing the amplitude of the temperature
fluctuations, thermal diffusion can weaken
the stabilizing buoyancy force
in the momentum equation \eqref{e:NS}.
This occurs when thermal diffusion acts faster than the
buoyancy force, that is when $\tau_\kappa < \tau_N$,
where $\tau_\kappa=l_{r}^2/\kappa$ is the thermal diffusion timescale
at the radial flow length scale $l_r$ and $\tau_N=1/N$ is the buoyancy timescale.
By equating these two timescales, we obtain the critical (radial) flow length scale
$l_\text{c}=(\kappa/N)^{1/2}$, or $l_\text{c}/d=\left(\text{Pr}\,\text{Re}\,\mathcal{N}\right)^{-1/2}$,
below which thermal diffusion weakens buoyancy.
To explore the effect of thermal diffusion
in our stratified simulations at $\text{Re}=5\times 10^4$ and $\Pm=1$, we
varied the Prandtl number $\text{Pr}$ at fixed $\mathcal{N}$.

\begin{figure}
\centering
\resizebox{\hsize}{!}{\includegraphics{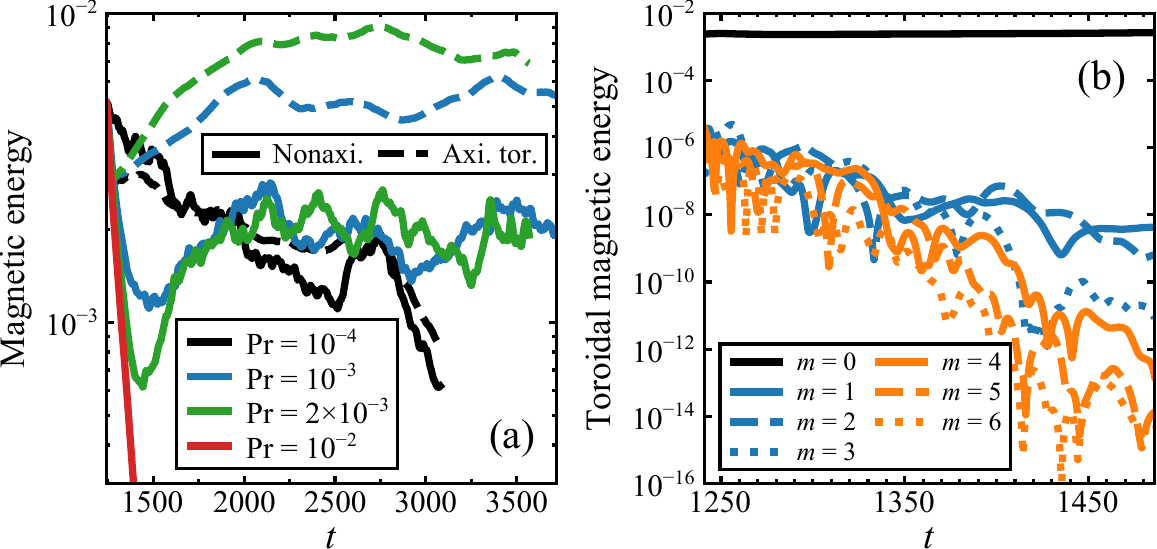}}
\caption{(a)~Nonaxisymmetric (solid lines) and axisymmetric toroidal (dashed lines)
magnetic energies as a function of time
in runs S3$-$S6 ($\Rey=5\times 10^4$, $\Pm=1$, $\mathcal{N}=10$
and different values of $\text{Pr}$ as indicated in the legend).
Only the nonaxisymmetric energy is shown for run S6 at $\text{Pr}=10^{-2}$.
(b)~Toroidal magnetic energy at radius $r/r_{\text{o}}=0.5$
as a function of time
for the azimuthal modes $m=0-6$ for the stable run S6.
}
\label{f:enerPr}
\end{figure}
\Figref{f:enerPr}a displays the temporal evolution
of the turbulent magnetic energy (solid lines) in four runs at $\mathcal{N}=10$
with $\text{Pr}$ increasing from $10^{-4}$ to $10^{-2}$.
While AMRI
is observed for $\Prt\leq 2\times 10^{-3}$, further increasing the Prandtl number to $10^{-2}$ (solid red line)
suppresses the instability. In this stable run the nonaxisymmetric modes
decay at a rate increasing with the azimuthal wavenumber $m$ as expected (\figref{f:enerPr}b).

The analysis of a few snapshots of the turbulent radial velocity $u_r^\prime$
shows that the radial flow length scale $l_r$, obtained from
a volume average of
$u_r^\prime\Big/\left|\er\cdot\nabla u_r^\prime\right|$,
is $\lesssim O(l_\text{c})$ for $\Prt\leq 2\times 10^{-3}$, which suggests
that buoyancy effects are limited by thermal diffusion in these runs.
In the run at $\text{Pr}=10^{-2}$, on the contrary,
$l_r /d \approx 0.05$ is several times larger than
the critical length scale $l_{\text{c}}/d=0.014$, hence thermal diffusion
cannot reduce the effect of stable stratification, which is too strong
and suppresses the instability.
Similarly, we observe AMRI turbulence at $\mathcal{N}=20$
when $\text{Pr}\leq 10^{-3}$, whereas the instability is suppressed
at the larger $\text{Pr}$ of $5\times 10^{-3}$ (\tabref{t:runs_dyn}).
A similar behavior where increasing buoyancy effects stabilize
the system is also typical of hydrodynamical shear instabilities in 
vertically stratified flow \citep{Lignieres99}.
We remark here that strong enough latitudinal gradients
of the differential rotation allow AMRI to develop
even when $l_{\text{c}}\ll l_r$ and
thermal diffusion does not limit stable stratification \citep{Jouve20}.

Finally, we note that all stratified AMRI runs
show either transient turbulence or cover periods
too short to test for dynamo action. For example, at $\mathcal{N}=10$,
we found decaying turbulence at $\text{Pr}=10^{-4}$ (\figref{f:enerPr}a, solid black line).
The decay rate of the turbulent fluctuations is
compatible with the one of the axisymmetric azimuthal field
which sustains the instability (\figref{f:enerPr}a, dashed black line).
In the two runs at $\text{Pr}=10^{-3}$ and $2\times 10^{-3}$,
the turbulence is sustained for 0.19 and 0.09 Ohmic diffusion times $\tau_\text{Ohm}$
respectively and it may decay on longer timescales not captured here (\tabref{t:runs_dyn}).
These periods correspond to 140 and 115 eddy turnover times
$\tau_{\text{eddy}}$,
which ensure a robust statistics for the turbulence.
Hereafter $\tau_{\text{eddy}}$ is defined as the time averaged ratio
$l^\prime_{u,\perp}/u^\prime_{\text{rms}}$.
\section{Angular momentum transport}
\label{s:AM}
In this section we examine the transport of AM
induced by AMRI in the unstratified and stratified simulations
described above.
An equation of conservation for the specific AM
$L=s\,\overline{u}_\phi=s^2\overline{\Omega}$
can be obtained by averaging the
azimuthal component of the momentum equation \eqref{e:NS} over longitude
and multiplying both members by $s=r\sin\theta$,
\beq
\label{e:am}
\frac{\partial L}{\partial t} - s\,\textbf{f}\cdot\ephi=-\Div \left( \Fmc + \Frs + \Fvd + \Fms + \Fmt \right),
\eeq
where 
\begin{align}
\Fmc &= \azavg{\bfu}_\text{M}\,L , \label{e:fmc}\\
\Frs &= s\,\azavg{\urf\upf}\,\er + s\,\azavg{\utf\upf}\,\etheta , \label{e:frs}\\
\Fvd &=-\frac{1}{\Rey}\,s^2\Grad \overline{\Omega} , \label{e:fvd}\\
\Fms &=-s\,\azavg{\Brf\Bpf}\,\er - s\,\azavg{\Btf\Bpf}\,\etheta , \label{e:fms}\\
\Fmt &=-\azavg{B}_\phi\,\azavg{\bfB}_\text{M} \label{e:fmt}
\end{align}
are the fluxes of the meridional circulation~(MC), Reynolds stresses (RS),
viscous diffusion (VD), Maxwell stresses (MS) and magnetic tension (MT), respectively.
Hereafter the subscript $_\text{M}$ indicates the
meridional component (e.g., $\overline{\bfu}_\text{M}=\overline{u}_r \er + \overline{u}_\theta \etheta$).
Note that there is no contribution from the magnetic pressure $B^2/2$ in \eqnref{e:am}
since the longitudinal
magnetic pressure gradient vanishes when integrated over $\phi$.

We decompose the flux terms~(\ref{e:fmc})-(\ref{e:fmt}) into the sum of their radial
and latitudinal contributions, that is
\begin{equation}
\textbf{F}^i = F_r^i\,\er + F_\theta^i\,\etheta
\end{equation}
where $i=\{\text{MC},\text{RS},\text{VD},\text{MS},\text{MT}\}$;
for example, the Reynolds stresses flux is $\Frs = F_r^{\text{RS}}\er + F_\theta^{\text{RS}}\etheta$,
where $F_r^{\text{RS}}=s\,\azavg{\urf\upf}$ and $F_\theta^{\text{RS}}=s\,\azavg{\utf\upf}$.
To assess the net AM transport in the radial and latitudinal
directions, following \cite{BrunToomre02}, we integrate $F_r^i$ and $F_\theta^i$ over spherical surfaces
of varying radii and over cones of varying inclination respectively,
\beq
\label{e:int_flx_r}
I_r^i(r,t)=\int_0^\pi F_r^i (r,\theta,t)\,r^2\sin\theta\, \text{d}\theta
\eeq
and
\beq
\label{e:int_flx_th}
I_\theta^i(\theta,t) = \int_{\rin}^{\rout} F_\theta^i (r,\theta,t)\,r\sin\theta\, \text{d}r .
\eeq
We then time average these integrated fluxes, obtaining
$\hat{I}_r^i$ and $\hat{I}_\theta^i$ respectively.
In the unstratified dynamo runs, the time averages
are calculated over quasi steady states generally covering
intervals $\Delta t$ of a few hundreds
of eddy turnover times $\tau_\text{eddy}$ (\tabref{t:runs_dyn}).
In the stratified runs the turbulence is transient and no steady
state is reached, so that the time averaging interval $\Delta t$
is defined by periods, after the initial transient,
where the turbulent magnetic energy does not decay
significantly. Such periods cover at least
$56\,\tau_\text{eddy}$ (\tabref{t:runs_dyn}), which still ensures a robust
statistics for the turbulence.
\subsection{Unstratified dynamo runs}
\label{s:am_unstrat}
\begin{figure}[b]
\centering
\includegraphics[width=\hsize]{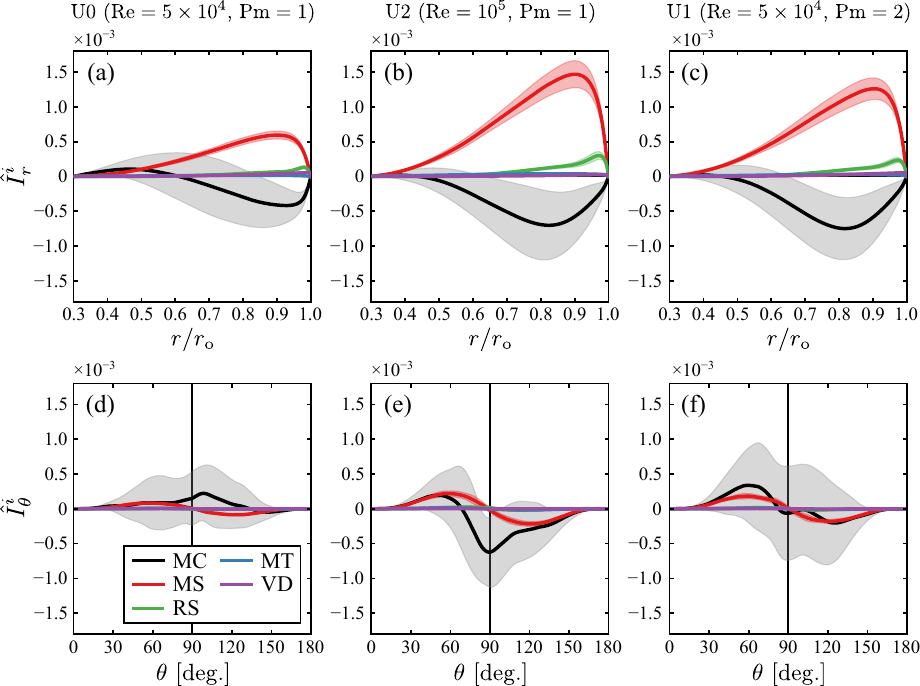}
\caption{
Time averaged integrated fluxes (a-c) $\hat{I}_r^i$ and (d-f) $\hat{I}_\theta^i$
for the unstratified runs U0, U2 and U1 (from left to right).
The shaded area around each flux contribution
indicates 1 standard deviation above and below the time average.
The scale of the vertical axis is the same in all panels.}
\label{f:am_flx_unstrat}
\end{figure}
\begin{figure*}[ht]
\centering
\includegraphics[width=0.9\hsize]{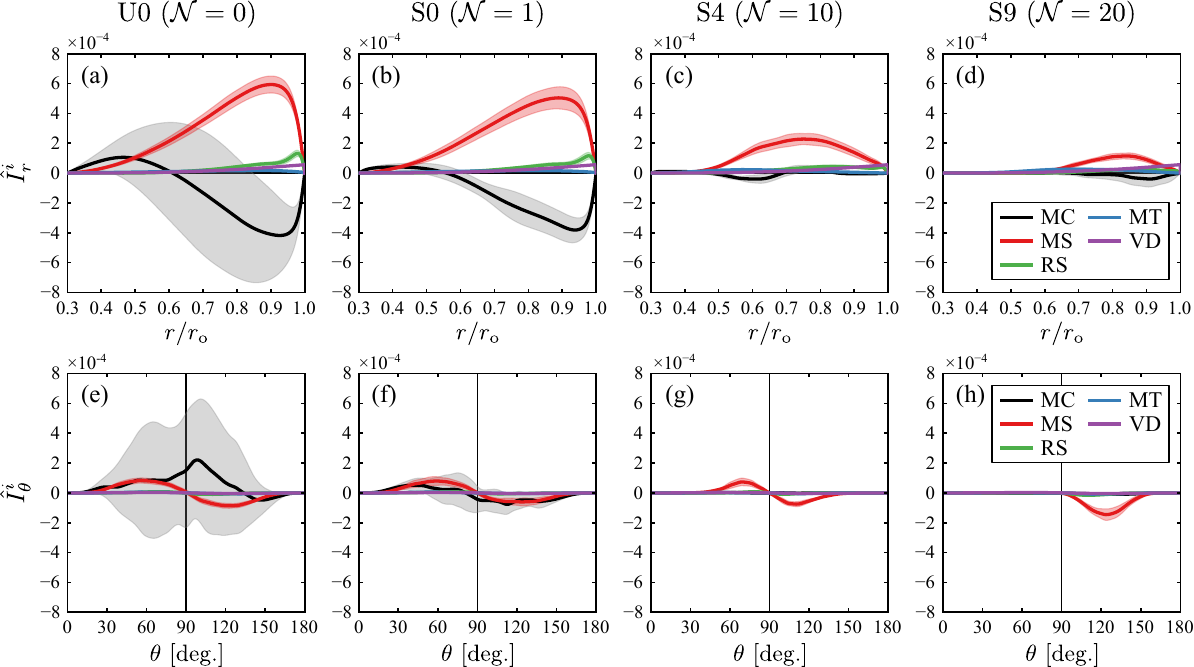}
\caption{Same as \figref{f:am_flx_unstrat} but for the
three stratified runs S0, S4 and S9 at $\text{Pr}=10^{-3}$
and $\mathcal{N}=1$, $10$ and $20$ respectively
(second to fourth from left panels).
The leftmost panels show the unstratified fiducial dynamo run U0
for the sake of comparison. All runs are at $\text{Re}=5\times 10^4$ and
$\text{Pm}=1$.
}
\label{f:am_flx_strat}
\end{figure*}
First, we analyze the AM transport in the unstratified dynamo runs
described in Sects.~\ref{s:dyn_fiducial}-\ref{s:Pm}.
\Figref{f:am_flx_unstrat}a,d present the time averaged
integrated fluxes $\hat{I}_r^i$ and $\hat{I}_\theta^i$
in the fiducial dynamo run U0.
In both directions, the Maxwell stresses (red lines)
and meridional circulation (black lines) fluxes
largely dominate over all the other contributions.
In the radial direction, the third largest contribution
is the Reynolds stresses flux $\hat{I}_r^{\text{RS}}$ (green line in \figref{f:am_flx_unstrat}a)
with a peak amplitude more than 4 times lower
than the one of the Maxwell stresses flux $\hat{I}_r^{\text{MS}}$.
Viscous diffusion and magnetic tension come next,
with peak amplitudes of $\hat{I}_r^{\text{VD}}$ and
$\hat{I}_r^{\text{MT}}$ about 10 and 30 times smaller
than the one of $\hat{I}_r^{\text{MS}}$.
In the latitudinal direction, the Reynolds stresses, viscous diffusion
and magnetic tension
fluxes all have comparable amplitudes
and, similarly to the radial fluxes, their peak values
are one order of magnitude lower than
$\hat{I}_\theta^{\text{MS}}$.

The color shaded areas in \figref{f:am_flx_unstrat}a,d
show 1 standard deviation intervals around the time averaged fluxes
and evidence that the temporal variability of the
meridional circulation fluxes is much higher than
the one of all the other terms.
This high variability is due to an oscillatory
behavior of the meridional flow (dotted red line in \figref{f:dyn_ener})
where the dominant large scale meridional circulation
cells characterizing the steady state solution (black isocontours in \figref{f:dyn_snap}c)
are regularly replaced by new ones of opposite sign.
These new meridional circulation cells
are generated by radial flow plumes
arising at mid and high latitudes
in the inner fluid regions.

Since the radial fluxes of the prevailing Maxwell stresses
and meridional circulation dominate over the respective
latitudinal contributions, the transport of AM mainly occurs in
the radial direction.
The Maxwell stresses transport AM radially outwards
since $\hat{I}_r^{\text{MS}}>0$.
The meridional circulation contributes to the outward transport
in the internal fluid regions where
$r/\rout\lesssim 0.6$, while it opposes in the external
regions at larger radii (black line in \figref{f:am_flx_unstrat}a).
It is not surprising that the meridional flow
contributes to the radial transport.
In fact, the radial length scales
of the meridional circulation cells
are comparable to the characteristic scale of the background shear
$l_{\Delta\Omega}\approx r_\text{o}$ (\figref{f:dyn_snap}c, black isocontours).

The larger amplitudes of $\hat{I}_r^{\text{MS}}$
with respect to $\hat{I}_r^{\text{MC}}$ suggest
a net outward AM transport.
This is confirmed by an additional numerical experiment that we performed
by stopping the forcing ($\textbf{f}=\mathbf{0}$)
during the quasi steady evolution of run U0 and
letting the azimuthal flow free to evolve.
We observe the cylindrical rotation profile flattening
over time,
with the azimuthal flow decelerating in the interior
and accelerating in the equatorial region to reach uniform rotation
on a few hundreds of rotation times $\tau_\Omega$
(Appendix~\ref{s:appendix_no_forc}).

Although negligible relative to the radial one,
the transport of AM in the latitudinal direction
is also dominated by the Maxwell stresses and
the meridional circulation.
The former produce an equatorward transport since
$\hat{I}_\theta^{\text{MS}}$ is equatorially
antisymmetric, with a positive (negative) contribution
in the northern (southern) hemisphere (\figref{f:am_flx_unstrat}d, red line).
$\hat{I}_\theta^{\text{MC}}$ is mostly positive
(\figref{f:am_flx_unstrat}d, black line), hence it contributes to the equatorward
transport by the Maxwell stresses in the northern hemisphere
and it opposes in the southern one.
The amplitudes of $\hat{I}_\theta^{\text{MC}}$ 
are generally comparable to the one of $\hat{I}_\theta^{\text{MS}}$, except
in the equatorial region and at low
latitudes in the southern hemisphere where the former is larger.

We now discuss how the transport varies
with $\Rey$ and $\Pm$.
The middle and right panels of \figref{f:am_flx_unstrat}
display the time averaged integrated fluxes
of the dynamo runs U2 and U1 respectively, which we
discussed in Sects.~\ref{s:dyn_onset} and \ref{s:Pm}.
Relative to the fiducial dynamo run U0, run U2 has a
larger $\Rey$ of $10^5$ and run U1 a larger $\Pm$ of 2.
As for the fiducial dynamo, the radial transport in these runs
dominates over the latitudinal one
and we therefore discuss only the former in the following (\figref{f:am_flx_unstrat}b,c,e,f).
The net transport occurs radially outwards
due to the prevailing radial Maxwell stresses (\figref{f:am_flx_unstrat}b).
The peak amplitudes of $\hat{I}_r^{\text{MS}}$
in runs U2 and U1 are, respectively, about 3 and 2
times higher than the one of run U0, which
suggest similar variations in the AM transport efficiency.
The time averaged rms turbulent field strenghts
$B_\text{rms}^\prime$ of runs U2 and U1 are
$50\%$ and $41\%$ larger than the one of run U0 respectively,
which contributes to explain the observed
variations in the Maxwell stresses flux (\tabref{t:runs_fields}).
The time averaged rms turbulent flow velocity $u_\text{rms}^\prime$
in these runs
is also a few tens of percent higher
than the one of run U0 and contribute to explain the increase of
the radial Reynolds stresses flux $\hat{I}_r^{\text{RS}}$ (green lines in \figref{f:am_flx_unstrat}b,c).

As for run U0, the meridional circulation in runs U2 and U1
opposes the radial transport by the Maxwell stresses
over most of the fluid domain
(black lines in \figref{f:am_flx_unstrat}b,c).
The meridional circulation in these runs is still large scaled
and its rms amplitude $\overline{u}_{\text{M,rms}}$
is about $20\%$ higher than the one of run U0 (\tabref{t:runs_fields}),
which can explain the observed
variations of $\hat{I}_r^{\text{MC}}$.
These faster meridional flows
may result from the higher Reynolds and Maxwell stresses
which contribute to the meridional circulation maintenance
\citep[see, e.g.,][]{Miesch05}.
As for run U0, viscous diffusion and
magnetic tension play a negligible role in the transport.
Similar variations of the flux contributions with $\text{Pm}$
are obtained at $\text{Re}=10^5$ by comparing run U2
with run U3 at $\text{Pm}=0.6$ (not shown).

The weaker dependence of the Reynolds and Maxwell stresses
on $\text{Pm}$ than on $\text{Re}$ that we just discussed
qualitatively agrees
with previous results obtained from local and global simulations
of unstratified MRI turbulence.
\cite{Guseva17a} showed that the transport coefficient, the sum
of the Reynolds and Maxwell stresses, of AMRI turbulence in
Taylor-Couette flow scales as $\text{Pm}^{1/2}\text{Re}$
when $\text{Rm}\gtrsim 10^2$.
Local shearing box simulations with zero net flux suggest
a similar scaling with $\text{Pm}$, although the variation
with $\text{Re}$ is much weaker than the linear one
obtained for Taylor-Couette flow \citep{Lesur07}.
\subsection{Stably stratified runs}
\label{s:am_strat}
By weakening radial motions, stable stratification
modifies the transport of AM.
Here we discuss how the transport varies
with stratification considering the three runs S0, S4 and S9
($\text{Pr}=10^{-3}$ and $\mathcal{N}=1$, 10 and 20 respectively)
discussed in \secref{s:stable_strat}.
\Figref{f:am_flx_strat} displays the time averaged integrated
fluxes $\hat{I}_r^i$ and $\hat{I}_\theta^i$
in these runs, together with the fiducial dynamo run U0
in the two leftmost panels for the sake of reference.
In the weakly stratified run S0, the transport occurs
similarly to the unstratified case.
Only a small decrease of the dominant Maxwell stresses and
meridional circulation fluxes is observed (red and black lines in \figref{f:am_flx_strat}b,f).
However, the meridional circulation shows
a much weaker variability, as demonstrated by the small
1 standard deviation intervals around
$\hat{I}_r^\text{MC}$ and $\hat{I}_\theta^\text{MC}$
(gray shaded regions in \figref{f:am_flx_strat}b,f),
since the oscillatory behavior observed for run U0
is absent.

Increasing $\mathcal{N}$ to 10 (run S4) produces a
sizeable decrease of the radial Maxwell stresses flux $\hat{I}_{r}^{\text{MS}}$
(red line in \figref{f:am_flx_strat}c).
The latitudinal Maxwell stresses flux $\hat{I}_{\theta}^{\text{MS}}$
is confined at low to intermediate latitudes
(red line in \figref{f:am_flx_strat}g), which correlates
with the locations where the instability
is active (\figref{f:snap_strat}e).
The transport by the meridional flow
becomes almost negligible (black lines in \figref{f:am_flx_strat}c,g)
since radial flow motions are
severely limited by the stabilizing buoyancy force.
The meridional flow amplitude $\overline{u}_{\text{M,rms}}$
is indeed 2.5 times smaller than the one of the
weakly stratified run S0 (\tabref{t:runs_fields})
and multiple meridional circulation cells elongated
in the latitudinal direction are observed
(black isocontours in \figref{f:snap_strat}k).
The horizontal flow length scale $l_{u,\perp}/d$
increases to $0.22$, while is smaller at $0.17$ in run S0 as expected.

In run S9 at $\mathcal{N}=20$, $\hat{I}_{r}^{\text{MS}}$
further weakens and its peak amplitude becomes
comparable to the respective latitudinal contribution $\hat{I}_{\theta}^{\text{MS}}$ (red lines in \figref{f:am_flx_strat}d,h).
The latitudinal Maxwell stresses flux
is concentrated in the southern hemisphere
at the locations where the axisymmetric azimuthal
field is strong enough to support AMRI (\secref{s:stable_strat}).
The transport by the meridional circulation remains
very weak as in run S4 (black lines in \figref{f:am_flx_strat}d,h)
and no significant variation in the meridional
flow amplitude is observed (\tabref{t:runs_fields}).
However, the meridional circulation cells
become thinner
in the radial direction as expected
(black isocontours in \figref{f:snap_strat}l).

\begin{figure}[ht]
\centering
\resizebox{0.95\hsize}{!}{\includegraphics{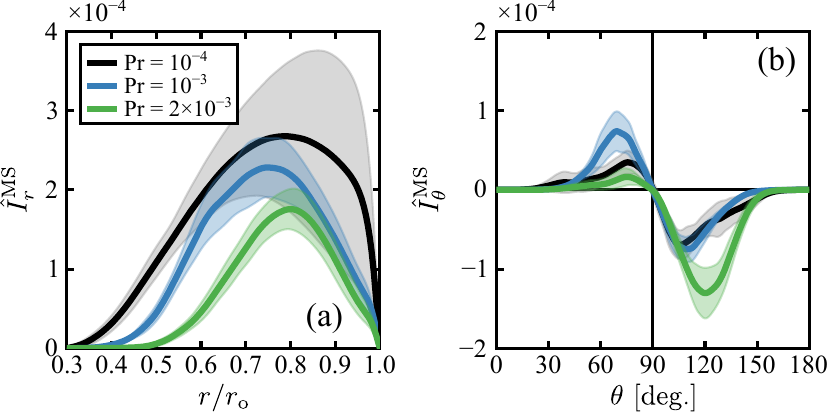}}
\caption{Time averaged integrated (a) radial and (b) latitudinal
Maxwell stresses fluxes in runs S3, S4 and S5
at $\mathcal{N}=10$ and $\text{Pr}=10^{-4}$, $10^{-3}$ and
$2\times 10^{-3}$ respectively.
All runs are at $\text{Re}=5\times 10^4$ and
$\text{Pm}=1$.
The shaded areas denote 1 standard deviation
intervals around the time averages.
}
\label{f:MaxwS_Pr}
\end{figure}
Similar results on the variation of the integrated fluxes
are obtained when strengthening buoyancy effects
by reducing thermal diffusion, that is when increasing $\text{Pr}$
at fixed $\mathcal{N}$.
Since the Maxwell stresses dominate the transport, we
discuss only their variation with $\text{Pr}$ here.
\Figref{f:MaxwS_Pr} presents $\hat{I}_{r}^{\text{MS}}$
and $\hat{I}_{\theta}^{\text{MS}}$ in runs S3, S4 and S5
at $\mathcal{N}=10$ and $\text{Pr}=10^{-4}$, $10^{-3}$ and
$2\times 10^{-3}$ respectively.
As for the runs discussed above, the radial Maxwell stresses flux
prevails over the latitudinal one.
The amplitude of $\hat{I}_r^\text{MS}$ lowers
when $\text{Pr}$ increases and the distribution
shifts towards the outer fluid regions where the instability
is active (\figref{f:MaxwS_Pr}a).
Similarly to what has been observed above for run S9, 
$\hat{I}_\theta^{\text{MS}}$ concentrates in the southern hemisphere
when buoyancy effects are stronger
at the larger $\text{Pr}$ of $2\times 10^{-3}$ (green line in \figref{f:MaxwS_Pr}b).

In all stratified runs explored here, Reynolds stresses,
viscous diffusion and magnetic tension contribute
very weakly to the AM transport in both the radial
and latitudinal directions.
Magnetic tension is the weakest of all contributions
since $\overline{B}_\text{p}$ is always small as we
discussed in \secref{s:stable_strat}.
\subsection{Turbulent viscosity}
\label{s:turbvisc}
We demonstrated above that the AM transport
induced by AMRI occurs radially outwards
and is dominated by the Maxwell stresses
when $\mathcal{N}>1$.
The meridional circulation contribution
is significant only in the unstratified and
weakly stratified runs at $\mathcal{N}=1$.
To quantify the transport efficiency, we therefore consider
the contribution by the radial Maxwell stresses only 
and we assume the classical turbulent viscosity hypothesis
in which AM is transported in the direction of slow rotation,
\begin{equation}
\label{e:nuT}
-\frac{1}{\mu_0\rho}\overline{\Brf\Bpf} = - \nu_\text{T}\,s\frac{\partial\overline{\Omega}}{\partial r}.
\end{equation}
Here $\nu_\text{T}$ is the radial turbulent viscosity
and $\overline{\Omega}=\Omega_\text{f}$.
We obtain the turbulent viscosity estimate $\hat{\nu}_\text{T}$
from the relation above
by taking first a volume average of both members,
\begin{equation}
\label{e:nuTest}
 \nu_\text{T} = (\mu_0\rho)^{-1}\langle\overline{\Brf\Bpf}\rangle\Big/\langle s\,\partial\Omega_\text{f}/\partial r\rangle,
\end{equation}
and then by time averaging \eqref{e:nuTest} over the intervals $\Delta t$
defined before and listed in \tabref{t:runs_dyn}.
In runs showing transient turbulence, $\hat{\nu}_\text{T}$
has to be interpreted as an upper estimate
for the transport efficiency since the instability
decays on timescales longer than $\Delta t$.
The dimensionless turbulent viscosity estimate
is $\widetilde{\nu}_\text{T}=\hat{\nu}_\text{T}/\Delta\Omega\,r_\text{o}^2$, where
we employed the shear timescale $\tau_{\Delta\Omega}=\Delta\Omega^{-1}$
and the shear length scale $l_{\Delta\Omega}\approx r_\text{o}$ as reference scales.

\begin{figure}
\centering
\resizebox{0.9\hsize}{!}{\includegraphics{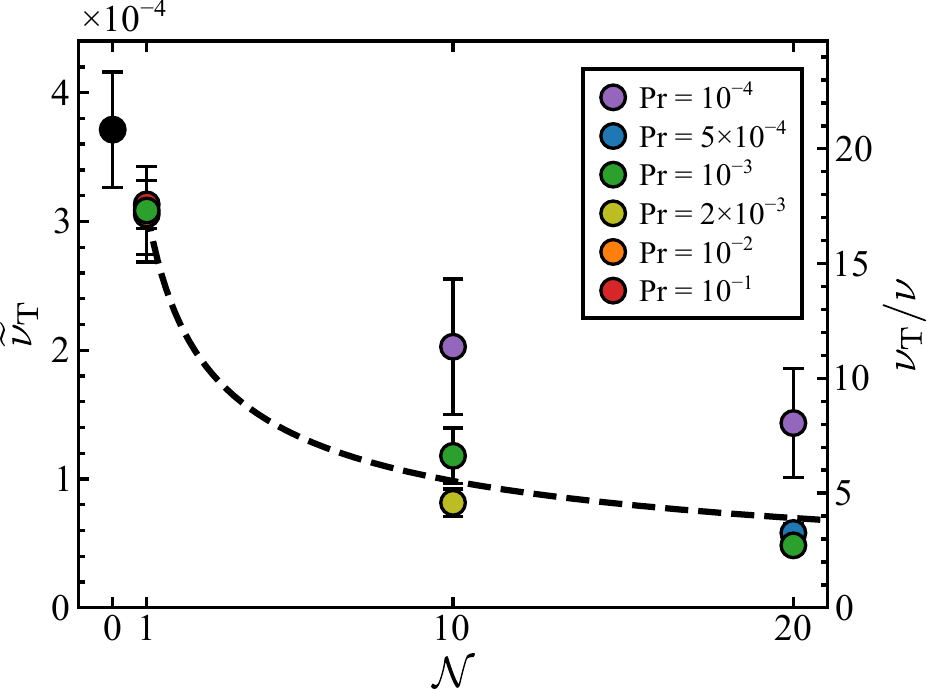}}
\caption{Turbulent viscosity $\widetilde{\nu}_{\text{T}}$
as a function of $\mathcal{N}$
for $\text{Re}=5\times 10^4$ and $\text{Pm}=1$.
The symbol color shows the Prandtl number
$\text{Pr}$ as indicated in the legend.
The black circle at $\mathcal{N}=0$ is the fiducial dynamo run U0.
The error bars show 2 standard deviations intervals around
the time averages, which are evaluated over
the periods $\Delta t$ listed in \tabref{t:runs_dyn}.
The dashed curve shows a power law fit $\widetilde{\nu}_\text{T}=a\,\mathcal{N}^{-\delta}$
of the stratified runs at $\text{Pr}=10^{-3}$.
The best fitting parameters are $a=3.1\times 10^{-4}$ and $\delta=0.50$.
The right vertical axis displays the turbulent viscosity
in units of the molecular viscosity $\nu$.
}
\label{f:turb_visc}
\end{figure}
\Figref{f:turb_visc} displays $\widetilde{\nu}_\text{T}$
as a function of $\mathcal{N}$ for all unstable runs
at $\text{Re}=5\times 10^4$ and $\text{Pm}=1$ explored here.
The turbulent viscosity lowers when buoyancy
effects strengthen, that is when $\mathcal{N}$ and/or
$\text{Pr}$ increase.
The largest value of $\widetilde{\nu}_\text{T}$ is $3.7\times 10^{-4}$
and is obtained for the unstratified fiducial dynamo run U0
as expected (black circle).
In the stratified runs at $\mathcal{N}>1$, the turbulent viscosity
approaches the unstratified value when $\text{Pr}$ decreases
since thermal diffusion limits
the stabilizing buoyancy force.
In the weakly stratified runs at $\mathcal{N}=1$ the effect
of stable stratification on the radial Maxwell stresses
is only marginal (\secref{s:am_strat})
and therefore $\widetilde{\nu}_\text{T}$ shows
only a weak decrease relative to the unstratified run.

\begin{figure}[b]
\centering
\resizebox{\hsize}{!}{\includegraphics{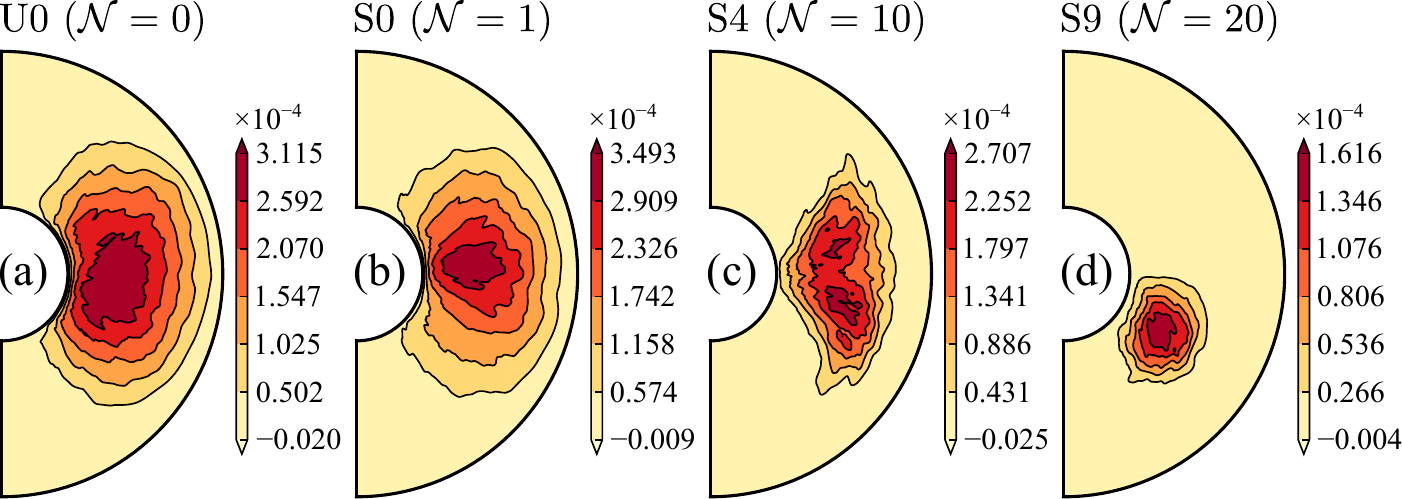}}
\caption{Time averaged radial Maxwell stresses fluxes $\hat{F}_r^{\text{MS}}$
for (a) the unstratified fiducial dynamo run U0 and (b-d) the stratified runs S0, S4 and S9 at $\text{Pr}=10^{-3}$
and $\mathcal{N}=1$, 10 and 20 respectively. All runs at $\text{Re}=5\times 10^4$ and $\text{Pm}=1$.}
\label{f:flx_MS_r}
\end{figure}
At fixed Prandtl number $\text{Pr}$, the turbulent
viscosity $\widetilde{\nu}_\text{T}$
lowers when increasing $\mathcal{N}$.
At $\text{Pr}=10^{-3}$, for example, $\widetilde{\nu}_\text{T}$ varies by
roughly a factor 6 in the range of stratification explored.
When $\mathcal{N}$ increases from 1 (run S0) to 10 (run S4),
such variation is mostly due to changes
in the turbulent field amplitudes.
In fact, $\widetilde{\nu}_\text{T}$ decreases by more
than a factor 2 from run S0 to run S4 and 
$(B^\prime_\text{rms})^2$ lowers
by a factor 1.6 (\tabref{t:runs_fields}).
The residual turbulent viscosity variation
may be an effect of volume averaging.
The unstable fluid regions of run S4 are indeed
more localized than those of run S0, as already discussed in \secref{s:stable_strat}.
This is also clearly evidenced by the distribution
of $\hat{F}_r^{\text{MS}}$, the time averaged radial Maxwell stresses flux,
which is shown in \figref{f:flx_MS_r}b,c for these two runs.
The smaller turbulent viscosity value
of run S9 at $\mathcal{N}=20$ instead originates from
the weaker spatial correlations of the turbulence,
together with a volume averaging effect.
In fact, runs S9 and S4 share a similar value of $B^\prime_\text{rms}$
of 0.018 (\tabref{t:runs_fields}) but the peak amplitudes
of $\hat{F}_r^{\text{MS}}$ are $40\%$ lower in the former run (\figref{f:flx_MS_r}c,d).

The variations of $\widetilde{\nu}_\text{T}$ with $\text{Pr}$,
observed when $\mathcal{N}>1$ is fixed,
likewise originate from changes in both
the size of the unstable regions and the spatial
correlations of the turbulent field.
In these runs, while $B^\prime_\text{rms}$
weakly changes with $\text{Pr}$ (\tabref{t:runs_fields}),
the integrated Maxwell stresses flux $\hat{I}_r^\text{MS}$
reduces when $\text{Pr}$ increases (\secref{s:am_strat}), which
suggests a decrease of the spatial correlations
of the turbulent field.
The distribution of $\hat{I}_r^{\text{MS}}$ also narrows
as $\text{Pr}$ increases, which yields
smaller values for the volume averaged Maxwell stresses
in our turbulent viscosity estimate.

A power law of the form
$\widetilde{\nu}_{\text{T}}=a\,\mathcal{N}^{-\delta}$, where $\delta >0$,
well describes the dependence of the turbulent
viscosity on stratification.
A least squares fit of the runs at $\text{Pr}=10^{-3}$
provides an exponent $\delta=0.50$
and a prefactor $a=3.1\times 10^{-4}$, which is shown by
the dashed line in \figref{f:turb_visc}.
For AMRI turbulence in Taylor-Couette flow, \cite{Spada16} suggest
a stronger scaling of $\mathcal{N}^{-1}$.
For a free shear, \cite{Jouve20} demonstrated that some of
the linear properties of AMRI, such as the instability threshold
and the range of linearly unstable modes,
depend only on the parameter combination $\mathcal{N}^2\text{Pr}$.
However, our simulations indicate that this
is not the case for the nonlinear evolution of AMRI
and for the AM transport efficiency.
\section{Transport of chemical elements}
\label{s:chemicals}
\subsection{Model formulation}
We assume the light stellar chemical elements
to contaminate the turbulent fluid flow 
as a passive scalar, that is their concentration
is so low that they have no dynamical influence
on the fluid motion itself.
The equation of evolution of the chemical concentration $c$ is
\begin{equation}
\label{eq:c}
    \frac{\partial c}{\partial t}+(\mathbf{u}\cdot\boldsymbol{\nabla}) c=\frac{1}{\text{Sc}\,\text{Re}}\boldsymbol{\nabla}^2 c,
\end{equation}
which is solved together with Eqs.~(\ref{e:NS})-(\ref{e:temp}). Here the Schmidt number
\begin{equation}
    \Sc=\frac{\nu}{D_c}
\end{equation}
is the ratio of the fluid kinematic viscosity $\nu$
to the molecular chemical diffusivity $D_c$.
We consider $\text{Sc}=1$ so that the diffusion timescale of the chemicals $d^2/D_c$
and the viscous timescale of the flow $\tau_\nu=d^2/\nu$ are equal.
We assume no sources nor sinks of the chemical elements
from the boundaries by imposing
a zero chemical concentration flux $\partial c/\partial r = 0$
at $r=\rin$ and $\rout$.
The initial distribution of the chemical concentration
is the spherically symmetric Gaussian
\beq
\label{e:chem_t0}
    c(r, t=t_0) = \frac{1}{C_0}\exp\left[{-\frac{(r-r_0)^2}{2\sigma_0^2}}\right].
\eeq
The distribution is
centered at mid depth in the fluid domain, $r_0=\left(1/\chi-1\right)^{-1}+1/2$,
and has a full width at half maximum of $\delta_0/d=1/10$, which corresponds to a variance
$\sigma_0^2=(\sigma_0^*/d)^2\approx 1.8\times 10^{-3}$.
The total mass of the chemicals in the fluid domain
$C_0=\int c(t=t_0)\,\text{d}V$
is a conserved quantity and serves as a normalization
constant here.

The chemical layer~(\ref{e:chem_t0}) is introduced
during the turbulent AMRI evolution of the
unstratified fiducial dynamo run U0
and of all the stratified runs
at the times $t=t_0$ listed in \tabref{t:chem}.
In the following sections we study the turbulent transport
of the chemicals, we estimate its efficiency and we compare
the results with the AM transport.
\begin{table}[b]
\caption{Input parameters and output diagnostics
of the simulations
where a passive scalar is introduced.
$t_0$ is the time of the simulation runs indicated
in the fourth column at which the passive scalar is introduced.
$\tau_\text{M}$ is the meridional
flow timescale calculated as explained in the main text.
$\widetilde{D}_\text{T}$ is the estimated
chemical turbulent diffusion and the error denotes 1 standard deviation
in time.
All runs are at $\text{Re}=5\times 10^4$ and $\Pm=1$.
}
\centering
\begin{tabular}{@{}lccccccl@{}}
\toprule
Name        & $\mathcal{N}$ & $\text{Pr}$ & Run & $t_0$ & $\tau_\text{M}/\tau_\Omega$ & $\widetilde{D}_\text{T}\times 10^5$\\ \midrule\midrule
\vspace{0.5ex}
C0  & 0  & --                 & U0 & $1764.1$ & $117.5$ & $18.9\pm 6.8$\\
C1  & 1   & $10^{-3}$   & S0 & $2151.3$ & $139.8$ & $16.6\pm 0.7$\\
C2  & 1   & $10^{-2}$   & S1 & $1811.2$ & $187.2$ & $12.0\pm 1.4$\\
\vspace{0.5ex}
C3  & 1   & $10^{-1}$   &S2 & $2092.9$ & $277.4$ & $8.3\pm 0.8$\\
C4  & 10   & $10^{-4}$   & S3 & $1400.2$ & $353.9$ & $10.6\pm 1.5$\\
C5  & 10 & $10^{-3}$   & S4 & $2625.1$ & $246.3$ & $5.4\pm 0.6$\\
\vspace{0.5ex}
C6  & 10  & $2\times 10^{-3}$ & S5 & $2466.5$ & $435.3$ & $3.9\pm 0.4$\\
C7  & 20   & $10^{-4}$   & S7 & $1755.6$ & $249.9$ & $7.8\pm 1.1$\\
C8  & 20   & $5\times 10^{-4}$  & S8 & $1902.6$ & $543.1$ & $3.3\pm 0.2$\\ 
C9  & 20   & $10^{-3}$ & S9 & $2282.4$ & $648.3$ & $2.7\pm 0.2$\\ \bottomrule
\end{tabular}
\label{t:chem}
\end{table}
\subsection{Estimate of the chemical turbulent diffusion}
First, we examine the temporal evolution of the azimuthally averaged (mean) chemical
concentration $\overline{c}$ in our numerical simulations
and analyze the turbulent transport.
If the turbulent transport occurs as a diffusive process, the radial distribution
of $\overline{c}$ remains Gaussian over time
and its variance $\sigma^2$ increases linearly.
The rate at which $\sigma^2$ grows defines
the radial turbulent chemical diffusivity
\begin{equation}
\label{e:Dturb}
D_\text{T}=\frac{1}{2}\frac{\text{d}\sigma^2}{\text{d}t}.
\end{equation}

\begin{figure}[t]
\centering
\resizebox{\hsize}{!}{\includegraphics{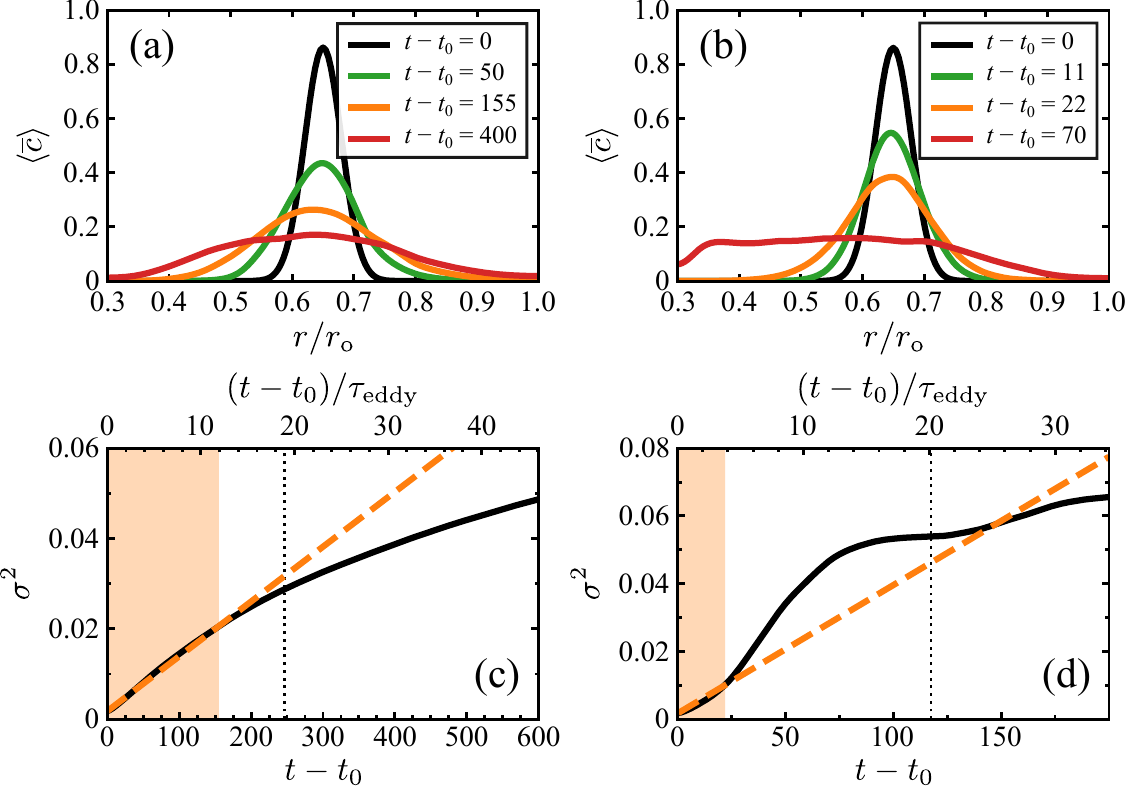}}
\caption{Chemical turbulent transport for
run C5 at $\mathcal{N}=10$ and $\text{Pr}=10^{-3}$ (left panels)
and for the unstratified run C0 (right panels).
(a,b)~Radial distribution of the horizontally averaged
mean chemical concentration $\langle \overline{c}\rangle$
at the times $t-t_0$ indicated in the legend.
The first three times cover the diffusive phase of the transport.
(c,d)~Variance $\sigma^2$ of the mean chemical
concentration $\overline{c}$ as a function of $t-t_0$.
The dashed line is defined by the time average of $\text{d}\sigma^2/\text{d}t$
evaluated over the diffusive phase of the transport (orange shaded background).
The vertical dotted line marks the time $t-t_0=\tau_\text{M}/\tau_\Omega$
after which the meridional flow is expected to dominate the transport.
Here $\tau_\text{M}$ is the meridional flow timescale
estimated as explained in the main text.
The upper horizontal axis displays time scaled in units
of the eddy turnover time $\tau_\text{eddy}$.
}
\label{f:chem_diff_evol}
\end{figure}
\begin{figure}[b]
\centering
\resizebox{\hsize}{!}{\includegraphics{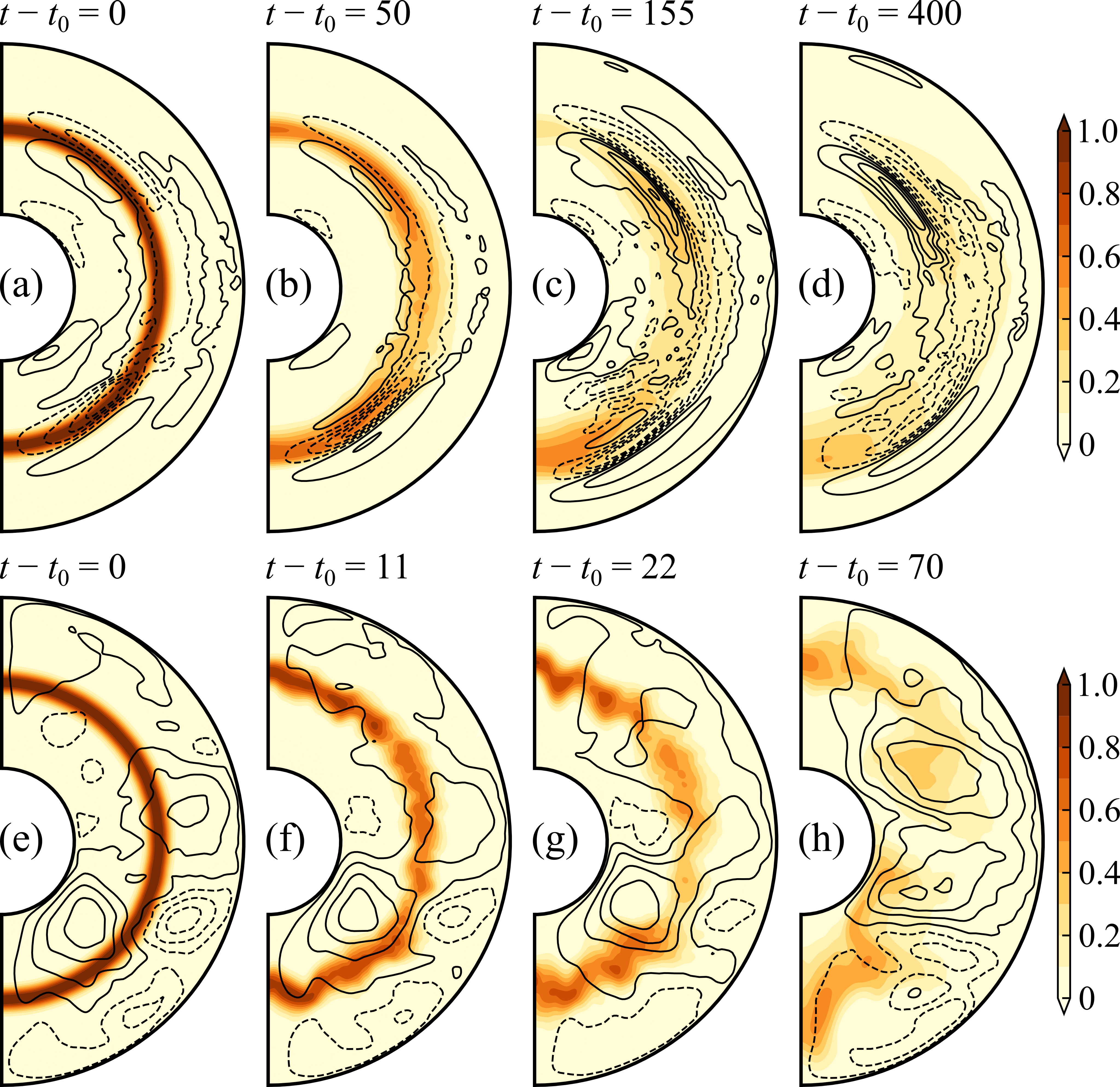}}
\caption{Snapshots of the mean chemical concentration $\overline{c}$
in runs C5 (top panels) and C0 (bottom panels)
at the times $t-t_0$ of \figref{f:chem_diff_evol}a,b. The mean concentration
$\overline{c}$ is normalized with its maximum value at $t=t_0$ here.}
\label{f:chem_snapsh}
\end{figure}
In all runs explored here, the turbulent transport of the mean
chemical concentration is diffusive during the initial
stages of its evolution.
\Figref{f:chem_diff_evol}a displays the evolution of the radial profile
of the horizontally averaged mean chemical concentration $\langle \overline{c}\rangle$
in run C5 ($\mathcal{N}=10$ and $\text{Pr}=10^{-3}$)
as an example.
The distribution remains Gaussian for a period
of about $150\,\Omega_\text{a}^{-1}$
after $t_0$ and its variance $\sigma^2$ increases
linearly as expected (\figref{f:chem_diff_evol}c; see also the snapshots
of $\overline{c}$ in \figref{f:chem_snapsh}a-c).
We calculated the chemical turbulent diffusivity $D_\text{T}$
from \eqref{e:Dturb} by performing a time average of $\text{d}\sigma^2/\text{d}t$
over this interval (dashed line in \figref{f:chem_diff_evol}c), which covers 12 eddy turnover times
$\tau_\text{eddy}$ (see the upper horizontal axis of \figref{f:chem_diff_evol}c).
This procedure yields a dimensionless chemical diffusivity
$\widetilde{D}_\text{T}=D_\text{T}/ \Delta\Omega\, r_\text{o}^2$ of $5.4\times 10^{-5}$.
The last column of \tabref{t:chem} reports the estimates $\widetilde{D}_\text{T}$
for all runs explored here.

In run C5, deviations from a diffusive transport
begin at time $t-t_0\approx 200$
when $\sigma^2$ starts to level off and does not evolve linearly anymore (\figref{f:chem_diff_evol}c).
At $t-t_0 = 400$ the distribution of the mean concentration
becomes decisively non-Gaussian and its
flanks reach the boundaries (\figref{f:chem_diff_evol}a, red line).
A snapshot of $\overline{c}$ at this time evidences
that the meridional circulation contributes, on such longer timescales, to expel
the chemicals from the central parts of the fluid domain, accumulating them
at high latitudes in the southern hemisphere (\figref{f:chem_snapsh}d).
We estimated a characteristic timescale for the meridional flow
in the region of the chemical layer
as $\tau_\text{M}=\int_{r_0-\sigma_0}^{r_0+\sigma_0} dr/\langle \overline{u}_\text{M}\rangle$,
where $\langle \overline{u}_\text{M}\rangle$ is the
horizontally averaged meridional velocity at $t=t_0$.
In run C5 $\tau_\text{M}=246\,\Omega_\text{a}^{-1}$ (vertical
dotted line in \figref{f:chem_diff_evol}c), which
agrees well with the timescale on which $\sigma^2$
starts to deviate from the initial diffusive evolution.

When buoyancy effects weaken by considering lower
$\mathcal{N}$ and/or $\text{Pr}$, the meridional flow timescale $\tau_\text{M}$
decreases as expected (\tabref{t:chem}) and therefore
the transport of the chemicals by the meridional flow
takes over the turbulent one earlier during the evolution.
Nonetheless, the shortest period of diffusive evolution
in our stratified simulations, obtained for run C1
($\mathcal{N}=1$ and $\text{Pr}=10^{-3}$),
covers about $8\,\tau_\text{eddy}$, which still
ensures a robust statistics for the turbulence.

In the unstratified run C0
the diffusive phase covers an even shorter
interval of $3.7\,\tau_\text{eddy}$ (orange shaded area in \figref{f:chem_diff_evol}d).
The meridional flow timescale $\tau_\text{M}$
reduces to $118\,\Omega_\text{a}^{-1}$
and $\langle\overline{c}\rangle$
becomes decisively non-Gaussian
already at $t-t_0=70$ (\figref{f:chem_diff_evol}c, red line; \figref{f:chem_snapsh}h).
The estimate of $\widetilde{D}_\text{T}$ in run C0
is therefore less accurate than all the other
runs where a reliable measurement is instead achieved.
In this run the standard deviation of $\widetilde{D}_\text{T}$
is as large as $31\%$ of the turbulent diffusivity value itself.
In the runs at $\mathcal{N}=10$ and $20$ and higher $\text{Pr}$,
the standard deviation is instead of only a few percent
of $\widetilde{D}_\text{T}$ since
the diffusive phase lasts more than $10\, \tau_\text{eddy}$ (\tabref{t:chem}).
For a diffusive transport, a certain degree of scale separation
between the initial chemical layer width $\delta_0$
and the characteristic radial length scale of the turbulence is required.
While in runs at $\mathcal{N}>1$ such a scale separation
is verified, in run C0 this becomes only marginally
satisfied, as evidenced by the strong latitudinal variations
of $\overline{c}$ (\figref{f:chem_snapsh}f,g), and may cause
additional accuracy loss in our estimate of the turbulent diffusivity.
\subsection{Chemical and angular momentum transport}
\begin{figure}
\centering
\includegraphics[width=\hsize]{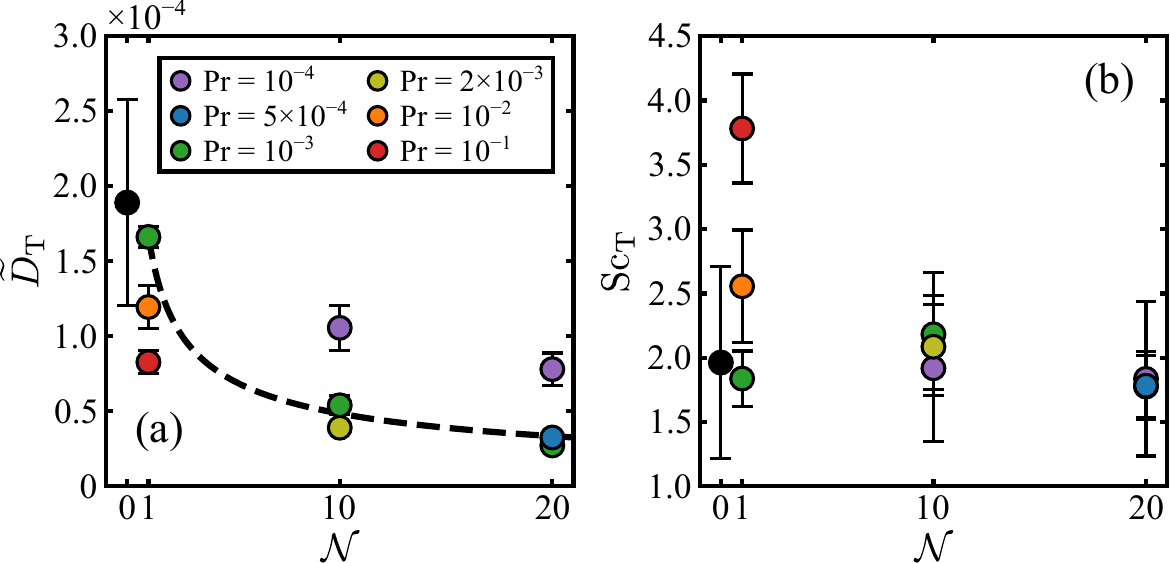}
\caption{(a)~Turbulent chemical diffusivity $\widetilde{D}_\text{T}$ and
(b)~turbulent Schmidt number $\text{Sc}_\text{T}$
as a function of $\mathcal{N}$.
The black circle shows the unstratified fiducial dynamo run U0.
The symbol color of the stratified runs codes
the Prandtl number $\text{Pr}$ as indicated in the legend in (a).
Error bars in (a) show 2 standard deviations intervals
around the time averaged turbulent diffusivity values.
The dashed line in (a) displays a power law fit
of the stratified runs at $\text{Pr}=10^{-3}$, that is
$\widetilde{D}_\text{T}=1.7\times 10^{-4}\,\mathcal{N}^{-0.54}$.
The errors in (b) are obtained by propagating the uncertainties
of the turbulent viscosity $\widetilde{\nu}_\text{T}$ and of the 
turbulent diffusivity $\widetilde{D}_\text{T}$.
}
\label{f:chem_turb_diff}
\end{figure}
We now discuss how the efficiency of the chemical turbulent
transport varies with the effect of stratification.
\Figref{f:chem_turb_diff}a shows that $\widetilde{D}_\text{T}$
lowers when buoyancy effects strengthen either by
increasing $\mathcal{N}$ and/or $\text{Pr}$.
The two most weakly stratified runs at $\mathcal{N}=1$ with
$\text{Pr}=10^{-3}$ and $10^{-2}$ have values
of $\widetilde{D}_\text{T}$ comparable to the unstratified case (black circle)
within their 2 standard deviations intervals.
In the runs at $\mathcal{N}=10$ and 20, as discussed in
the previous section, the turbulent diffusivity
estimates become more accurate due to
the longer diffusion phase and
the uncertainties are generally smaller
than the symbol sizes themselves.
As for the turbulent viscosity, a power law well describes
the dependence of the chemical diffusivity on stratification.
A power law fit of the stratified runs at $\text{Pr}=10^{-3}$
provides $\widetilde{D}_\text{T}=1.7\times 10^{-4}\,\mathcal{N}^{-0.54}$
(dashed line in \figref{f:chem_turb_diff}a) and the
exponent of $\mathcal{N}$ is very similar
to the one obtained for the
turbulent viscosity (\secref{s:turbvisc}).

The turbulent Schmidt number
\begin{equation}
\text{Sc}_\text{T}=\frac{\widetilde{\nu}_\text{T}}{\widetilde{D}_\text{T}}
\end{equation}
measures the efficiency of the AM transport
relative to the one of the chemical elements.
\Figref{f:chem_turb_diff}b displays $\text{Sc}_\text{T}$
as a function of $\mathcal{N}$ for all simulation runs
and shows that this is always larger than 1
and does not vary much with increasing buoyancy effects
in the ranges of $\mathcal{N}$ and $\text{Pr}$ explored.
The value $\text{Sc}_\text{T}=1.96\pm 0.75$ of the unstratified
run C0 (black circle) encompasses all the
other estimates obtained for the stratified runs, except for
the case at $\mathcal{N}=1$ and $\text{Pr}=10^{-1}$
which is larger at $\text{Sc}_\text{T}=3.8$.
Our simulations therefore suggest that
AMRI turbulence induces a transport of AM
which is systematically stronger than the one of chemical elements.
This is expected since the Maxwell stresses, which dominate
the transport of AM in our runs and do not directly influence
the one of chemical elements, are generally stronger
than the Reynolds stresses (Sects.~\ref{s:am_unstrat} and \ref{s:am_strat}),
which instead regulate the chemical transport.
However, we note that $\text{R}=(\mu_0\rho)^{-1}\langle B_r^\prime B_\phi^\prime\rangle/\langle u_r^\prime u_\phi^\prime\rangle$,
the ratio of the volume averaged
radial Maxwell stresses to the volume averaged
radial Reynolds stresses,
is always larger than 5 (\tabref{t:runs_fields}) and
overestimates the relative transport efficiency
measured by $\text{Sc}_\text{T}$.
\section{Summary and discussion}
\label{s:conclusions}
Additional transport
processes beyond atomic diffusion and standard hydrodynamical
mechanisms, such as meridional circulation and shear turbulence,
are required to explain the observed rotational and chemical evolution
of low mass stars. MHD turbulence is regarded as
one of the primary processes to enhance the transport
in radiative stellar interiors.
In this work we investigated numerically
the transport of AM and chemical elements
due to azimuthal MRI in a spherical shell
where cylindrical differential rotation is forced.

We first considered an unstratified flow
at a magnetic Prandtl number $\text{Pm}$ of 1
and explored the stability of
purely axisymmetric toroidal field configurations
to weak nonaxisymmetric perturbations.
The parameter regime where we observed AMRI
agrees well with predictions
obtained from a local linear stability analysis
and with the global linear analysis results of \cite{Guseva17a}
who analyzed Taylor-Couette flow
with an imposed field.
Our AMRI runs are characterized by values of
$\text{Le}_\phi^\text{max}$
not smaller than about $5\times 10^{-3}$,
otherwise diffusive effects stabilize the system,
and not larger than about 0.5, or else the nature
of the instability changes and TI is found.
Here $\text{Le}_\phi^\text{max}$ is the maximum value
in the fluid domain of the ratio of
the Alfv\'en frequency of the axisymmetric azimuthal field
$\omega_{\text{A}\phi}=\overline{B}_\phi\big/(\mu_0\rho)^{1/2}d$ 
to the reference rotation rate $\Omega_\text{a}$.

Next, we explored the nonlinear evolution of the
instability in these unstratified runs.
At $\text{Pm}=1$, we observed self-sustained dynamo action
when $\Rey \geq 5\times 10^4$
and the azimuthal field strength of the perturbed
axisymmetric solution is large enough (runs U0 and U2).
At the larger $\text{Re}$ of $10^5$, we found
dynamo action down to $\Pm=0.6$ (run U3).
These simulations are the first global MRI dynamos
ever reported at such low values of Pm.
We obtained transient turbulence in all the other unstable runs.

We then examined the effect of thermal stable stratification
on unstratified AMRI turbulence 
at $\text{Re}= 5\times 10^4$ and $\text{Pm}=1$.
To this end, we varied $\mathcal{N}=N/\Omega_\text{a}$,
the ratio of the Brunt-V\"{a}is\"{a}l\"{a} frequency to the reference rotation
rate, from 1 to 20 and the Prandtl number $\Pr$ 
in the range $10^{-4}-10^{-1}$.
When increasing buoyancy effects, the turbulence
becomes less isotropic and homogeneous.
However, when $\Pr$ is too large, thermal diffusion cannot limit
the stabilizing buoyancy force
anymore and AMRI is suppressed.
In our most stratified unstable runs,
we observe instability structures
elongated in the latitudinal direction and unstable regions
localized in the southern hemisphere where the
axisymmetric azimuthal field is strong enough to support AMRI.
All the stratified runs show transient turbulence,
although in a few cases the numerical
integration time is too short to test for dynamo action.
We argued that, by limiting radial flow motions, stable stratification
reduces the effective magnetic Reynolds number $\text{Rm}_\text{eff}$
below the critical value for dynamo onset, which is of about $800$
based on the unstratified runs.
Nonetheless, the magnetic
and kinetic energies show an oscillatory behavior typical
of $\alpha\Omega$-dynamos
that has been reported before \citep[e.g.,][]{ReboulSalze22}.

We explored the transport of AM in the unstratified
dynamo solutions and in the stratified runs,
showing that it occurs radially outwards and
is largely dominated by the Maxwell stresses when
$\mathcal{N}\geq 10$.
The meridional circulation opposes to the radial
transport by the Maxwell stresses only 
when no stratification is present or is weak at $\mathcal{N}=1$.
We quantified the radial AM transport by estimating
the turbulent viscosity $\widetilde{\nu}_\text{T}=\nu_\text{T}\big/\Delta\Omega\, r_\text{o}^2$,
where $\Delta\Omega$ is the global rotation contrast and
$r_\text{o}$ the outer boundary radius, and we showed that
it decreases when buoyancy effects strengthen.
Within the explored range of parameters, the turbulent
viscosity variations are well described 
by the power law $\widetilde{\nu}_\text{T}=a\, \mathcal{N}^{-1/2}$,
where $a=3.1\times 10^{-4}$.

Finally, we investigated the turbulent transport of a passive scalar.
Having demonstrated that this occurs as a diffusive process,
we estimated the diffusion coefficient $\widetilde{D}_\text{T}$.
In the range of parameters explored, $\widetilde{D}_\text{T}$
varies with buoyancy effects similarly to the turbulent viscosity $\widetilde{\nu}_\text{T}$
but its magnitude is sistematically lower.
A power law $\widetilde{D}_\text{T}=1.7\times 10^{-4}\,\mathcal{N}^{-0.54}$
well describes our simulation data so that
the turbulent Schmidt number
$\text{Sc}_\text{T}=\widetilde{\nu}_\text{T}/\widetilde{D}_\text{T}$
is always of about $2$ in the range of parameters explored.
When buoyancy effects are high enough
and the AM transport is dominated by the turbulent
magnetic field fluctuations and the one
of the chemicals by the flow fluctuations,
we tested whether
the ratio of the radial Maxwell stresses
to the radial Reynolds stresses
is a good proxy for $\text{Sc}_\text{T}$.
We found that this ratio generally overestimates $\text{Sc}_\text{T}$
by roughly a factor 3 in our simulations.

As far as an application to stars is concerned, there are several
limitations to the setup explored here.
First, we examined MRI of purely toroidal fields
whereas magnetic field configurations with both toroidal
and poloidal components are expected in stellar interiors.
Secondly, the background differential rotation profile
and its amplitude may not be representative
of the shear in stellar interiors.
We also ignored the additional stabilizing effect
of chemical buoyancy in the momentum equation.
Relaxing one or more of these assumptions
may yield to turbulent diffusion coefficients that differ
from those estimated here.

However, other fluid properties employed
in our simulations approach those expected
in certain regions of stellar interiors. 
For example, the values of the molecular diffusivity ratios
and of the stable stratification are close
to those of the electron-degenerate cores of red giants
and of the nondegenerate radiative outer regions of these stars, respectively.
In fact, the Prandtl number Pr is typically $\sim 10^{-3}$
in degenerate red giant cores \citep{Garaud15}
and the magnetic Prandtl number $\text{Pm}$
ranges between $10^{-1}$ and $10$ \citep{Ruediger15}.
The nondegenerate radiative regions
immediately above the hydrogen burning shell
are less dense than the degenerate zones below
and expand during their evolution, hence the thermal component
of the Brunt-V\"{a}is\"{a}l\"{a} frequency $N_T$ is relatively
low at $\sim 10^{-3}\,\text{s}^{-1}$ in the subgiant phase
and further decreases to about $5\times 10^{-4}\,\text{s}^{-1}$
on the red giant branch \citep{Talon08}.
The typical angular rotation frequency $\Omega/2\pi$
of the cores of these stars is roughly
in the range $600-700\,\text{nHz}$ \citep{Deheuvels14,Gehan18}, which
yields $N_T/\Omega\approx 100-200$,
that is only from 5 to 10 times larger than the highest value
of $\mathcal{N}=20$ employed in our simulations.
Such moderate values of $N_T/\Omega$
also charaterize the interior of pre-MS stars \citep{Gouhier21}.

However, in both the radiative regions of red giants above and below
the hydrogen burning shell, the values
of the molecular diffusivity ratios and $N_T/\Omega$
cannot be captured simultaneously.
In the degenerate cores, $N_T/\Omega$ is as large as $\sim 10^3$ during the subgiant phase
and increases to $10^4$ in evolved red giants \citep{Talon08}.
In the outer nondegenerate layers, where the diffusivities are
dominated by radiative and collisional contributions only,
$\Pr$ and $\Pm$ drop to about $10^{-7}$ and $10^{-2}$
respectively \citep{Garaud15,Ruediger15}.

We showed that the turbulent viscosity in our simulations
scales with $\mathcal{N}^{-1/2}$, which is slower than
the $\mathcal{N}^{-1}$ scaling suggested for
AMRI in Taylor-Couette flow
with imposed current-free magnetic fields \citep{Spada16}.
Assuming that our scaling prediction
$\widetilde{\nu}_\text{T}=3.1\times 10^{-4}\mathcal{N}^{-1/2}$
is confirmed for larger values of the stable stratification
than those explored here, we obtain a turbulent viscosity
$\nu_\text{T}=\widetilde{\nu}_\text{T}\,r_\text{o}^2\Delta\Omega$
in the range $2-7\times 10^8\,\text{cm}^2/\text{s}$
for the degenerate cores of subgiants and red giants.
For $\mathcal{N}$, here we used the values of $N_T/\Omega$
discussed above; for the shear contrast $\Delta\Omega/2\pi$
we considered the typical angular velocity difference between the core and
the envelope of subgiants of $900\,\text{nHz}$,
and for the radius of the radiative region $r_\text{o}=0.05\,R_\odot$,
which is $2\%$
of the typical subgiant radius $R=2.5R_\odot$ \citep{Deheuvels14}.
Such a high turbulent viscosity value enforces
solid body rotation in the degenerate core on
a timescale of around $2000$ years, which
is not incompatible with the observations.
It is indeed likely that the observed radial shear
of subgiants is localized around the hydrogen burning shell,
hence above the degenerate core which may be
in rigid rotation instead \citep{Deheuvels14}.
This observational evidence is further supported
by the fact that stable stratification peaks
at the hydrogen burning shell, where the transport
of AM would be the slowest \citep[e.g.,][]{Fuller19}.
In regions around the hydrogen burning shell,
chemical stratification largely dominates
over the thermal contribution.
The thermal Brunt-V\"{a}is\"{a}l\"{a} frequency $N_T$
is lower in these regions
than in the core so that our scaling
predicts even larger turbulent viscosity values, hence faster transport
than the one estimated above.
Stellar evolution models that reproduce the rotational
evolution of low mass evolved stars suggest a significantly
smaller turbulent diffusion coefficient which increases
monotonically from $10^2\,\text{cm}^2/\text{s}$
in early subgiants to almost $10^6\,\text{cm}^2/\text{s}$
at the end of the red giant phase \citep{Spada16,Moyano22}.

Our simulations show that the transport of AM due to
AMRI may be more efficient than the one induced by TI.
The theory of \cite{Fuller19} for TI-driven dynamo action
predicts that the turbulent viscosity scales as $(N_T/\Omega)^{-2}$,
whereas here we obtain a power exponent of $-1/2$ for AMRI.
TI dynamo action may have been recently identified
in global numerical simulations \citep{Petitdemange23},
but robustly testing the proposed turbulent viscosity scalings
requires further study.

The magnetic field strengths recently measured in red
giant cores can be directly compared to our numerical
simulation results in terms of the
radial Lenhert number $\text{Le}_{r}^{\text{rms}}$, the ratio
of the local Alfv\'en frequency based on the rms radial field $B_{r}^{\text{rms}}/(\mu_0\rho)r$
to the rotation rate $\Omega$.
In the 13 red giants where magnetic field measurements
exist so far, $\text{Le}_{r}^{\text{rms}}$
ranges between 0.05 and 0.25 at the hydrogen
burning shell, the region of maximum sensitivity for the
asteroseismic inversions \citep{Li23}.
Such strong field amplitudes are incompatible
with those expected from TI dynamo action
which
predicts $\text{Le}_{r}^{\text{rms}}\sim(N/\Omega)^{-5/3}\sim 10^{-9}$,
where $N$ here is the Brunt-V\"{a}is\"{a}l\"{a} frequency
including both thermal and compositional contributions \citep{Fuller19,Li22,Li23}.
For AMRI, our numerical results indicate that $\text{Le}_{r}^{\text{rms}}$
decreases less steeply with thermal stratification.
However, the typical values we obtain at a moderate stratification
of $\mathcal{N}=20$ are $\text{Le}_{r}^{\text{rms}}\sim 10^{-3}$, which is
already one order of magnitude lower than
the smallest observed value.

Although AMRI turbulence seems unable to explain
the strong magnetic fields observed in the core,
we cannot exclude that weaker fields
could be generated in the outer radiative regions
of red giants, in particular above the hydrogen burning shell
where $N_T/\Omega$ is lower at $\sim 10^2$
as discussed before.
AMRI could also be triggered during earlier stages of the
evolution of low mass stars,
for example during the core contraction phase
immediately after the main sequence.
Numerical simulations suggest that large scale fields,
relic of a core convective dynamo during the main sequence,
can destabilize differential rotation produced
by core contraction through axisymmetric MRI \citep{Gouhier22}.

Recent stellar evolution models indicate that
MHD turbulence is required to reconcile
the internal rotation of the Sun
with its surface Li abundance
and to reproduce the observed Li depletion
of pre-main sequence and solar type stars \citep{Eggenberger22}.
These models employ prescriptions for the transport by
TI dynamo action where the turbulent Schmidt number scales as
$\text{Sc}_\text{T}\sim (N/\Omega)^{2}\gg 1$.
We showed that AMRI turbulence transports
AM more efficiently than chemical elements, as suggested
by these stellar evolution models, but $\text{Sc}_\text{T}$
is too low at values of about 2 and does not show
significant variations for the
moderate degrees of thermal stratification explored.

In conclusion, we confirm that AMRI turbulence
can strongly enhance the transport of AM and
chemical elements in radiative stellar interiors and
its efficiency may be higher than the one of TI.
The turbulent viscosity moderately decreases with stable stratification
and the chemical turbulent diffusion coefficient follows
a similar scaling but is weaker in amplitude.
While our numerical simulations capture
the mild molecular diffusivity ratios of degenerate stellar cores,
extensive parameter investigations are needed to robustly
extrapolate the results at the extreme degrees of stratification
that characterize stellar interiors.
Further numerical studies in a spherical geometry
exploring MRI-induced transport for different shear profiles
and amplitudes than those examined here
and considering chemical buoyancy
could give additional insights on the rotational
and chemical evolution of low mass stars.
\begin{acknowledgements}
The authors acknowledge support from the project
BEAMING ANR-18-CE31-0001
of the French National Research Agency (ANR).
LJ acknowledges support from the Institut Universitaire de France.
This work was performed using high performance computing resources
from CALMIP (Grants 2021A-P1118, 2021B-P1118, 2022A-P1118, 2022B-P1118
and 2023A-P1118)
and from GENCI–IDRIS (Grants 2022-A0110410970 and 2023-AD0100413776).
The authors are indebted to S\'ebastien Deheuvels and Fran\c{c}ois Rincon
for helpful discussions and thank an anonymous referee whose insightful comments
have improved the quality of the manuscript.
\end{acknowledgements}

%
%
\bibliographystyle{aa}
\bibliography{biblio_AM_MRI}

\begin{appendix}
\section{Tayler instability} 
\label{s:appendix_Tayler}
As mentioned in \secref{s:unstrat_stability}, we observe TI in two of our unstratified
simulation runs at $\text{Pm}=1$.
Here we provide evidence of this instability
in run U4 at $\text{Re}=5\times 10^3$ and
$\text{Ha}_{\phi}^\text{max}=5012$.

TI is a pinch-type instability of purely toroidal fields
expected to occur in radiative stellar regions \citep{Tayler73,Spruit99}.
In spherical coordinates, \cite{Goossens80} studied the adiabatic stability
of weak axisymmetric toroidal fields $B_\phi(r,\theta)$
when no rotation is present using an energy method.
The toroidal field $B_\phi$ has arbitrary radial dependence and
angular dependence $(1-x^2)^{1/2}\text{d}P_\ell/\text{d}x$,
where $x=\cos\theta$ and $P_\ell$ is the Legendre polynomial of degree $\ell$.
The field is weak in the sense that the magnetic pressure is
much smaller than the hydrostatic pressure, so that it does not
modify the basic stellar structure.
A necessary and sufficient condition for nonaxisymmetric instability is \citep{Goossens81}
\begin{equation}
\label{e:GT80}
B_\phi^2\left(m^2 -2\cos^2\theta\right) - \sin\theta\cos\theta \frac{\partial B_\phi^2}{\partial\theta}<0 .
\end{equation}
The instability is most likely for azimuthal modes $|m|=1$
and for large latitudinal gradients of $B_\phi^2$.
A Taylor expansion easily shows that
regions close to the poles ($|x| \ll 1$) are always unstable for $|m|=1$.
Away from the poles, instability can also occur
but only if $B_\phi^2$
increases sufficiently rapidly with $x$.
The growth rate of the most unstable mode $m_\text{max}^\text{TI}=1$
is of the order of the toroidal Alfv\'en frequency
$\omega_\text{A}=B_\phi\big/(\mu_0\rho)^{1/2}r \sin\theta$ \citep{Tayler73,Goossens81}.
Rotation has a stabilizing influence on TI.
Although not removing the instability, it reduces
its maximum growth rate to \citep{Pitts85,Bonanno13b}
\begin{equation}
\label{e:gamTI}
\gamma_\text{max}^\text{TI}=\omega_\text{A}^2/\Omega .
\end{equation}

\begin{figure}[t]
\centering
\resizebox{0.75\hsize}{!}{\includegraphics{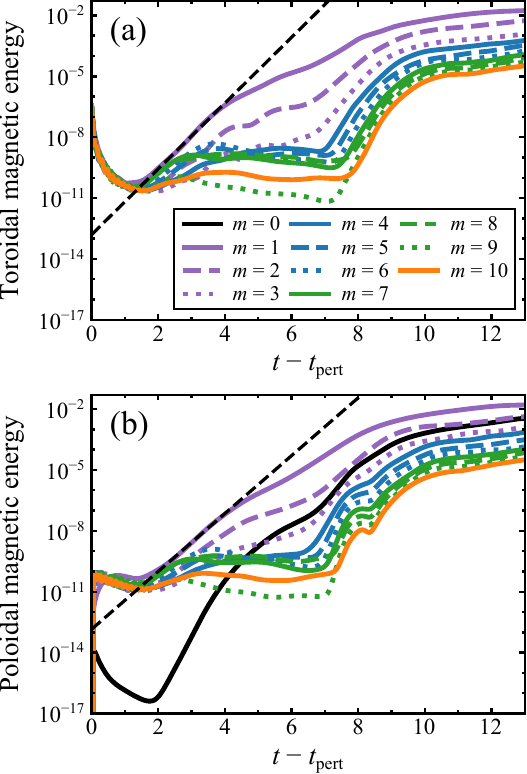}}
\caption{(a)~Toroidal and (b)~poloidal magnetic energy evolution
of the azimuthal modes $m=0-10$ of run U4
($\mathcal{N}=0$, $\text{Re}=5\times 10^3$, $\text{Pm}=1$ and
$\text{Ha}_{\phi}^\text{max}=5012$).
The black dashed lines show linear fits of the
energies of the most unstable mode $m=1$
over the interval $2\leq t-t_\text{pert}\leq 4$
covering the linear phase of the instability growth.
The growth rates $\gamma_\text{max}/\Omega_\text{a}$
provided by these fits are
1.9 and 1.6 based on the toroidal and poloidal energies respectively.
}
\label{f:Tay_ener}
\end{figure}
\begin{figure*}[t]
\centering
\includegraphics[width=15cm]{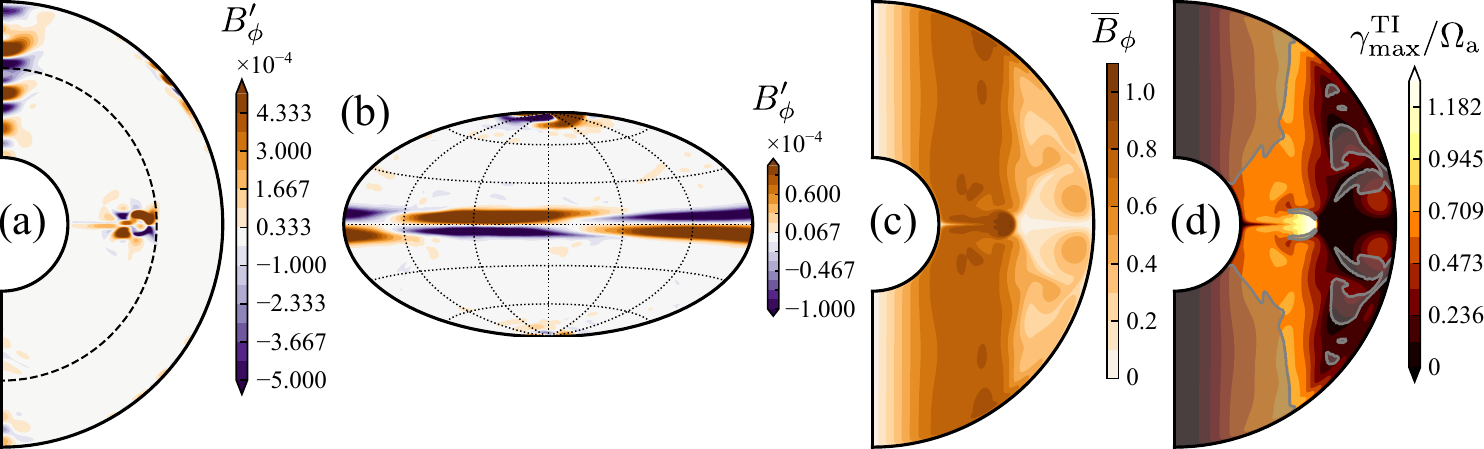}
\caption{(a)~Meridional cut and (b) surface projection of
the nonaxisymmetric azimuthal field $B_\phi^\prime$
at time $t-t_\text{pert}=3.5$ during the linear phase of the instability growth.
The surface projection is taken at radius $r/r_\text{o}=0.7$, which
is indicated by the dashed line in (a).
(c)~Axisymmetric azimuthal field $\overline{B}_\phi$ at the
perturbation time $t_\text{pert}=33.2$.
(d)~Theoretical TI growth rate $\gamma_\text{max}^\text{TI}=\omega_\text{A}^2/\Omega$.
The grey shaded areas show the locations where the
instability condition (\ref{e:GT80}) for the $m=1$ mode
and the axisymmetric azimuthal field solution in (c) is fulfilled.
}
\label{f:Tay_snap}
\end{figure*}
\Figref{f:Tay_ener} displays the temporal evolution of the
toroidal and poloidal magnetic energies
of the spherical harmonic orders $0\leq m \leq 10$ for run U4.
The nonaxisymmetric perturbations
grow exponentially after about 2 system rotations
from the perturbation time $t_\text{pert}=33.2$.
The most unstable azimuthal mode is $m=1$ (solid purple lines)
as expected for TI.
Its linear growth rate, calculated over the time interval
$t-t_\text{pert}=2-4$ of the toroidal and poloidal
magnetic energy evolution, is $\gamma_\text{max}/\Omega_\text{a}\approx 1.9$
and 1.6 respectively (black dashed lines).

The instability saturates at $t-t_\text{pert}\approx 12$
when the rms nonaxisymmetric toroidal field strength
is roughly 5 times lower than the one of the
axisymmetric toroidal field $\overline{B}_\phi$
and about 2 times higher than the nonaxisymmetric poloidal one.
We observe the generation of axisymmetric poloidal field $\overline{B}_\text{p}$
due to the flow and field instability fluctuations at a rate which
is roughly twice the one of the $m=1$ mode (solid black line in \figref{f:Tay_ener}b).
After saturation, the instability decays on longer
times not shown in \figref{f:Tay_ener}.

During its linear growth, the instability is localized
close to the poles,
but also around the central equatorial regions
of the fluid domain (\figref{f:Tay_snap}a,b).
For the axisymmetric azimuthal
field configuration $\overline{B}_\phi$ that we perturbed (\figref{f:Tay_snap}c),
the instability condition (\ref{e:GT80})
predicts, in addition to the classical wedge-shaped regions
around the poles, several other unstable locations dispersed
at lower latitudes, which originate from locally high latitudinal gradients of the field
(gray shaded areas in \figref{f:Tay_snap}d).
The unstable central equatorial regions
correlate with the locations where the expected
growth rates $\gamma_\text{max}^\text{TI}$ from Eq.~(\ref{e:gamTI})
are larger (\figref{f:Tay_snap}d)
and where the instability is also observed.
At these locations $\gamma_\text{max}^\text{TI}/\Omega_\text{a}\approx 1.3$,
which agrees well with the growth rate $\gamma_\text{max}$
of the $m=1$ mode observed in the simulation run.

TI is also observed in the unstratified run at $\text{Re}=10^4$
and $\text{Ha}_\phi^\text{max}=9787$ (\figref{f:ReHa}),
for which we obtained similar results in the analysis of
the linear instability growth that we therefore do not
discuss here.
\newpage
\section{Free azimuthal flow evolution of the fiducial dynamo run U0} 
\label{s:appendix_no_forc}
Here we confirm that a net outward
transport of AM is induced by AMRI, as already
anticipated by the analysis of \secref{s:AM},
performing a numerical experiment
where the azimuthal flow is free to evolve.
\begin{figure}[h]
\centering
\resizebox{0.65\hsize}{!}{\includegraphics{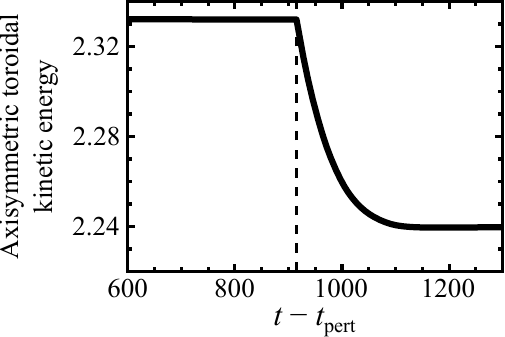}}
\caption{
Axisymmetric toroidal kinetic energy as a function of time
for the unstratified fiducial dynamo run U0
when the azimuthal flow is free to evolve.
The azimuthal flow forcing is stopped ($\textbf{f}=\textbf{0}$) at time $t-t_\text{pert}=915.0$
which is marked by the vertical dashed line.}
\label{f:noforc_ekin}
\end{figure}

\begin{figure}
\centering
\resizebox{0.9\hsize}{!}{\includegraphics{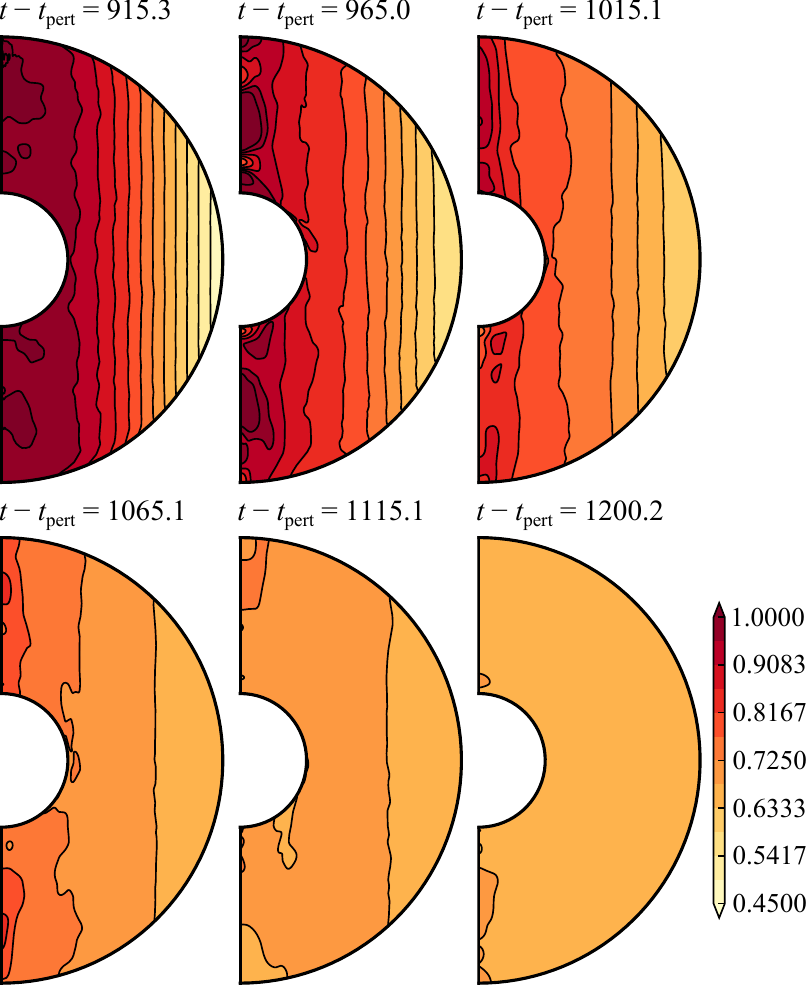}}
\caption{
Snapshots of the azimuthally averaged angular velocity $\overline{\Omega}$
during the free azimuthal flow evolution of run U0 ($t-t_\text{pert}>915$) shown in \figref{f:noforc_ekin}.
}
\label{f:noforc_snapsh}
\end{figure}

\Figref{f:noforc_ekin} displays part of the temporal evolution of
the axisymmetric toroidal kinetic energy
in the quasi steady state of the unstratified fiducial dynamo run U0
($t-t_\text{pert}<915$).
At $t-t_\text{pert}=915.0$ (vertical dashed line), the
azimuthal body force $\textbf{f}$
in the momentum equation (\ref{e:NS}) is set to zero.
The azimuthal flow relaxes to a state of uniform rotation
in a few hundreds of rotation times $\tau_\Omega=1/\Omega_\text{a}$.
\Figref{f:noforc_snapsh} demonstrates that the initial cylindrical profile
of the azimuthally averaged angular velocity
$\overline{\Omega}$ flattens over time under the influence of AMRI.
The flow decelerates
in the interior and accelerates
in the outer regions, producing a net outward transport of AM, and reaches
a state of almost rigid rotation
at $t-t_\text{pert}\approx 1200$.
\end{appendix}

\end{document}